%% file: main.tex
\documentclass[pra,aps,floatfix,twocolumn,amsmath,amssymb]{revtex4-2}
\usepackage{graphicx}
\usepackage{color}
\usepackage{bm}
\usepackage{hyperref}
\hypersetup{colorlinks=true,citecolor=blue,linkcolor=blue,urlcolor=blue}
\usepackage[T1]{fontenc}
\usepackage{newtxtext,newtxmath}
\allowdisplaybreaks

\begin{document}

\setcitestyle{super}

\title{%
Tensor-network study of correlation-spreading dynamics
in the two-dimensional Bose-Hubbard model}

\author{Ryui Kaneko}
\email{rkaneko@phys.kindai.ac.jp}
\affiliation{%
Department of Physics, Kindai University, Higashi-Osaka, Osaka 577-8502, Japan}

\author{Ippei Danshita}
\email{danshita@phys.kindai.ac.jp}
\affiliation{%
Department of Physics, Kindai University, Higashi-Osaka, Osaka 577-8502, Japan}

\date{\today}
 
\begin{abstract}
Recent developments in analog quantum simulators based on cold atoms and
trapped ions call for cross-validating the accuracy of
quantum-simulation experiments with use of quantitative numerical
methods; however, it is particularly challenging for dynamics of systems
with more than one spatial dimension.
Here we demonstrate that a tensor-network method running on classical
computers is useful for this purpose.
We specifically analyze real-time dynamics of the two-dimensional
Bose-Hubbard model after a sudden quench starting from the Mott
insulator by means of the tensor-network method based on infinite
projected entangled pair states.
Calculated single-particle correlation functions are found to be in good
agreement with a recent experiment.
By estimating the phase and group velocities from the single-particle
and density-density correlation functions, we predict how these
velocities vary in the moderate interaction region, which serves as a
quantitative benchmark for future experiments and numerical simulations.
\end{abstract}


\maketitle


\section{Introduction}

State-of-art experimental platforms of cold atoms and trapped ions as
analog quantum simulators have offered unique opportunities for studying
far-from-equilibrium dynamics of isolated quantum many-body systems.
Thanks to their high controllability and long coherence time, these
platforms have already addressed a variety of intriguing phenomena that
are in general difficult to simulate with classical computers, such as
correlation spreading~\cite{cheneau2012,jurcevic2014,takasu2020} and
relaxation \cite{trotzky2012,langen2015,kaufman2016} after a quantum
quench, many-body localization in a disorder
potential~\cite{schreiber2015,choi2016,smith2016}, and quantum scar
states~\cite{bernien2017,turner2018}. Nevertheless, accurate numerical
methods using classical computers are highly demanded at the current
stage of the studies of quantum many-body dynamics, since the classical
computation still has complementary advantages over the quantum
simulation in that it is free of noise and much more accessible owing to
its wide dissemination. In this sense, it is important to cross-check
the validity of quantum-simulation experiments and some numerical
methods by comparing them with each other.

\begin{figure}[t]
\centering
\includegraphics[width=.95\columnwidth]{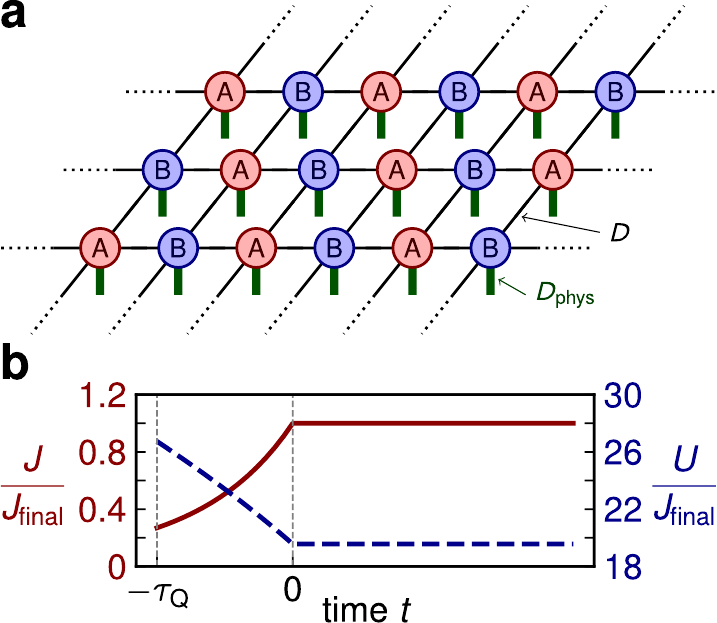}
\caption{%
Setup for numerical simulations of quench dynamics.
(a) Schematic figure of 
infinite projected entangled pair state
(iPEPS)
with a two-site unit cell.
Sublattice sites are represented by $A$ and $B$.
A rank-five tensor at each site is represented as a circle
with four thin lines and one thick line.
The former lines correspond to the virtual degrees of freedom
with the bond dimension $D$,
while the latter line corresponds to the physical degrees of freedom
with the
dimension of the local Hilbert space $D_{\rm phys}$.
The wave functions become more accurate as $D$ increases.
(b) Time dependence of the hopping
$J$ (a red solid line)
and the onsite interaction
$U$ (a blue dashed line)
with a finite-time quench.
The parameter $U/J$ is varied
from $\sim 99.4$ to $\sim 19.6$ for $-\tau_{\rm Q}<t<0$
with $\tau_{\rm Q}$ being a finite quench time.
In the case of a sudden quench, we discard the region
$-\tau_{\rm Q}<t<0$.
}
\label{fig:quench_peps}
\end{figure}

In particular, direct comparisons between experimental and numerical
outputs have been made for dynamical spreading of two-point spatial
correlations of the Bose-Hubbard
model~\cite{cheneau2012,takasu2020,barmettler2012,nagao2019,
nagao2021,mokhtarijazi2021},
which can be realized experimentally with ultracold bosons in optical
lattices~\cite{greiner2002}. The correlation spreading has attracted much
theoretical
interest~\cite{lauchli2008,schuch2011,barmettler2012,carleo2014,
cevolani2018,fitzpatrick2018,despres2019,nagao2019,nagao2021,
wang2020,kuwahara2021,mokhtarijazi2021,
yin2021_arxiv,faupin2021a_arxiv,faupin2021b_arxiv}
in the sense that
it is closely related to fundamental phenomena, including the
propagation of quantum information and the thermalization. In one
spatial dimension, quasi-exact numerical methods based on matrix product
states (MPSs) have been used to validate the performance of the quantum
simulators~\cite{cheneau2012,takasu2020,barmettler2012}. In two
dimensions (2D), by contrast, accurate numerical simulations are
challenging. Indeed, the comparisons with respect to a single-particle
correlation have shown that a few types of the truncated Wigner
approximation (TWA) fail to capture the real-time evolution accurately
enough to extract the propagation velocity of the
correlation~\cite{takasu2020,nagao2021}. 
Moreover, while the propagation velocities obtained by a two-particle
irreducible strong-coupling (2PISC) approach quantitatively agree with
the experimental value,
and the approach
is applicable to much weaker interaction than in the experiment,
it does not necessarily provide the exact value of the correlation
itself~\cite{mokhtarijazi2021}.

In this paper,
we present quantitative numerical analyses of the
correlation-spreading dynamics of the 2D Bose-Hubbard model starting
from a Mott insulating initial state with unit filling. To this end, we
employ the tensor-network method based on
the infinite projected entangled pair state
(iPEPS)~\cite{martin-delgado2001,verstraete2004a_arxiv,verstraete2004b,
verstraete2008,jordan2008,phien2015,orus2014,orus2019}
or the tensor product state~\cite{hieida1999,okunishi2000,nishino2001,
maeshima2001,nishio2004_arxiv}, which is an extension of MPS to 2D systems
[see Fig.\ref{fig:quench_peps}(a)].
The iPEPS studies on real-time dynamics of
isolated~\cite{kshetrimayum2017,
czarnik2019,hubig2019,
kshetrimayum2020,kshetrimayum2021,hubig2020,
alhambra2021,schmitt2021_arxiv,dziarmaga2021}
and open~\cite{kshetrimayum2017,
czarnik2019,weimer2021,keever2021,kilda2021}
quantum systems
in 2D have begun very recently.
Previous simulations suggest that
iPEPS can represent
relatively low-entangled states in short-time dynamics
for simple spin $1/2$
systems~\cite{kshetrimayum2017,
czarnik2019,hubig2019,
kshetrimayum2020}
and some itinerant electron
systems~\cite{hubig2020}.
This observation may be valid for real-time dynamics
in Bose-Hubbard systems;
however, little is known about it until now.
We find that the single-particle correlation computed with iPEPS, as
well as the estimated propagation velocity of the correlation front,
agrees very well with the experimental result~\cite{takasu2020},
demonstrating that iPEPS can be useful for actual
quantum-simulation experiments.
We also conduct numerical simulations in a moderate interaction region,
which has not been addressed by the previous
experiments~\cite{cheneau2012,takasu2020}. From the real-time evolution
of the single-particle and density-density correlations, we show that
the phase and group velocities approach each other when the interaction
decreases.

\section{Results}

\subsection{Model}

We consider the Bose-Hubbard model
on a square lattice~\cite{fisher1989,Jaksch1998}.
The Hamiltonian is given as
\begin{equation}
 \hat{H} =
 -J \sum_{\langle ij \rangle}
 (\hat{a}^{\dagger}_i \hat{a}_j + \hat{a}^{\dagger}_j \hat{a}_i)
 + \frac{U}{2} \sum_{i}
 \hat{n}_i (\hat{n}_i - 1)
 - \mu \sum_{i} \hat{n}_i,
\end{equation}
where $\hat{a}^{\dagger}_i$ and $\hat{a}_i$
are the creation and annihilation operators at site $i$,
$\hat{n}_i=\hat{a}^{\dagger}_i\hat{a}_i$ is
the number operator,
$J$ is the strength of the hopping between nearest-neighbor sites,
$U$ is the strength of the onsite interaction,
and $\mu$ is the chemical potential.
The notation $\langle ij \rangle$ indicates that
sites $i$ and $j$ are nearest neighbors.
For simplicity, we ignore the effects
of the trap potential and the Gaussian envelopes
of optical lattice lasers,
which do not affect short-time dynamics.
We set the lattice spacing $d_{\rm lat}$ to be unity.
The ground state at the commensurate filling is
the Mott insulating (superfluid) state
for $U\gg J$ ($U\ll J$).
Hereafter,
we will consider a sudden quench and a quench with a short time
[see Fig.~\ref{fig:quench_peps}(b)
and Supplementary Note 1 for details].

\subsection{Quench starting from the Mott insulator:
Comparison with the exact diagonalization and the experiment}

\begin{figure}[t]
\centering
\includegraphics[width=.95\columnwidth]{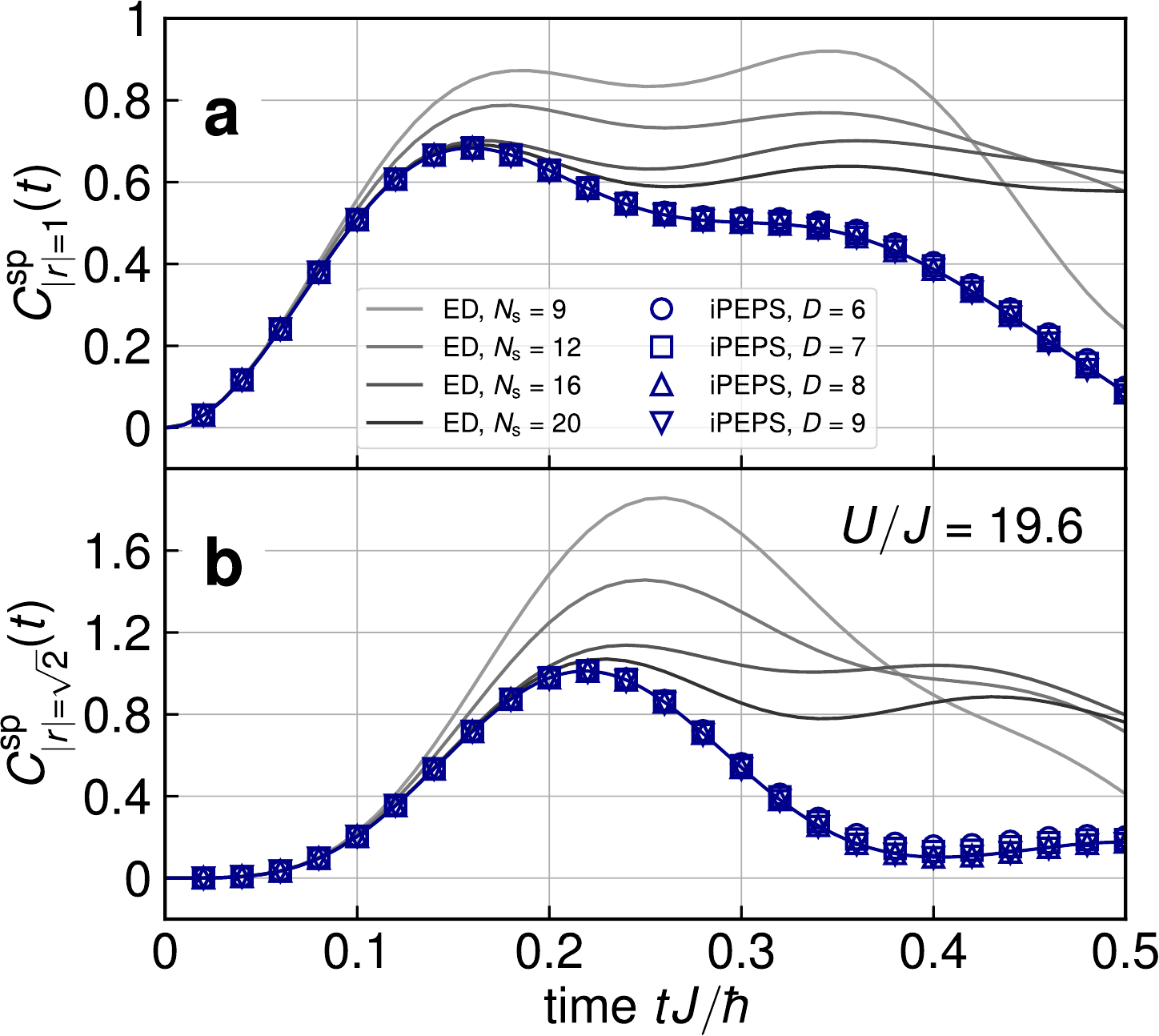}
\caption{%
Single-particle correlation functions $C^{\rm sp}_{\bm{r}}(t)$
in the case of a sudden quench.
Comparison is made
between the infinite projected entangled pair state algorithm
(iPEPS, blue lines with symbols)
and the exact diagonalization method (ED, gray lines).
The unit of time is taken as
the inverse of the strength of the hopping $J$.
$U$ is the strength of the interaction.
$D$ is the bond dimension.
$N_{\rm s}$ is the system size.
The correlations at distances
(a) $|\bm{r}|=1$
and
(b) $|\bm{r}|=\sqrt{2}$
are shown.
Both results overlap in a short time.
}
\label{fig:comp_ed}
\end{figure}

Let us first focus on the case of a sudden quench.
We compare our results of iPEPS with those of the exact diagonalization
(ED) method
and obtain consistent results in a short time.
In the ED simulations using the QuSpin
library~\cite{weinberg2017,weinberg2019},
we choose the system sizes $L_x\times L_y$ up to $5\times 4$
and use the periodic-periodic boundary condition.
We examine to what extent the energy is conserved
in the iPEPS simulations.
The grand potential density $\langle \hat{H}\rangle$
at $T=0$ starting from the Mott insulator
$\otimes_i |n_i=1\rangle$ should remain constant.
They well converge for the bond dimensions $D\ge 6$
and remain nearly constant up to
$t\sim 0.4\hbar/J$
(see Supplementary Note 2
for the time dependence of the grand potential density).
We also investigate how the single-particle correlations converge
with increasing bond dimensions.
The equal-time single-particle correlation function
at a distance $\bm{r}=(x,y)$ for the system size $N_{\rm s}$
is defined as
\begin{equation}
 C^{\rm sp}_{\bm{r}}(t) =
 \frac{1}{2N_{\rm s}}
 \sum_{i,j}{\vphantom{\sum}}'
 \langle \hat{a}^{\dagger}_{i}(t) \hat{a}_{j}(t)
 + \hat{a}^{\dagger}_{j}(t) \hat{a}_{i}(t) \rangle.
\end{equation}
Here $\sum_{i,j}'$ denotes
the summation over $(i,j)$ that satisfies
$|x_j-x_i|=x$ and $|y_j-y_i|=y$.
In the iPEPS simulations,
$1/N_{\rm s}\times\sum_{i,j}'$ is replaced by
$1/2\times\sum_{i=A,B}\sum_{j}'$
with $A$ and $B$ being sublattice sites
because of the translational invariance.
As shown in Fig.~\ref{fig:comp_ed},
$C^{\rm sp}_{|\bm{r}|=1}(t) :=
[C^{\rm sp}_{\bm{r}=(1,0)}(t) +
C^{\rm sp}_{\bm{r}=(0,1)}(t)]/2$
exhibits a peak at $t\sim 0.15\hbar/J$ in both results,
and they overlap
in this short time.
For $t\gtrsim 0.15\hbar/J$,
the correlation functions of ED
start to exhibit a significant finite-size effect,
whereas those of iPEPS converge for $D\ge 6$.
We observe similar behavior
for $C^{\rm sp}_{|\bm{r}|=\sqrt{2}}(t) := 
C^{\rm sp}_{\bm{r}=(1,1)}(t)$.
The iPEPS results are better simulated up to a longer time
(see also Supplementary Note 3 for other interaction parameter regions).

\begin{figure}[t]
\centering
\includegraphics[width=.95\columnwidth]{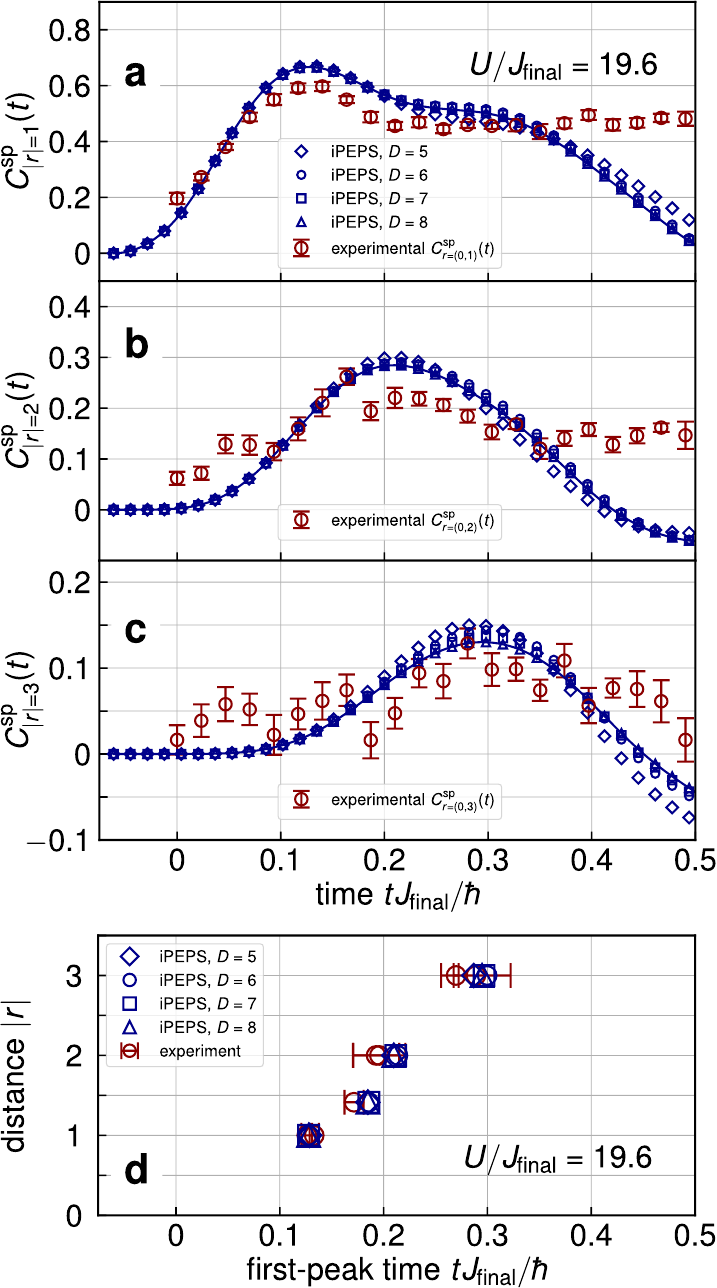}
\caption{%
Single-particle correlation functions $C^{\rm sp}_{\bm{r}}(t)$
in the case of a finite-time quench.
Comparison is made
between the infinite projected entangled pair state algorithm
(iPEPS, blue lines with symbols)
and the experiment (red circles with error bars).
The unit of time is taken as
the inverse of the strength of the hopping $J_{\rm final}$
after the quench.
$U$ is the strength of the interaction.
$D$ is the bond dimension.
The correlations at distances
(a) $|\bm{r}|=1$,
(b) $|\bm{r}|=2$,
and
(c) $|\bm{r}|=3$ 
are shown.
The error bars represent the standard error of
five independent measurements~\cite{takasu2020}.
(d) Comparison of the first-peak time
between the iPEPS and experimental results.
The error bars represent the fitting errors~\cite{takasu2020}.
The iPEPS and experimental results agree within the experimental errors
in all cases.
}
\label{fig:comp_exp}
\end{figure}

Next, we compare the correlations of iPEPS with
those of the experiment~\cite{takasu2020}
for a finite quench time.
In the experiment, a quench to the Mott insulating region
has been investigated so far.
Figures~\ref{fig:comp_exp}(a--c) show
the time evolution of correlations at distances
$|\bm{r}|=1$, $2$, and $3$, respectively.
Qualitative behavior is essentially equivalent
to the case of the sudden quench,
although the correlation function
shifts to an earlier time.
For $|\bm{r}|=1$, both data show a peak
at $t\sim 0.12\hbar/J_{\rm final}$.
Similarly,
the first-peak times are consistent with each other
for $|\bm{r}|=2$ and $3$,
and they become longer with increasing distances.
When the energy is approximately conserved
(namely, for $t\lesssim 0.4\hbar/J_{\rm final}$,
see the time dependence of the grand potential density
in Supplementary Note 4),
the intensities of correlations also overlap very well.
They are also consistent with
those obtained by
TWA~\cite{nagao2019,takasu2020,nagao2021},
while the iPEPS simulations can deal with a slightly longer time
and capture the correlation peaks more clearly
(see also Supplementary Note 5
for a detailed comparison with the TWA results).
To see how well they match more quantitatively,
we also compare the first-peak position of iPEPS
with that of the experiment~\cite{takasu2020}
as shown in Fig.~\ref{fig:comp_exp}(d).
Both iPEPS and experimental results agree very well.

\subsection{Estimates of group and phase velocities
in the moderate interaction region}

Having confirmed the applicability of iPEPS simulations
to real-time evolution of the Bose-Hubbard model,
we study how information propagates by a sudden quench
in the moderate interaction region.
There are two kinds of velocity that are relevant to the correlation spreading.
One is the group velocity $v_{\rm gr}$,
which corresponds to the propagation of the envelope
of the wave packet and is a suitable quantity
to characterize the spreading of correlations. 
In non-relativistic quantum many-body systems,
$v_{\rm gr}$ is bounded above, and the upper bound is
known as the Lieb-Robinson bound~\cite{lieb1972,
hastings2010_arxiv,wang2020,kuwahara2021,
yin2021_arxiv,faupin2021a_arxiv,faupin2021b_arxiv}.
Notice that the Lieb-Robinson bound for the Bose-Hubbard model
has not been rigorously derived with a few exceptions
for limited situations~\cite{schuch2011,wang2020,kuwahara2021,
yin2021_arxiv,faupin2021a_arxiv,faupin2021b_arxiv}.
The phase velocity $v_{\rm ph}$
is the other characteristic quantity,
which corresponds to the propagation of the first peak
of the wave packet,
and does not have to obey the Lieb-Robinson bound.

Although the exact Lieb-Robinson bound is not known
for the Bose-Hubbard model,
there are some values that can be used as a guide.
As discussed in
previous studies~\cite{cheneau2012,takasu2020,barmettler2012}
in the weak interaction region,
the single-particle dispersion
up to constant is approximately given as
$\epsilon_{U\ll J}(\bm{k}) \sim -2J\sum_{\alpha}\cos k_{\alpha}$
($\alpha=x,y$ in 2D),
which is equivalent to the dispersion of free particles.
The velocity of the correlation spreading
would be well characterized by
the group velocity of the single-particle excitation.
The largest velocity of a single quasiparticle
(along the horizontal or vertical direction $\alpha$)
is described by the maximal slope of the dispersion
and is given by
$v = \max_{k_{\alpha}}
[ d|\epsilon_{U\ll J}(\bm{k})|/dk_{\alpha} ] / \hbar = 2J/\hbar$.
Because both doublon and holon quasiparticles
propagate with the group velocity $v$,
the front of the correlation function
moves at the speed of $v_{\rm front}$,
which should be smaller than
$v_{\rm max} = 2v = 4J/\hbar$.
Therefore, this speed $v_{\rm max}$ can be regarded as
the Lieb-Robinson-bound-like value.
Likewise, in the strong interaction region,
the doublon and holon dispersions
up to constant are approximately given as
$\epsilon^{(d)}_{U\gg J}(\bm{k}) \sim
-4J\sum_{\alpha}\cos k_{\alpha}$
and
$\epsilon^{(h)}_{U\gg J}(-\bm{k}) \sim
-2J\sum_{\alpha}\cos k_{\alpha}$,
respectively.
Because the doublons and holons propagate
with respective velocities $4J/\hbar$ and $2J/\hbar$,
$v_{\rm front}$ should be smaller than these two sum
$v_{\rm max} = 6J/\hbar$.
Although we know the approximate limit values,
the intermediate interaction region is yet to be explored.

\begin{figure}[t]
\centering
\includegraphics[width=.95\columnwidth]{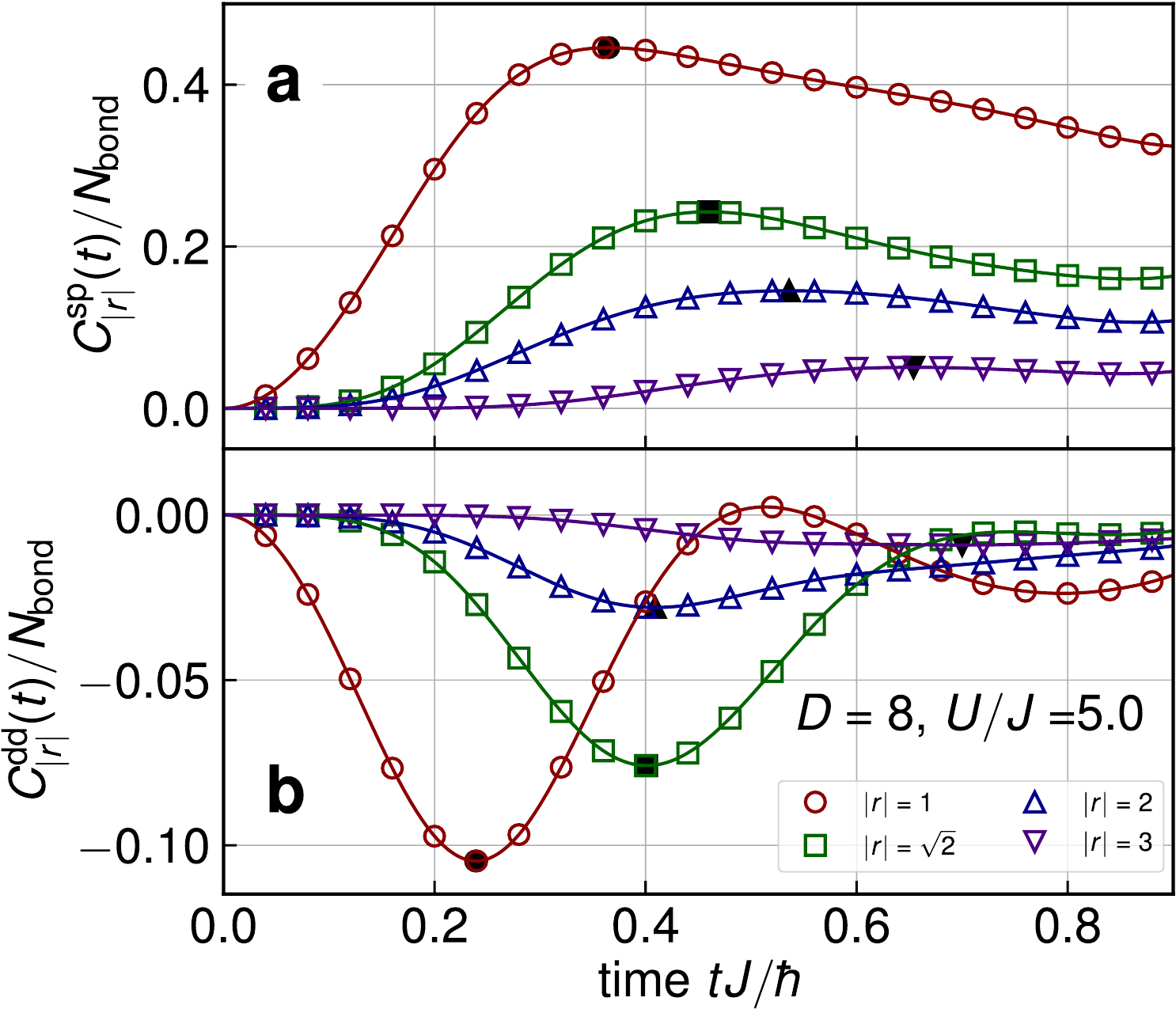}
\caption{%
Single particle and density-density correlation functions used to
extract the propagation velocities.
(a) Single-particle [$C^{\rm sp}_{\bm{r}}(t)$]
and (b) density-density [$C^{\rm dd}_{\bm{r}}(t)$]
correlation functions per bond 
at the interaction strength $U/J=5$
for the bond dimension $D=8$.
The unit of time is taken as
the inverse of the strength of the hopping $J$.
The normalization factor at a distance
$\bm{r}=(x,y)$ is given as
$N_{\rm bond}=2$ for $x\not=y$
($|\bm{r}|=1$, $2$, and $3$),
while it is
$N_{\rm bond}=4$ for $x=y$
($|\bm{r}|=\sqrt{2}$).
The black symbol corresponds to the first peak in the correlation
function obtained by cubic spline interpolation of data points.
The propagation velocities
along the horizontal or vertical axis
are extracted by the data at $|\bm{r}|=1$, $2$, and $3$.
The velocity estimated from
the density-density correlation functions is slower
than that from the single-particle correlation functions.
}
\label{fig:correlations}
\end{figure}

To estimate the group velocity from the single-particle correlations,
long-time simulations are required in general.
However, it is challenging in the iPEPS simulations.
To circumvent the difficulty,
we estimate the group velocity by
the density-density correlation.
It is known that the propagation velocity of the first peak
of this correlation agrees very well with the group
velocity~\cite{cheneau2012,barmettler2012}.
The equal-time density-density correlation function
at a distance $\bm{r}=(x,y)$
for the system size $N_{\rm s}$ is defined as
\begin{equation}
 C^{\rm dd}_{\bm{r}}(t) = 
 \frac{1}{N_{\rm s}}
 \sum_{i,j}{\vphantom{\sum}}'
 \langle \hat{n}_i(t) \hat{n}_j(t) \rangle_{\rm c},
\end{equation}
where
$\langle\cdots\rangle_{\rm c}$ denotes
a connected correlation function.
In our simulations,
$\langle \hat{n}_i(t) \hat{n}_j(t) \rangle_{\rm c}
= \langle \hat{n}_i(t) \hat{n}_j(t) \rangle - 1$
because $\langle \hat{n}_i(t) \rangle = 1$
for all sites and time steps.
As in $C^{\rm sp}_{\bm{r}}(t)$,
the summation is replaced by that within
sublattice sites in the iPEPS simulations.
The parity-parity correlation closely related to
the density-density one can be measured in experiments
by using the quantum-gas microscope techniques~\cite{cheneau2012}.

We extract the propagation velocities from the first peak
in both correlations for $|{\bm r}|=1$, $2$, and $3$.
For simplicity, we consider the sudden quench hereafter.
When the interaction becomes weaker,
we have confirmed that the energy is conserved in a longer time frame;
typically, $t\lesssim 0.9\hbar/J$ for $U/J\sim 5$
(see Supplementary Note 6
for the time dependence of the grand potential density
in the weaker interaction region).
All the correlation peaks for $|\bm{r}|\le 3$ appear in this time frame
(see Fig.~\ref{fig:correlations}).
The first peak of the single-particle correlation appears
at $t\sim 0.35\hbar/J$ for $|\bm{r}|=1$,
while it appears at $t\sim 0.65\hbar/J$ for $|\bm{r}|=3$.
By contrast,
the first peak of the density-density correlation appears
at $t\sim 0.25\hbar/J$ for $|\bm{r}|=1$,
while it appears at $t\sim 0.7\hbar/J$ for $|\bm{r}|=3$.
It takes a long time for propagation in the latter case.
(See also the correlations for other interaction parameters
given in Supplementary Note 7.
Extraction of propagation velocities
in the intermediate and strong interaction regions is
summarized in Supplementary Notes 8 and 9, respectively.)
The first-peak time is almost a linear function of the
distance,
and the system exhibits the light-cone-like spreading of
correlations
(see the time dependence of distance summarized in
Supplementary Note 8).

\begin{figure}[t]
\centering
\includegraphics[width=.95\columnwidth]{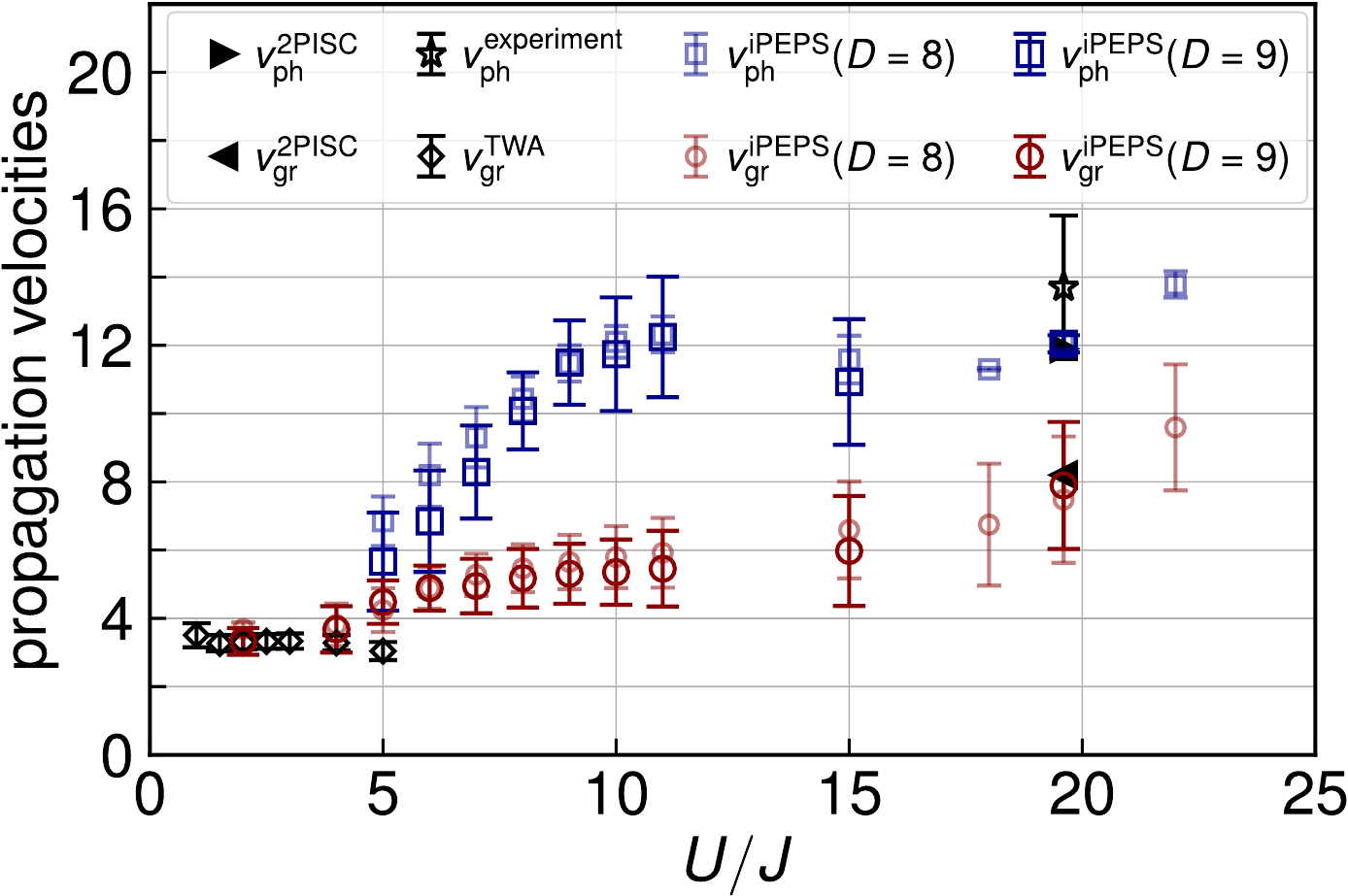}
\caption{%
Propagation velocities
as functions of the ratio between
the interaction ($U$) and hopping ($J$) strengths.
The group
($v_{\rm gr}$, red circles)
and phase
($v_{\rm ph}$, blue squares)
velocities are estimated from the density-density and single-particle
correlation functions
for the bond dimensions $D=8$ and $D=9$
using the infinite projected entangled pair state (iPEPS) algorithm.
The data for $D=8$ and $D=9$ overlap within the error bars.
The velocities and their error bars are obtained by extrapolation
of the distance dependence of the peak time.
The results obtained by
the two-particle irreducible strong-coupling (2PISC)
approach~\cite{mokhtarijazi2021} (triangles),
the truncated Wigner approximation
(TWA)~\cite{nagao2019} (diamonds),
and the experiment~\cite{takasu2020} (a star)
are
shown.
Both velocities gradually merge with decreasing interaction.
}
\label{fig:velocity}
\end{figure}

We summarize the interaction dependence of the group and phase
velocities in Fig.~\ref{fig:velocity}.
In the weak interaction region,
the estimated group velocities are $v_{\rm gr}\sim 4J/\hbar$.
They are similar to those obtained by the TWA
at filling factor $\nu=10$~\cite{nagao2019}.
They are also consistent with the group velocity
$v_{\rm gr}(U=0) = 4J/\hbar$ of a single
particle~\cite{nagao2019}.
In the strong interaction region,
the estimated group velocity
$v_{\rm gr}\sim (8\pm 2)J/\hbar$ at $U/J=19.6$
coincides with that obtained by the 2PISC
approach~\cite{mokhtarijazi2021}
within the error bar of extrapolation.
It is also comparable to the group velocity
$v_{\rm gr}(U\gg J) = 6J/\hbar
\times [1+\mathcal{O}(J^2/U^2)]$
of a quasi-particle in the large $U$
limit~\cite{cheneau2012,barmettler2012,takasu2020}.
Similarly,
the estimated phase velocity agrees very well
with the results of the 2PISC
approach~\cite{mokhtarijazi2021}
and the experiment~\cite{takasu2020}.
In the intermediate region,
the estimated group velocity is closer to
the single-particle group velocity in the superfluid region,
whereas it is comparable to the 2PISC result
near and above the critical point
$U_{\rm c}/J\sim 16.7$~\cite{elstner1999,capogrosso-sansone2008,
krutitsky2016}.

In all parameter regions, no anomalies appear in the propagation
velocities.
As for the real-time dynamics after a sudden quench,
there is no sign of the superfluid-Mott insulator quantum phase
transition.
This is because
non-universal high-energy excitations come into play
during the time evolution.
The quantum phase transition at zero temperature
does not have to affect the time-evolved states.

Both group and phase velocities gradually converge to
the same value as $U/J$ is decreased.
This phenomenon can be understood in terms of
the separation of the energy scales.
When the interaction $U$ is much stronger than the hopping $J$,
the correlation function oscillates rapidly as a function of
time~\cite{barmettler2012,despres2019}.
The correlation function exhibits the envelope of the wave packet.
The time scale of the period of oscillation is $\sim 1/U$,
which determines the phase velocity $v_{\rm ph} \sim U$.
On the other hand,
the time scale of the period of the envelope is $\sim 1/J$,
which determines the group velocity $v_{\rm gr} \sim J$.
Hence,
the group and phase velocities differ as long as $U\gg J$.
When the interaction $U$ becomes comparable to the hopping $J$,
they start to coincide by slowing down the vibration.
Note that this phenomenon occurs
irrespective of the presence or absence of phase transitions.

\section{Conclusions}

We have studied real-time dynamics of the 2D
Bose-Hubbard model after a sudden quench starting from the Mott
insulator with unit filling.
We have employed the 2D tensor-network method
based on the
iPEPSs,
which are the 2D extension of
the well-known
MPSs
in one dimension.
Calculated single-particle correlation functions
reproduce the recent experimental results very well.
The iPEPS algorithm can simulate real-time dynamics
long enough for extracting the propagation velocities
from correlations.
This fact suggests that,
for the quench dynamics starting from the Mott insulator
in the 2D Bose-Hubbard model,
time-evolved states are not so highly entangled
before and 
even slightly after
the time 
at which
the correlation front is reached.
This finding raises questions about
our understanding of how quantum states get entangled
with real-time evolution.

We have also estimated the group and phase velocities
in the moderate interaction region,
in which the 2PISC
approach and the TWA
are not applicable.
The estimated group velocities are continuously
connected without singularity in the middle.
Our findings would be useful
in the future analog quantum simulation
and
in the future examination of
the rigorous Lieb-Robinson bound
of Bose-Hubbard systems.
The ability of the tensor-network method that accurately calculates
the real-time dynamics of 2D quantum many-body systems
opens up the possibility of applying it
to other quantum-simulation platforms,
such as Rydberg atoms, trapped ions, and superconducting circuits.

\section{Methods}

\subsection{Real-time evolution by
infinite projected entangled pair states}

We prepare iPEPS with a two-site unit cell
[see Fig.~\ref{fig:quench_peps}(a)].
The symbols $D$ and $D_{\rm phys}$
denote
the virtual bond dimension
and
the dimension of the local Hilbert space,
respectively.
The former improves the accuracy of the wave function,
whereas the latter corresponds to
the maximum particle number 
$n_{\rm max}$ as $D_{\rm phys} = n_{\rm max}+1$.
Although $n_{\rm max}$ can take infinity in Bose-Hubbard systems,
it is practically bounded above
in the presence of interaction~\cite{huber2007,davidson2015}.
We can choose finite $D_{\rm phys}$
in the simulations of real-time dynamics.
In the case of a sudden quench to the Mott
insulating region
($U/J > U_{\rm c}/J \sim
16.7$~\cite{elstner1999,capogrosso-sansone2008,
krutitsky2016}),
we set the dimension of the local Hilbert space
as $D_{\rm phys}=3$
because the number of particles deviates only slightly
from unity~\cite{huber2007,davidson2015}.
For $U/J < U_{\rm c}/J$, we choose $D_{\rm phys}=5$
so that the wave functions can further take into account
the effect of particle fluctuations.
When $U$ is close to zero
(at $U/J=2$ in our simulations),
we use slightly larger $D_{\rm phys}=7$
(see Supplementary Note 10
for the details of
the choice of the dimensions of the local Hilbert space). 
The initial Mott insulating state $\otimes_i |n_i=1\rangle$
can be represented 
with the bond dimension $D=1$.
As for static properties,
the Bose-Hubbard model was investigated by
finite PEPS or iPEPS,
and the phase transition between
the Mott insulating and superfluid phases
was reproduced~\cite{murg2007,jordan2009,kshetrimayum2019,
jahromi2019,jahromi2020,schmoll2020,tu2020,wu2020,
vlaar2021}.

The wave function at each time
$|\Psi(t)\rangle = e^{-it\hat{H}/\hbar}|\Psi(0)\rangle$
is obtained by real-time evolving
iPEPS~\cite{kshetrimayum2017,czarnik2019,hubig2019}.
The real-time evolution operator
in a small time step $dt$
can be approximated by the Suzuki-Trotter
decomposition~\cite{trotter1959,suzuki1966,suzuki1976}
as 
$e^{-idt\hat{H}/\hbar} \sim \prod_{\langle ij \rangle} e^{-idt\hat{H}_{ij}/\hbar}$,
where
$\hat{H}_{ij} =
 - J
 (\hat{a}^{\dagger}_i \hat{a}_j + \hat{a}^{\dagger}_j \hat{a}_i)
 +  U
 [ \hat{n}_i (\hat{n}_i - 1)
 + \hat{n}_j (\hat{n}_j - 1) ]
 / (2z)
 - \mu
 ( \hat{n}_i + \hat{n}_j )
 / z$
with the coordination number $z=4$
is the local Hamiltonian satisfying
$\hat{H} = \sum_{\langle ij \rangle} \hat{H}_{ij}$.
After applying the two-site gate
$e^{-idt\hat{H}_{ij}/\hbar}$
to neighboring tensors,
we approximate the local tensors
by the singular value decomposition
in such a way that
the virtual bond dimension of iPEPS remains $D$.
In the actual simulations,
the second-order Suzuki-Trotter decomposition
is used for this simple update
algorithm~\cite{jiang2008,jordan2008},
and the tensor-network library
TeNeS~\cite{motoyama2021_arxiv,tenes,ptns}
is adopted.
The wave functions are optimized
up to the bond dimension $D=9$.
Qualitative behavior of correlation functions is
found to be
nearly the same for $D\ge 6$.
When extracting the propagation velocities,
we mainly use the data for $D=8$ and $D=9$ to ensure
sufficient convergence of physical quantities.
We do not preserve the $\rm U(1)$ symmetry during the calculation.
Even without respecting the symmetry, we
have numerically found that at these values of $D$,
the number of particles is nearly conserved
during the real-time evolution starting from the Mott insulator.

Physical quantities in the thermodynamic limit
are calculated by
the corner transfer matrix renormalization group (CTMRG)
method~\cite{nishino1996,nishino1997,
nishino1999,okunishi2000,
orus2009,corboz2010,
corboz2011,corboz2014,phien2015,orus2014,orus2019}.
The bond dimension of the environment tensors
is chosen as $\chi=2D^2$ to ensure that
physical quantities are well converged.

To compare our results obtained by iPEPS
with the experiment~\cite{takasu2020},
we consider a quench with a short time
$\tau_{\rm Q}=
0.1\mathrm{ms}$~\cite{nagao2019,nagao2021}
[see Fig.~\ref{fig:quench_peps}(b)].
For $-\tau_{\rm Q}<t<0$,
both $J$ and $U$ are controlled.
The wave function is updated as
$|\Psi(t+dt)\rangle \sim e^{-idt\hat{H}(t)/\hbar} |\Psi(t)\rangle$
with the time-dependent Hamiltonian $\hat{H}(t)$
in this region.
For $t>0$, both parameters are fixed.
We take $J_{\rm final}=J(t=0)
\sim 0.0612
\hbar/\tau_{\rm Q}$ as the unit of energy.
The discrete time step for the real-time evolution
is set to be
$dt/(\hbar/J_{\rm final})
=\tau_{\rm Q}/(\hbar/J_{\rm final})/15
\sim 0.00408$
for all $t$.
To compare the iPEPS results with the exact real-time dynamics
in finite-size systems,
we also consider a sudden parameter change
and set the time step as $dt/(\hbar/J)=0.005$.
We have checked that
the simulations with doubled and halved $dt$
do not change the results significantly.


\section{Data availability}

The data obtained by the iPEPS and ED simulations
in this paper are available
at \texttt{https://doi.org/10.5281/zenodo.6085592}.
The experimental data~\cite{takasu2020}
and the data obtained by the TWA simulations~\cite{nagao2019}
in this paper are available from the authors upon request.

\section{Code availability}

The codes in this paper are available from the authors upon request.


\def\bibsection{\section*{\refname}}

\input{main.bbl.back}



 

\begin{acknowledgments}
We acknowledge fruitful discussions with S.\ Goto and K.\ Nagao.
We thank Y.\ Takahashi and Y.\ Takasu for useful discussions and the experimental data.
This work was financially supported by JSPS KAKENHI
(Grants Nos.\ JP18H05228, JP21H01014, and JP21K13855),
by JST CREST (Grant No.\ JPMJCR1673), by JST FOREST (Grant No.\ JPMJFR202T),
and by MEXT Q-LEAP (Grant No.\ JPMXS0118069021).
The numerical computations were performed on computers at
the Yukawa Institute Computer Facility and on computers at
the Supercomputer Center, the Institute for Solid State Physics,
the University of Tokyo.
\end{acknowledgments}


\section{Author contributions}

R.K.\ and I.D.\ designed and coordinated the studies.
R.K.\ performed the numerical simulations.
R.K.\ and I.D.\ contributed to the writing of the paper.

\section{Competing interests}

The authors declare no competing interests.

\section{Additional information}

{\bf Supplementary Information}
The online version contains supplementary material available at
\texttt{https://doi.org/10.1038/s42005-022-00848-9}.

\onecolumngrid

\end{document}


\title{%
Supplementary Information:
Tensor-network study of correlation-spreading dynamics
in the two-dimensional Bose-Hubbard model}

\author{Ryui Kaneko}
\email{rkaneko@phys.kindai.ac.jp}
\affiliation{%
Department of Physics, Kindai University, Higashi-Osaka, Osaka 577-8502, Japan}

\author{Ippei Danshita}
\email{danshita@phys.kindai.ac.jp}
\affiliation{%
Department of Physics, Kindai University, Higashi-Osaka, Osaka 577-8502, Japan}

\date{\today}


\maketitle

\tableofcontents


\section{Time dependence of the hopping and
the interaction during a finite-time quench}

We briefly review the detailed setup
in the experiment~\cite{takasu2020} and the time dependence
of the hopping and the interaction.
In the experiment, ultracold Bose gas of $^{174}{\rm Yb}$ atoms
is confined in the optical lattice with the lattice spacing
$d_{\rm lat}=266{\rm nm}$.

At the time $t = -\tau_{\rm Q} = -0.1{\rm ms}$,
the Mott insulator at $\nu=1$ is prepared
in the optical lattice with the depth $V_0=15E_{\rm R}$.
The interaction and the hopping are
$U(V_0=15E_{\rm R}) \sim 0.648 E_{\rm R}$
and
$J(V_0=15E_{\rm R}) \sim 0.00652 E_{\rm R}$,
respectively,
and their ratio is $U/J\sim 99.4$.
Here $E_{\rm R}/(2\pi\hbar) = 4021.18{\rm Hz}$
is the recoil energy of the system.

For $-\tau_{\rm Q}<t<0$,
the lattice depth is decreased rapidly to $V_0=9E_{\rm R}$.
The change of the depth is nearly a linear function of time,
namely, $V_0(t)/E_{\rm R} = 9 - 6t/\tau_{\rm Q}$.
Finally,
the interaction and the hopping become
$U(V_0=9E_{\rm R}) \sim 0.474 E_{\rm R}$
and
$J(V_0=9E_{\rm R}) \sim 0.0242 E_{\rm R}$,
respectively.
Their ratio becomes $U/J\sim 19.6$,
which corresponds to the Mott insulating region
close to the critical point
$U_{\rm c}/J \sim 16.7$~\cite{elstner1999,capogrosso-sansone2008,
krutitsky2016}.
In the iPEPS simulations,
we take $J_{\rm final} = J(V_0=9E_{\rm R})$
as the unit of energy.
The corresponding unit of timescale is
$\tau_{\rm unit} = \hbar/J_{\rm final} \sim 1.63 {\rm ms}$,
and the length of the quench time is expressed as
$\tau_{\rm Q} = 0.1{\rm ms} \sim 0.0612 \tau_{\rm unit}$.

For $t>0$, the hopping and the interaction are fixed.
The single-particle correlation functions are obtained
by measuring the time-of-flight interference pattern.

\begin{figure}[!]
\centering
\includegraphics[width=.95\columnwidth]{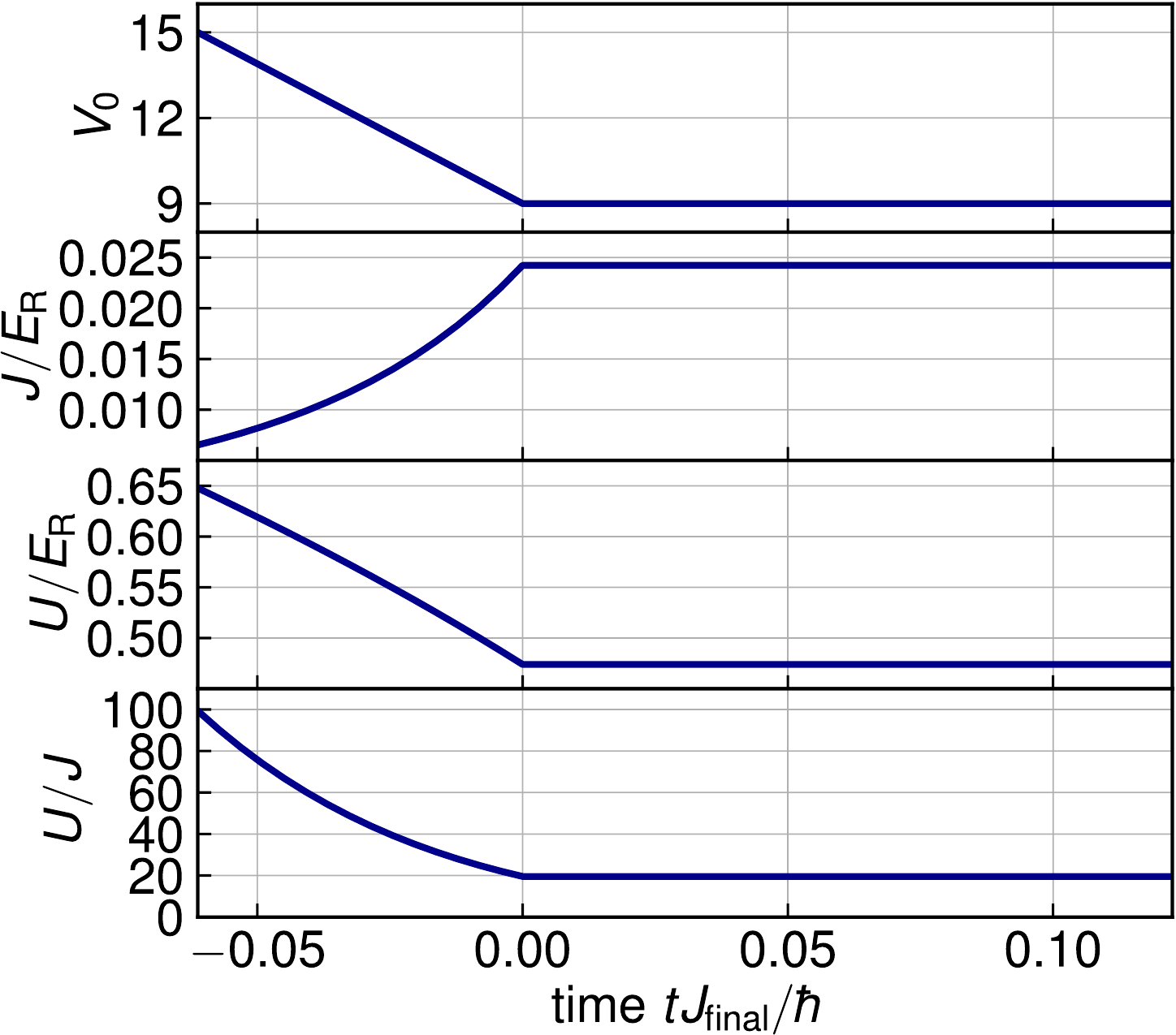}
\caption{Time dependence of
the depth $V_0$ of the optical lattice,
the hopping $J/E_{\rm R}$,
the interaction $U/E_{\rm R}$,
and the ratio $U/J$.
The recoil energy is $E_{\rm R}/(2\pi\hbar) = 4021.18{\rm Hz}$.
}
\label{fig:time_dep_param}
\end{figure}

All the time dependence of the hopping $J(t)$ and
the interaction $U(t)$ are summarized in
Fig.~\ref{fig:time_dep_param}.

\section{Further comparison with the exact diagonalization
for a sudden quench
to stronger interaction}

\begin{figure}[!]
\centering
\includegraphics[width=.95\columnwidth]{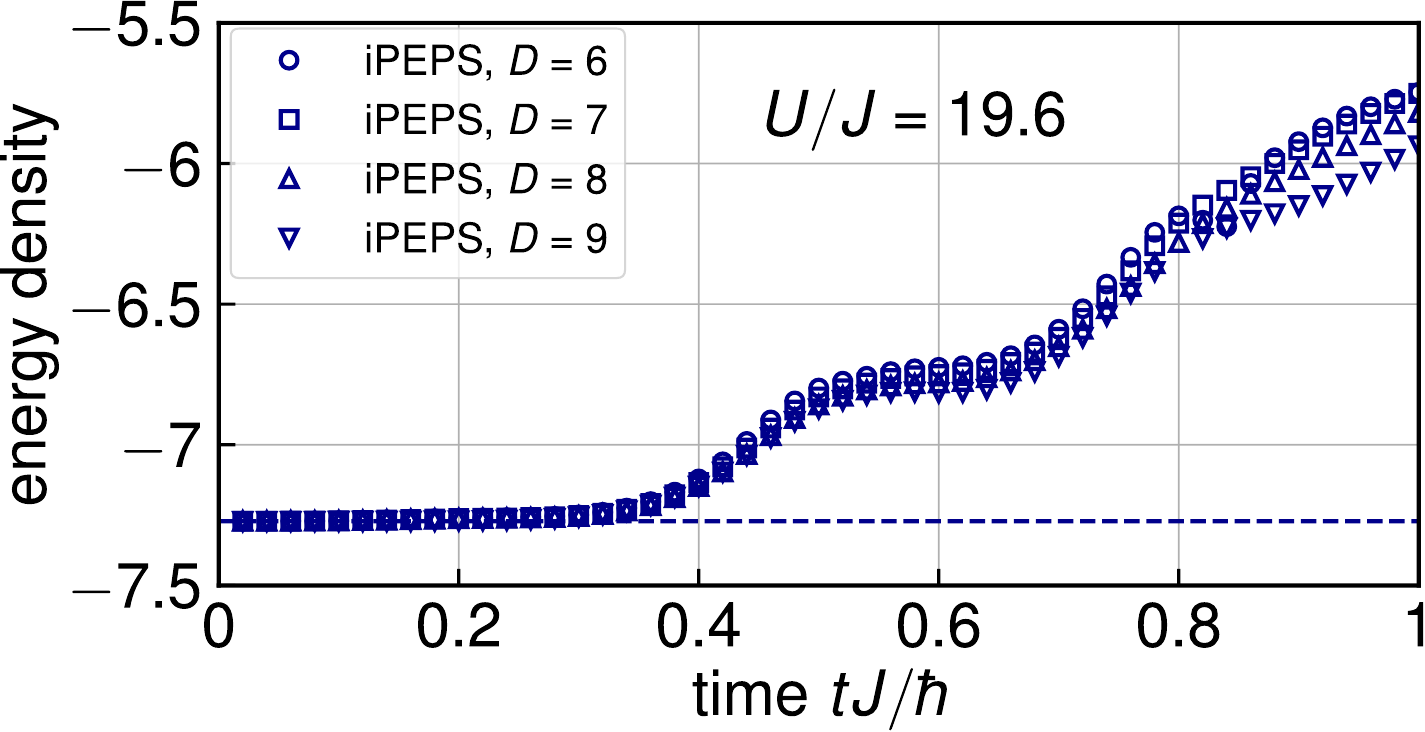}
\caption{Time dependence of the grand potential density
in the unit of hopping energy $J$
for a sudden quench
at $U/J=19.6$.
The energy density at $t=0$ is given as a dashed line.
The energy density is nearly conserved
in a short time $t\lesssim 0.4\hbar/J$.}
\label{fig:cons_ene}
\end{figure}

To examine to what extent the energy is conserved
in the case of a sudden quench at $U/J=19.6$,
we calculate the time dependence of the grand potential density
at $T=0$ obtained by iPEPS
in Fig.~\ref{fig:cons_ene}.
It is nearly constant
($e_{\rm MI}/J=-\tilde{\mu}U/J$ 
with
$\tilde{\mu}=\mu/U
=-0.371$~\cite{elstner1999,capogrosso-sansone2008,krutitsky2016})
for $t\lesssim 0.4\hbar/J$.
The data almost converge for the bond dimensions $D\ge 6$
in this short time.

\begin{figure}[!]
\centering
\includegraphics[width=.95\columnwidth]{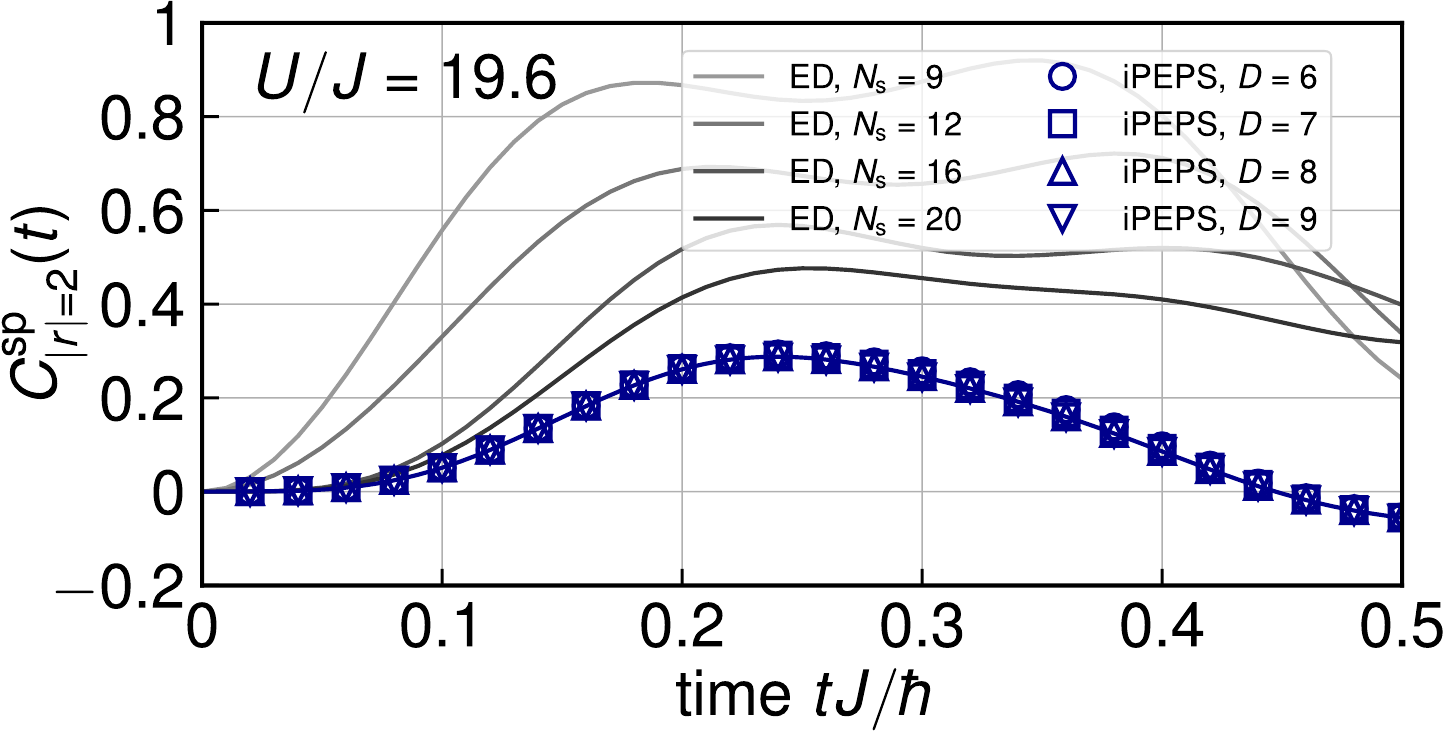}
\caption{Comparison of the single-particle correlation functions
in the case of a sudden quench
between iPEPS
(blue lines with symbols)
and ED
(gray lines).
The correlations at a distance $|\bm{r}|=2$ is shown.}
\label{fig:comp_ed_exp_r_2}
\end{figure}

We also compare the single-particle correlation functions
at a far distance $|\bm{r}|=2$
obtained by iPEPS with those obtained by ED
in Fig.~\ref{fig:comp_ed_exp_r_2}.
In contrast to the ED data at distances
$|\bm{r}|=1$ and $|\bm{r}|=\sqrt{2}$,
they do not converge for the system sizes $N_{\rm s}\le 20$.
This is because a distance $|\bm{r}|=2$
reaches a half the length of the lattice,
and consequently, the boundary effect is significant.
The data of iPEPS simulations, in which the quantities
at the thermodynamic limit are
obtained directly, converge very well for $D\ge 6$.
The first peaks in the ED simulations
gradually converge to those in the iPEPS simulations
as the system size increases.
This observation suggests that the iPEPS result
represents the correlation functions
in the thermodynamic limit fairly well.


\section{Further comparison with the exact diagonalization
for a sudden quench
to weaker interaction}

To demonstrate the applicability of the iPEPS method
for a relatively weaker interaction region,
we also compare the single-particle correlation functions
obtained by the iPEPS and ED methods.
As examples,
we show the correlation functions for $U/J=10$ and $4$.

\begin{figure}[!]
\centering
\includegraphics[width=.95\columnwidth]{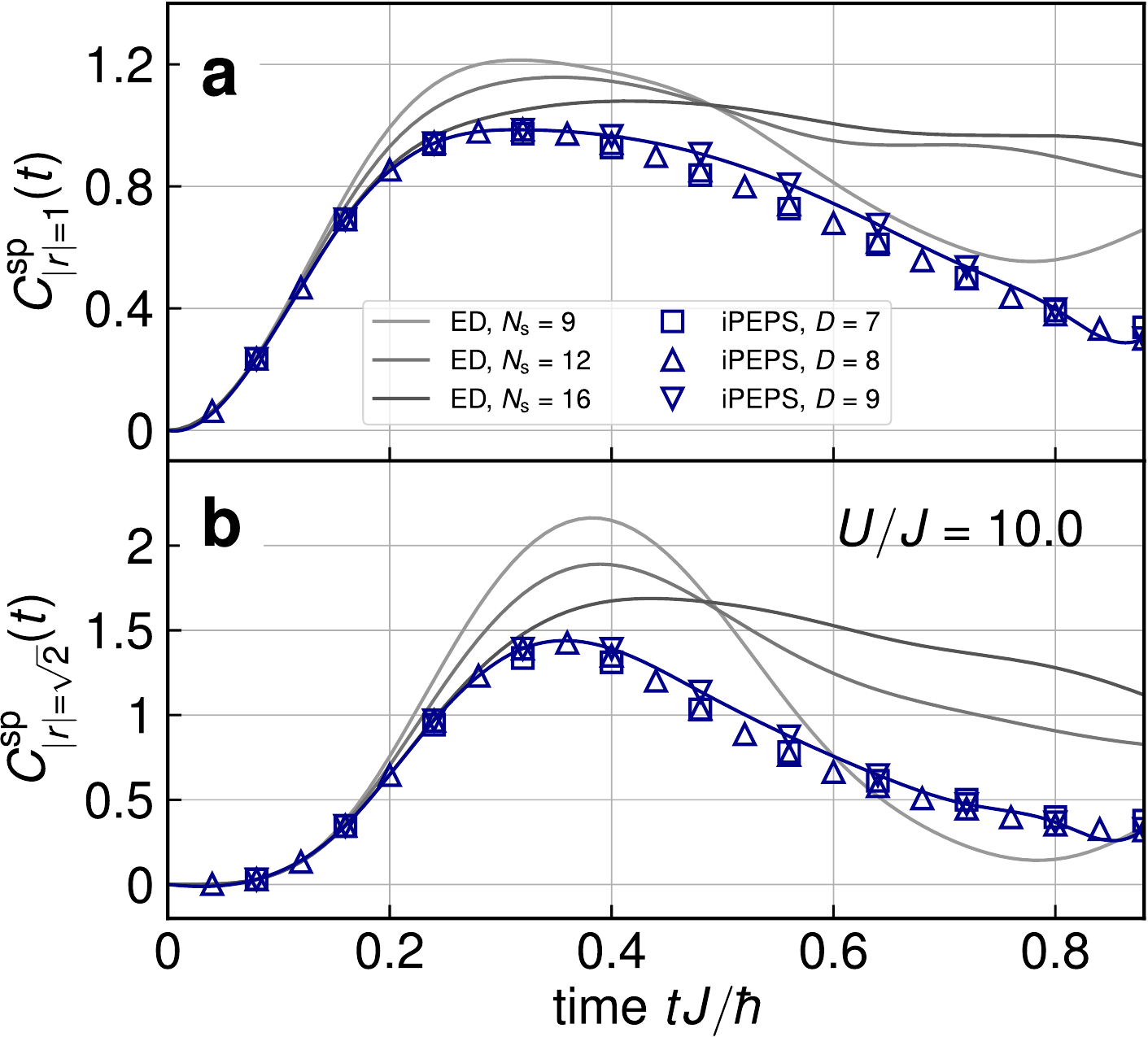}
\caption{Comparison of the single-particle correlation functions
in the case of a sudden quench
between iPEPS
(blue lines with symbols)
and ED
(gray lines)
for $U/J=10$.}
\label{fig:comp_ed_u10}
\end{figure}

For $U/J=10$,
the ED results show relatively large size dependence
for $N_{\rm s}\le 16$
(see Fig.~\ref{fig:comp_ed_u10}).
The iPEPS results are well converged for $D\ge 7$.
The data obtained by iPEPS and those by ED for $N_{\rm s}=16$
overlap very well within a short time frame
($tJ/\hbar\lesssim 0.3$).

\begin{figure}[!]
\centering
\includegraphics[width=.95\columnwidth]{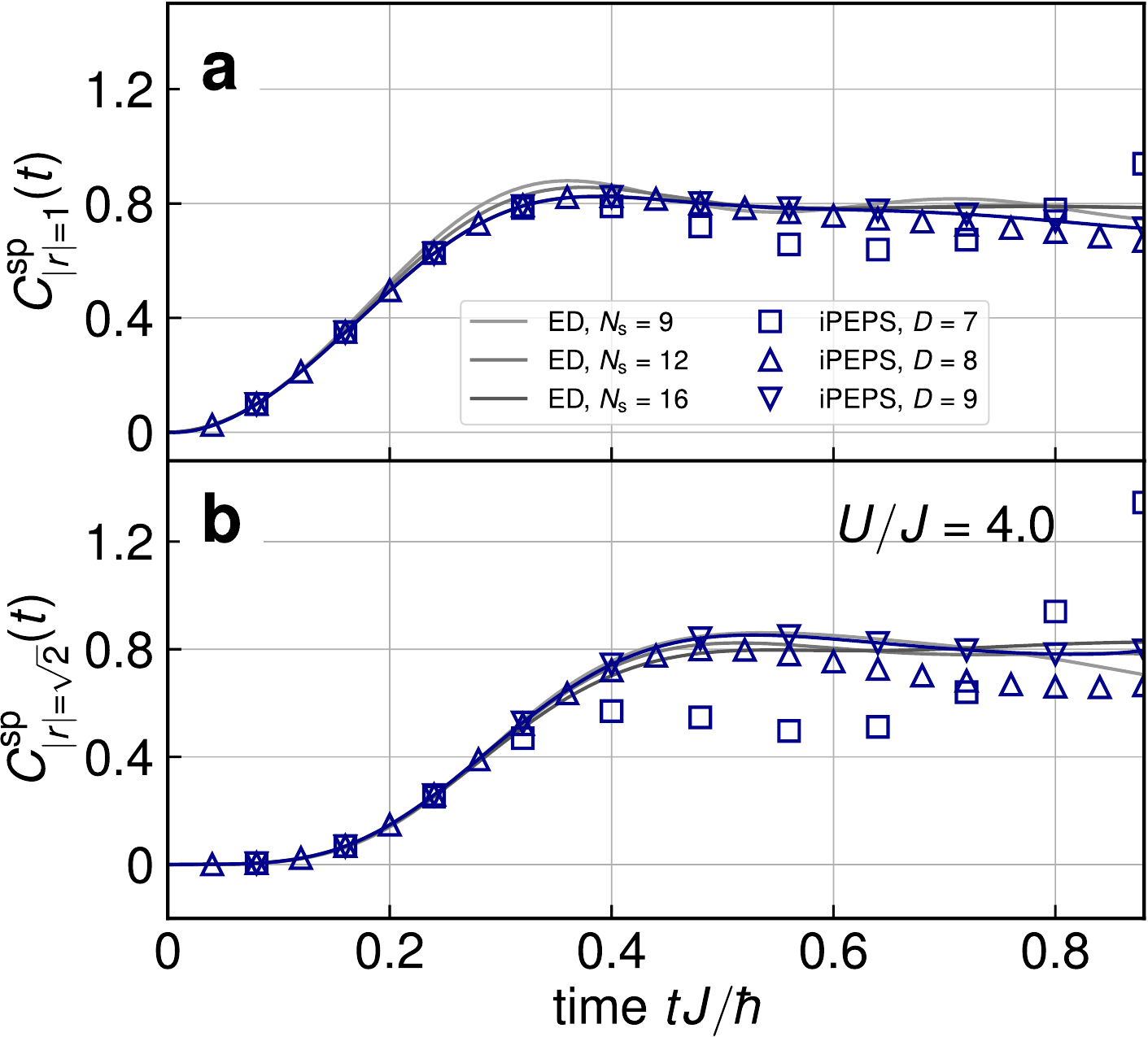}
\caption{Comparison of the single-particle correlation functions
in the case of a sudden quench
between iPEPS
(blue lines with symbols)
and ED
(gray lines)
for $U/J=4$.}
\label{fig:comp_ed_u4}
\end{figure}

On the other hand,
for $U/J=4$,
the ED results are nearly converged
for $N_{\rm s}\ge 9$
(see Fig.~\ref{fig:comp_ed_u4}).
Likewise,
the iPEPS results are nearly converged for $D\ge 8$
when $tJ/\hbar\lesssim 0.6$.
As in the case of $U/J=10$,
the data obtained by iPEPS for $D=9$ and those by ED
overlap very well
for $tJ/\hbar\lesssim 0.8$.


\section{Further comparison with the experiment
for a finite-time quench}

\begin{figure}[!]
\centering
\includegraphics[width=.95\columnwidth]{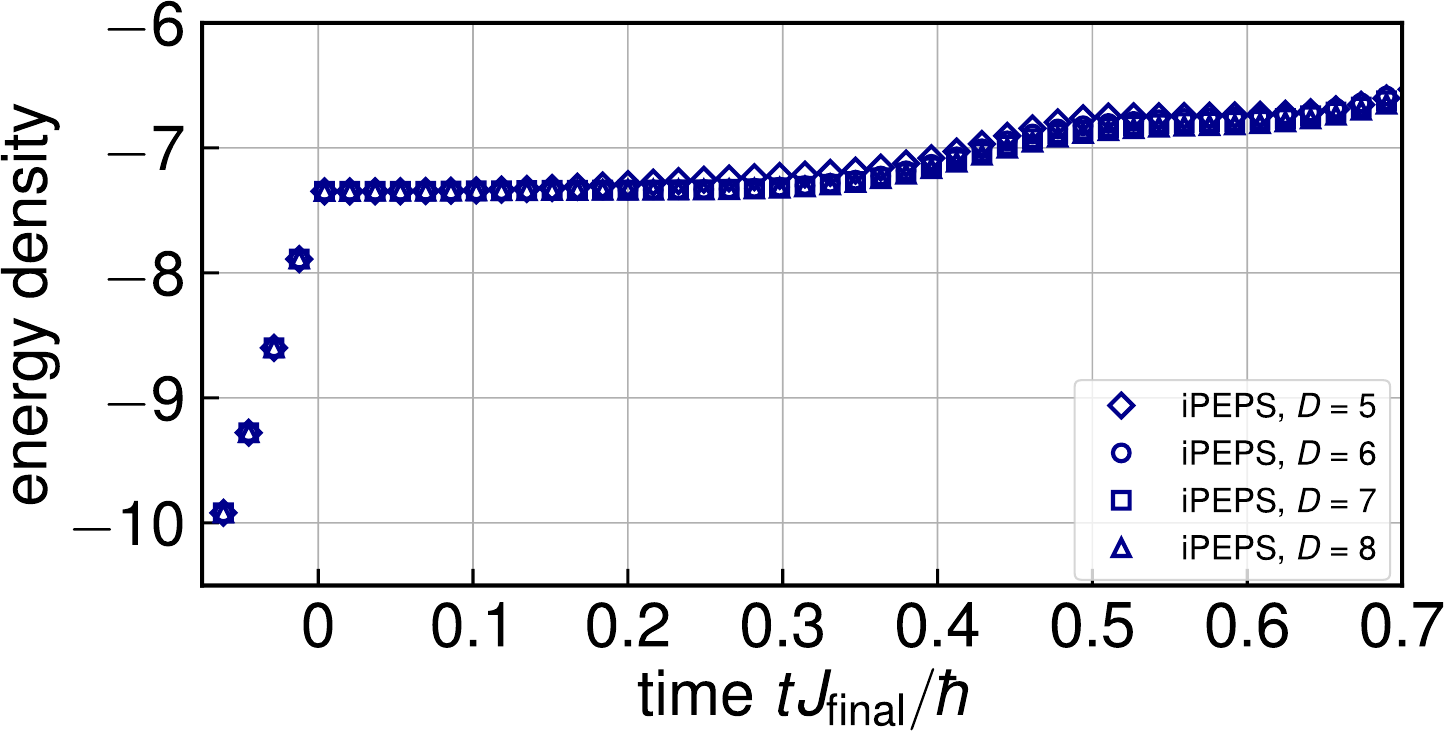}
\caption{Time dependence of the grand potential density
in the unit of hopping energy $J_{\rm final}$
for a finite-time quench
at $U/J_{\rm final}=19.6$.
The energy density is nearly conserved
in a short time $0<t\lesssim 0.4\hbar/J_{\rm final}$.}
\label{fig:cons_ene_exp}
\end{figure}

As in the sudden quench case,
we show the time dependence of the grand potential density
at $T=0$ obtained by iPEPS
in the case of a finite-time quench at $U/J_{\rm final}=19.6$
in Fig.~\ref{fig:cons_ene_exp}.
For $-\tau_{\rm Q}<t<0$,
both parameters $U$ and $J$ are controlled,
and the energy also varies.
At $t=-\tau_{\rm Q}$, it is
$e_{\rm MI}(t=-\tau_{\rm Q})/J_{\rm final}
= -\tilde{\mu}U(t=-\tau_{\rm Q})/J_{\rm final}
\sim -0.371\times 26.7 \sim -9.91$.
On the other hand, for $t>0$,
it is $e_{\rm MI}(t>0)/J_{\rm final} \sim -0.371\times 19.6 \sim -7.27$
as in the case of the sudden quench.
The energy density is nearly conserved
for $t \lesssim 0.4\hbar/J_{\rm final}$.

\begin{figure}[!]
\centering
\includegraphics[width=.95\columnwidth]{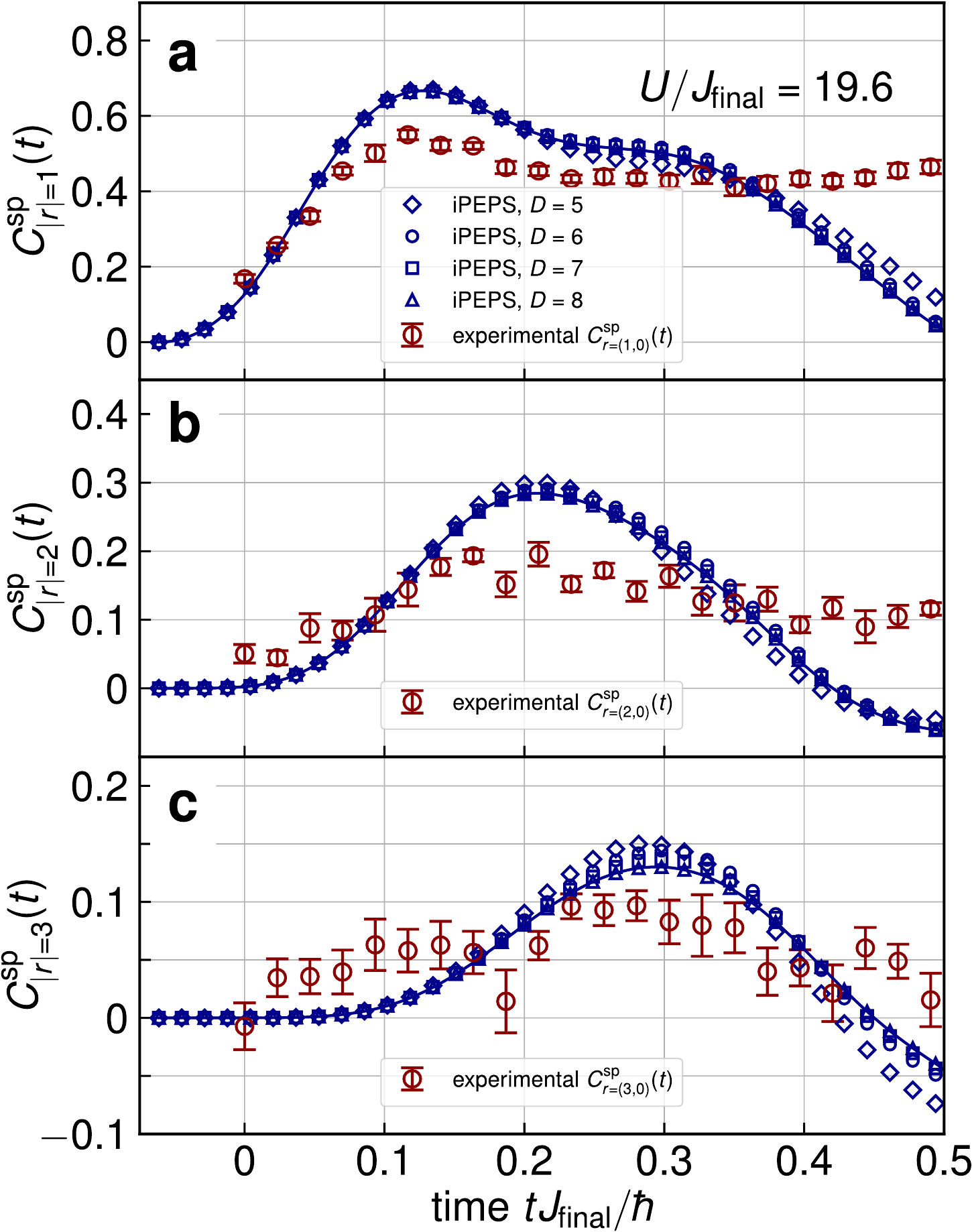}
\caption{Comparison of the single-particle correlation functions
in the case of a finite-time quench
between iPEPS
(blue lines with symbols)
and the experiment
(red circles with error bars).
The correlations at distances
(a) $|\bm{r}|=1$,
(b) $|\bm{r}|=2$,
and
(c) $|\bm{r}|=3$
are shown.
Here the experimental results for
$\bm{r}=(x,0)$ are given,
while those for
$\bm{r}=(0,y)$ are given in the main text.}
\label{fig:comp_peps_exp_x0}
\end{figure}

In the experiment~\cite{takasu2020},
the correlation functions along the horizontal
[$\bm{r}=(x,0)$]
and vertical
[$\bm{r}=(0,y)$]
directions are observed separately.
We have compared the results for $\bm{r}=(0,y)$
in the main text.
Here we show the results for $\bm{r}=(x,0)$ 
(see Fig.~\ref{fig:comp_peps_exp_x0}).
The intensities along the horizontal direction
are slightly smaller than those along the vertical
direction in the experiment.
Nevertheless, the intensity obtained by iPEPS
is almost consistent with that in the experiment.
The first-peak times for
$|\bm{r}|=1$, $2$, and $3$ obtained by iPEPS
also agree very well with the experimental results.

\begin{figure}[!]
\centering
\includegraphics[width=.95\columnwidth]{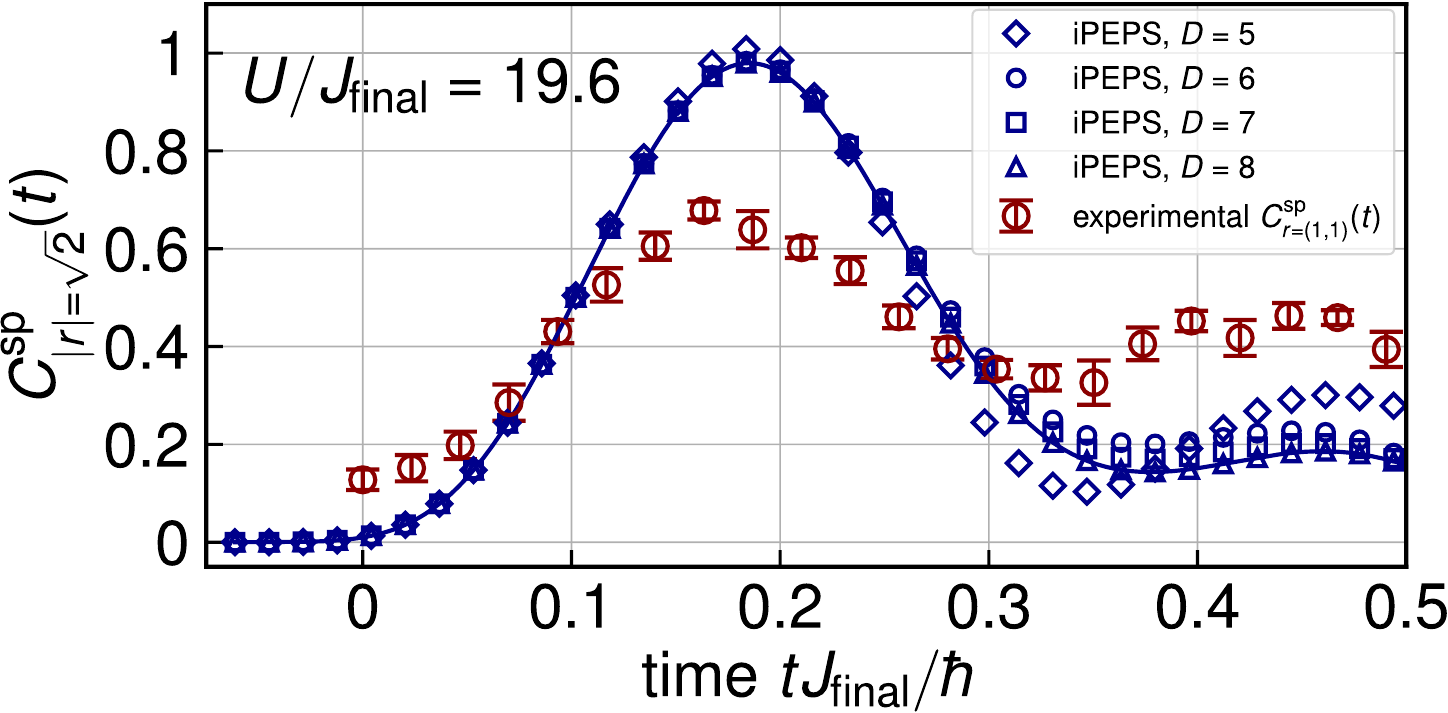}
\caption{Comparison of the single-particle correlation functions
in the case of a finite-time quench
between iPEPS
(blue lines with symbols)
and the experiment
(red circles with error bars).
The correlations at a distance $|\bm{r}|=\sqrt{2}$ is shown.}
\label{fig:comp_ed_exp_r_sqrt2}
\end{figure}

We also compare the single-particle correlation function
at a distance $|\bm{r}|=\sqrt{2}$
in Fig.~\ref{fig:comp_ed_exp_r_sqrt2}. 
The first-peak times are nearly the same
between the experiment and iPEPS simulations.
However, the intensity obtained by iPEPS is nearly $1.5$ times larger
than that in the experiment,
and the difference is at more than a few-sigma level.
Note that the intensity of
correlations at a distance $|\bm{r}|=\sqrt{2}$
by iPEPS overlaps perfectly with those
by ED (see the main text)
and by TWA (see the next section)
within the time shorter than $\sim 0.2\hbar/J_{\rm final}$.
At the moment, we do not know the origin of the difference
between the experimental and theoretical results.


\section{Comparison with
the truncated Wigner approximation
for a finite-time quench}

\begin{figure}[!]
\centering
\includegraphics[width=.95\columnwidth]{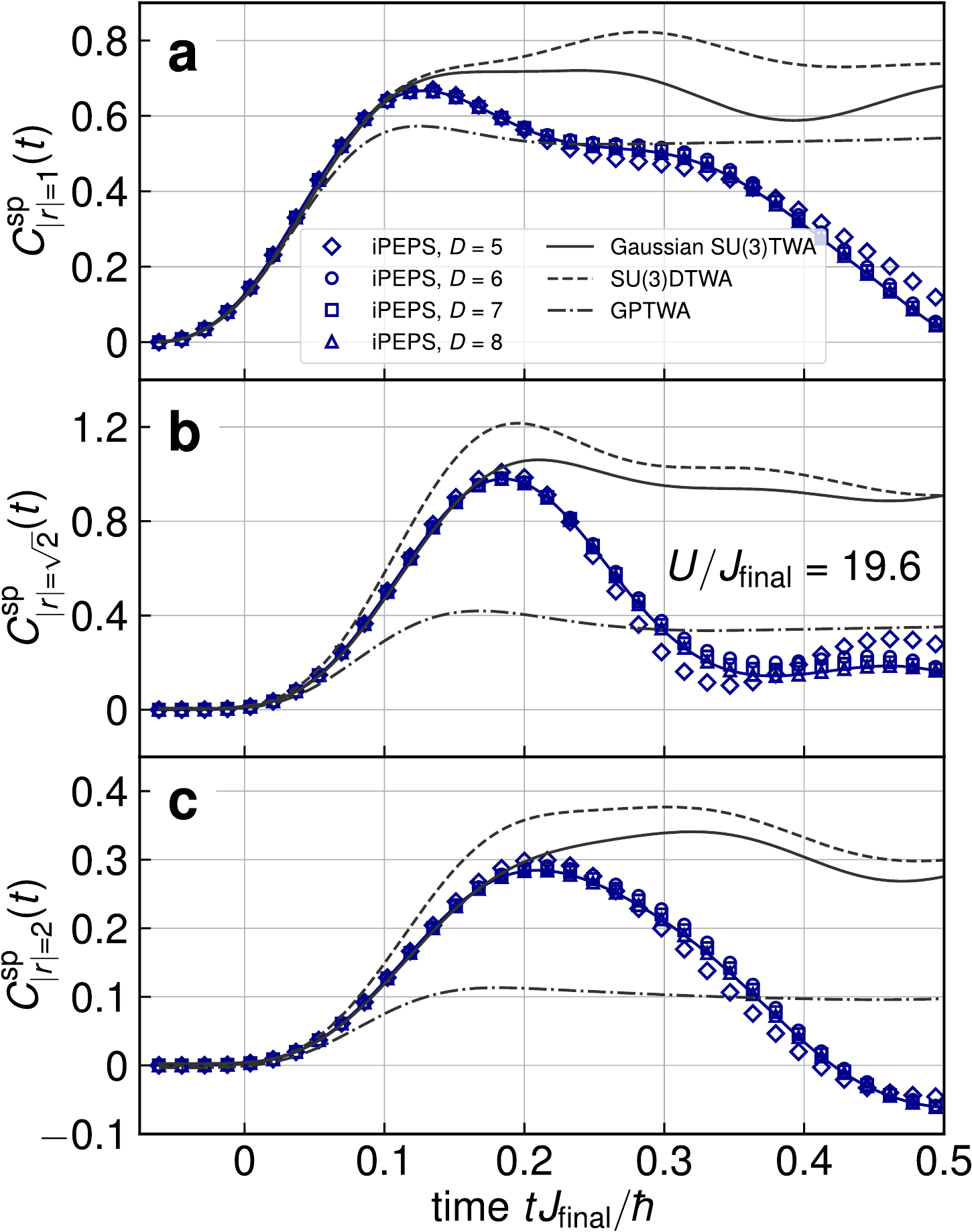}
\caption{Comparison of the single-particle correlation functions
in the case of a finite-time quench
between iPEPS
(blue lines with symbols)
and TWA
(gray lines).
The correlations at distances
(a) $|\bm{r}|=1$,
(b) $|\bm{r}|=\sqrt{2}$,
and
(c) $|\bm{r}|=2$
are shown.
Results of iPEPS and Gaussian SU(3)TWA are in good agreement
in a short time
slightly before the appearance of the first peaks obtained by iPEPS.
}
\label{fig:comp_twa}
\end{figure}

We compare the single-particle correlation functions
of iPEPS and TWA~\cite{nagao2021} in the case of a finite-time quench
in Fig.~\ref{fig:comp_twa}.
When $|\bm{r}|=1$, in a short time $t\lesssim 0.1\hbar/J_{\rm final}$,
the results of iPEPS, Gaussian SU(3)TWA, and discrete SU(3)TWA [SU(3)DTWA]
agree very well,
whereas the intensity obtained by the Gross-Pitaevskii TWA (GPTWA)
is slightly smaller.
On the other hand, the first-peak time of iPEPS
agrees very well with that of GPTWA,
whereas the peaks of Gaussian SU(3)TWA and SU(3)DTWA are very broad.
Remarkably, when $|\bm{r}|=\sqrt{2}$ and $|\bm{r}|=2$,
in a short time slightly before the appearance of the first peak
in iPEPS simulations,
the results of iPEPS and Gaussian SU(3)TWA overlap perfectly.
The intensity obtained by SU(3)DTWA (GPTWA) is slightly larger
(much smaller) than that obtained by iPEPS.
As in the case of $|\bm{r}|=1$,
the first peaks of SU(3)TWA and SU(3)DTWA are broad,
and it is hard to extract the correct first-peak times
from these TWA data.
For $|\bm{r}|=2$, the first-peak times obtained by
SU(3)TWA and SU(3)DTWA are longer than
those estimated from the experiment and iPEPS.
The iPEPS can simulate real-time dynamics
in a slightly longer time
and capture the first-peak structures more clearly.


\section{Time dependence of energy in the moderate interaction region}

\begin{figure}[!]
\centering
\includegraphics[width=.95\columnwidth]{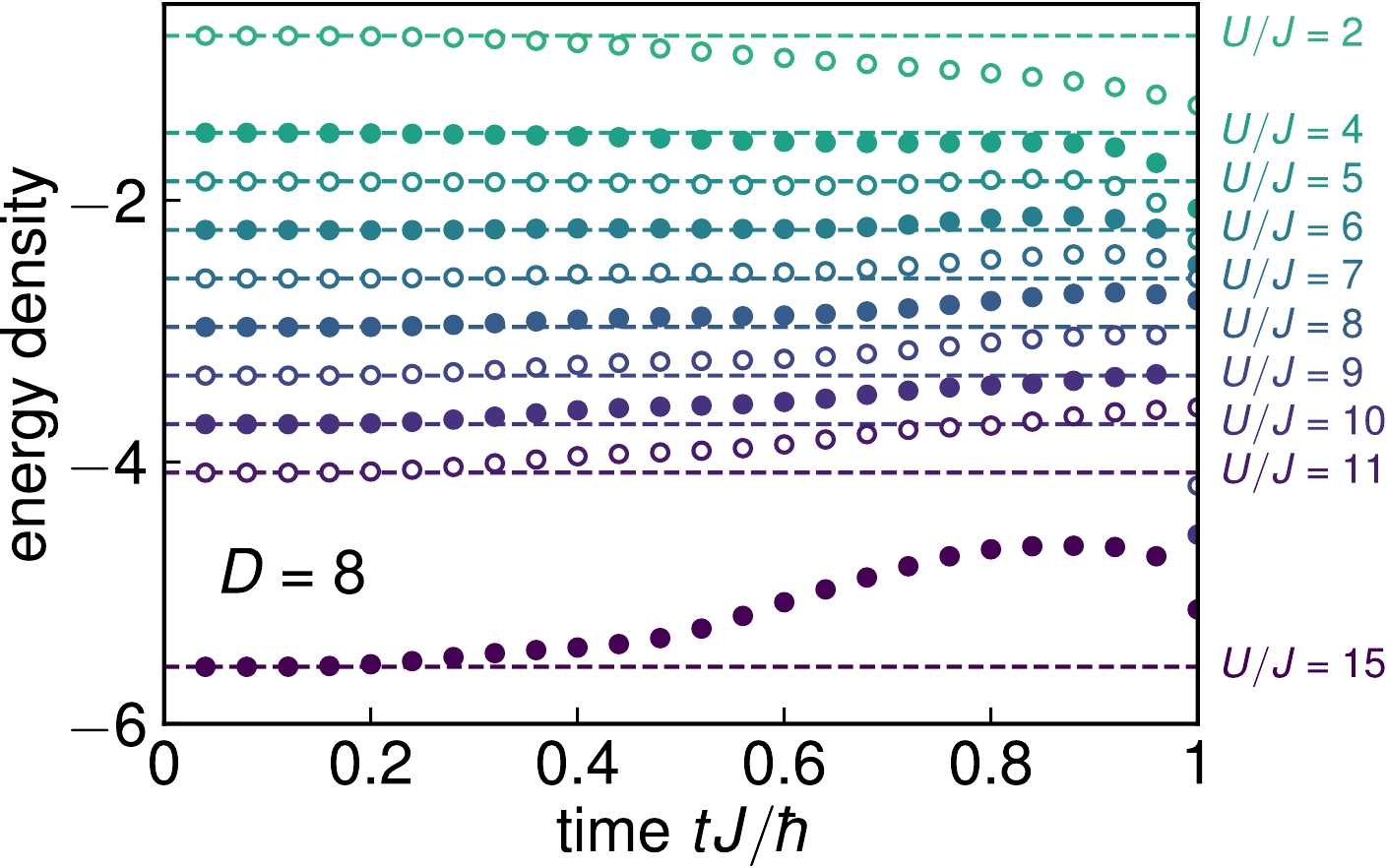}
\caption{Time dependence of the grand potential density
in the unit of hopping energy $J$
for a sudden quench at each $U/J$.
The energy densities at $t=0$ are given as dashed lines.
When $U/J\sim 5$,
the energy density is nearly conserved for
a rather longer time $t\lesssim 0.9\hbar/J$.}
\label{fig:cons_ene_many_U}
\end{figure}

Figure~\ref{fig:cons_ene_many_U}
shows the time dependence of the grand potential density
at each $U/J$.
When $U/J\sim 10$, the energy gradually increases
with increasing time.
This tendency is similar to what we have found
at $U/J=19.6$.
By contrast, when $U/J\sim 2$,
the energy decreases with increasing time.
In the intermediate region, especially for $U/J\sim 5$,
the behavior of energy going up and going down
cancel each other out.
Remarkably, it is conserved for $t\lesssim 0.9\hbar/J$,
much longer than the time at $U/J=19.6$.
This indicates that
the real-time evolution can be simulated in a longer time
in the moderate interaction region.

\section{Single-particle and density-density
correlation functions for several interaction parameters}

To investigate the propagation velocities,
we use the single-particle and density-density
correlation functions.
Here we demonstrate how they behave for small and large
interaction regions.

\begin{figure}[!]
\centering
\includegraphics[width=.95\columnwidth]{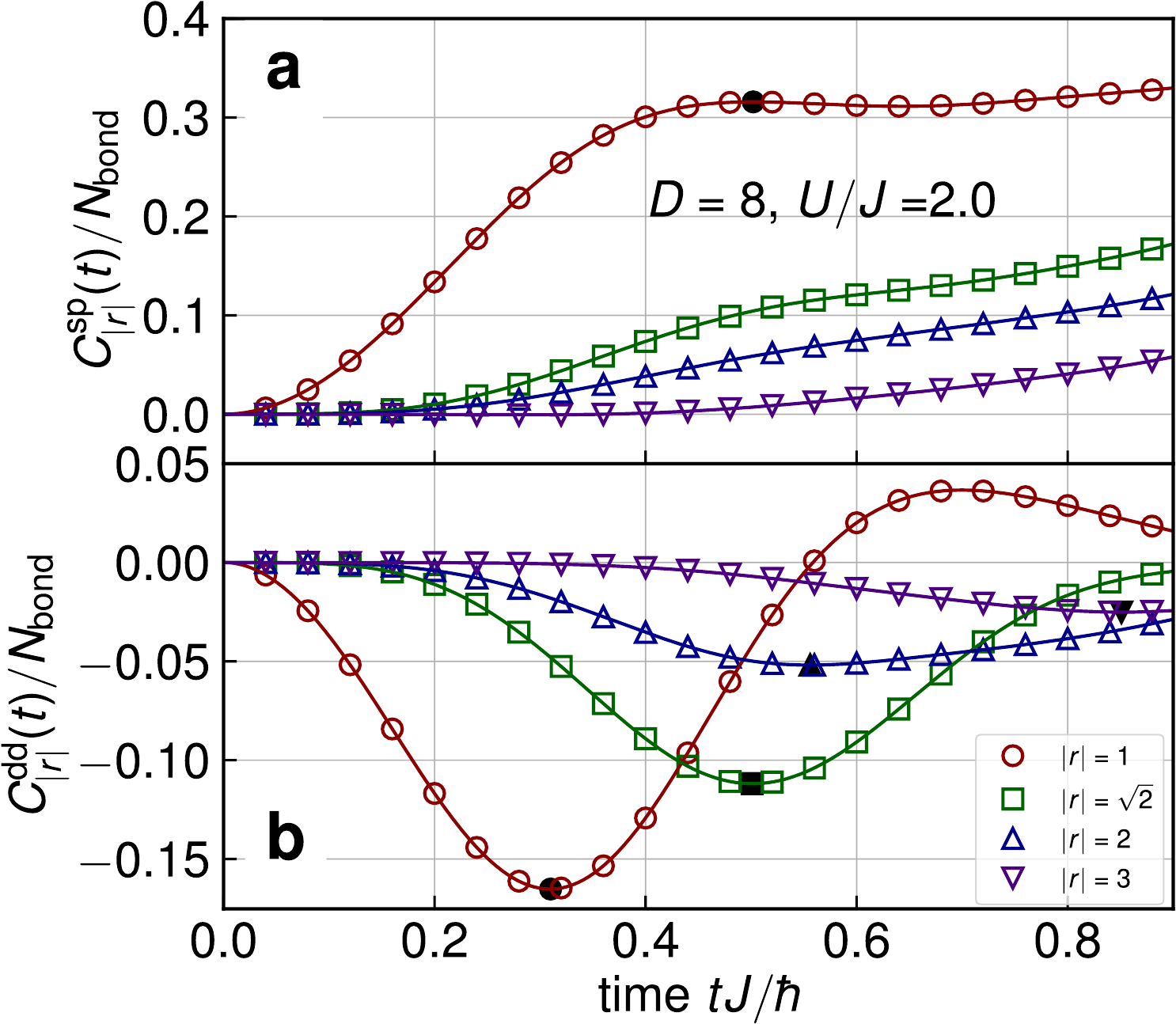}
\caption{(a) Single-particle and (b) density-density
correlation functions per bond at $U/J=2$
for the bond dimension $D=8$.}
\label{fig:correlations_U2}
\end{figure}

When the interaction is small ($U/J=2$),
it is difficult to follow the correlation peaks
in the single-particle correlation functions
within accessible simulation time,
as shown in Fig.~\ref{fig:correlations_U2}(a).
The first peak at $|\bm{r}|=1$ is broad,
while peaks do not appear
at $|\bm{r}|=\sqrt{2}$, $2$, and $3$
for $t\lesssim 0.9\hbar/J$.
This behavior persists until $U/J \sim 4$.
Because of the lack of data points
of correlation peaks,
we can extract the phase velocities
only for $U/J\gtrsim 5$.
On the other hand,
we observe clear first peaks in the density-density
correlation functions
[see Fig.~\ref{fig:correlations_U2}(b)].
The first-peak times are nearly consistent with
those obtained by TWA at high filling $\nu=10$~\cite{nagao2019}.

\begin{figure}[!]
\centering
\includegraphics[width=.95\columnwidth]{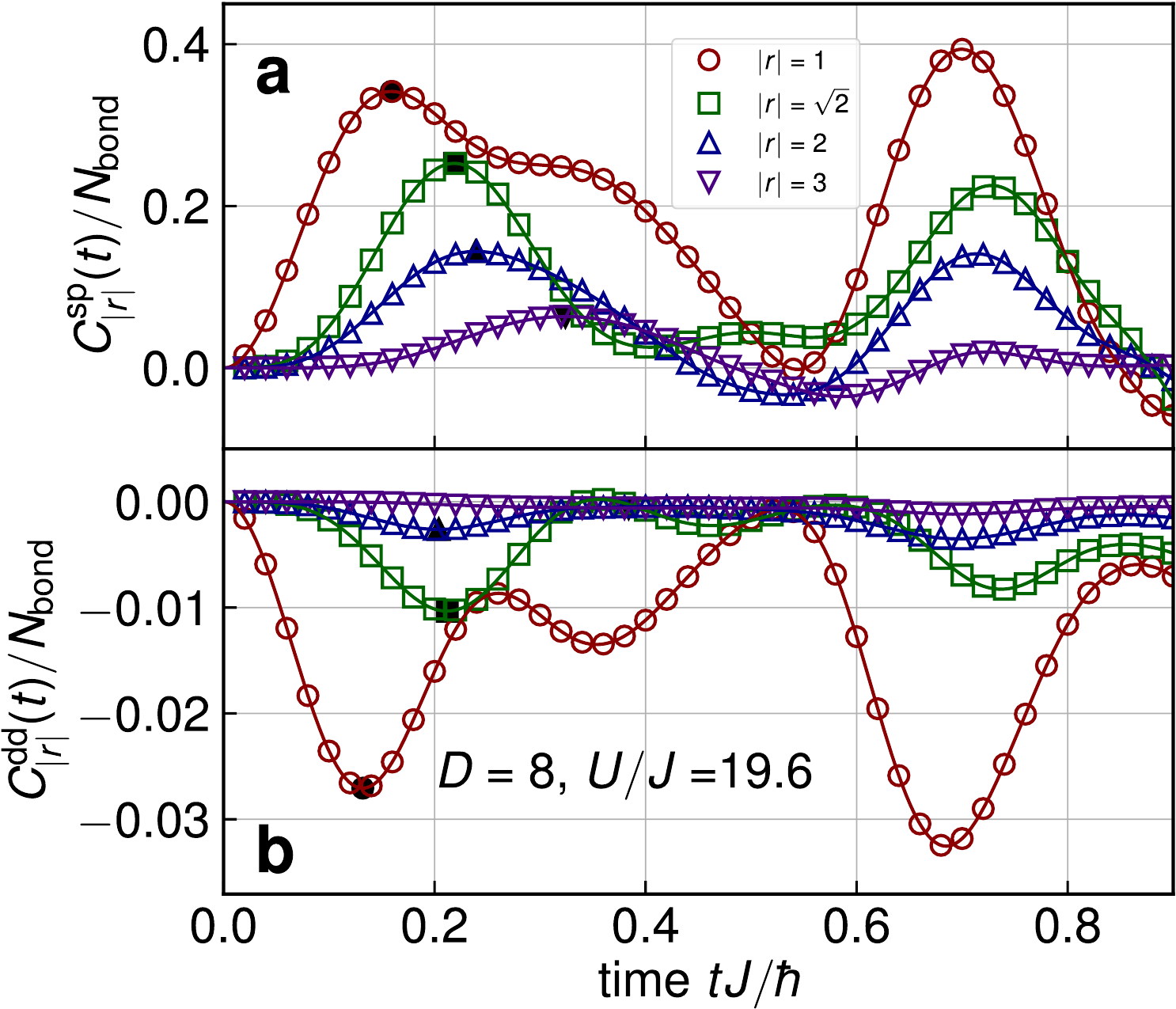}
\caption{(a) Single-particle and (b) density-density
correlation functions per bond at $U/J=19.6$
for the bond dimension $D=8$.}
\label{fig:correlations_U20}
\end{figure}

When the interaction is strong ($U/J=19.6$),
we successfully capture all the first peaks up to $|\bm{r}|\le 3$
within the reliable simulation time $t\lesssim 0.4\hbar/J$
(see Fig.~\ref{fig:correlations_U20}).
The maximum intensity in the single-particle correlation function
($\max |C^{\rm sp}_{|\bm{r}|}(t)/N_{\rm bond}| \sim 0.35$)
is comparable to that for the small interaction
[compare Figs.~\ref{fig:correlations_U2}(a)
and \ref{fig:correlations_U20}(a)].
By contrast, the maximum intensity of the density-density
correlation function
($\max |C^{\rm dd}_{|\bm{r}|}(t)/N_{\rm bond}| \sim 0.03$)
is much smaller than that
for the small interaction
[compare Figs.~\ref{fig:correlations_U2}(b)
and \ref{fig:correlations_U20}(b)].
The particle fluctuation is strongly suppressed
in the large $U$ region,
and the peak intensity decays rapidly as a function of a distance.

\begin{figure}[!]
\centering
\includegraphics[width=.95\columnwidth]{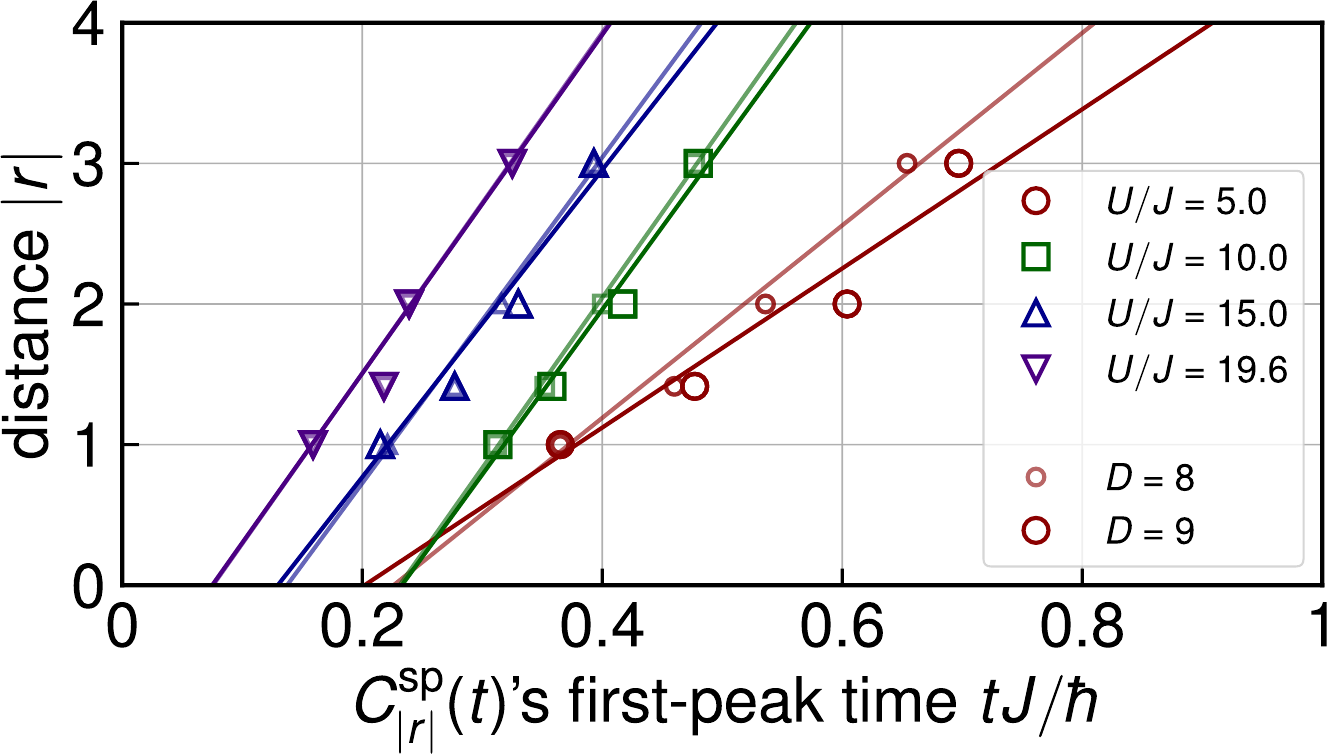}
\caption{Peak-time dependence of distance
obtained from the single-particle correlation functions
for $D=8$ and $D=9$.}
\label{fig:extract_v_sp}
\end{figure}

\begin{figure}[!]
\centering
\includegraphics[width=.95\columnwidth]{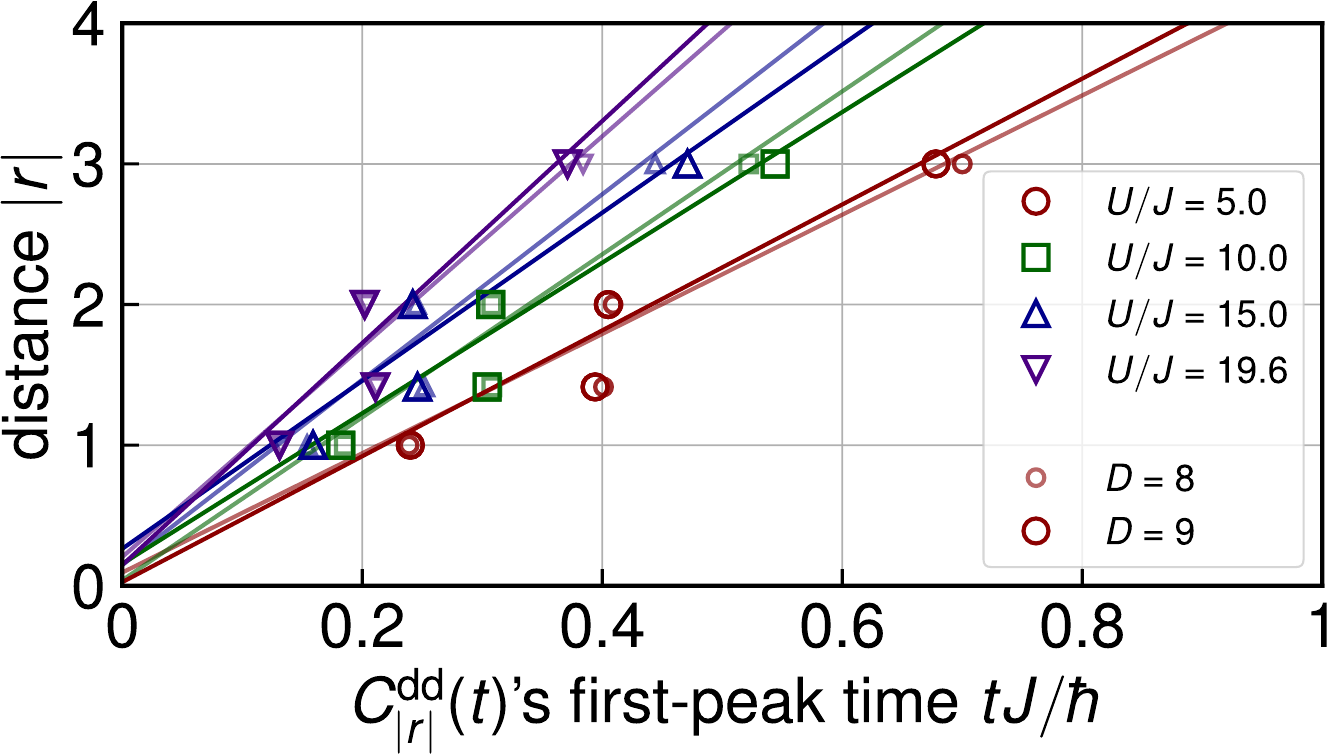}
\caption{Peak-time dependence of distance
obtained from the density-density correlation functions
for $D=8$ and $D=9$.}
\label{fig:extract_v_dd}
\end{figure}


\section{Extraction of propagation velocities
in the moderate interaction region}

We show the peak-time dependence of distance
obtained from the single-particle correlation functions
$C^{\rm sp}_{|\bm{r}|}(t)$ in Fig.~\ref{fig:extract_v_sp}.
For all the values of $U/J$ that we have
taken in our analyses,
the first-peak times of the correlations at $|\bm{r}|=3$
are well below the time
when the energy density is no longer conserved
(see Fig.~\ref{fig:cons_ene_many_U}).
At $U/J=5$, the data points for the bond dimensions
$D=8$ and $D=9$ deviate a little.
On the other hand, for $U/J\ge 10$,
they are nearly the same and are well converged.
The phase velocity along the horizontal or vertical axis
is estimated by fitting data points at $|\bm{r}|=1$, $2$,
and $3$.
The slopes are found to be nearly the same for $D=8$ and $D=9$.

We also show the peak-time dependence of distance
obtained from the density-density correlation functions
$C^{\rm dd}_{|\bm{r}|}(t)$ in Fig.~\ref{fig:extract_v_dd}.
In this case,
the first-peak times of the correlations at $|\bm{r}|=3$
is comparable to or slightly less than the time
when the energy density is no longer conserved
(see Fig.~\ref{fig:cons_ene_many_U}).
Therefore, data points up to $|\bm{r}|=3$
reach near the limit of what we can do in our iPEPS simulations.
For all $U/J$,
the data points for $D=8$ and $D=9$ are similar.
The group velocity along the horizontal or vertical axis
is estimated by fitting data points at $|\bm{r}|=1$, $2$,
and $3$.

Note that
the peak at $|\bm{r}|=\sqrt{2}$ does not have to be
on the line connecting data points
at $|\bm{r}|=1$, $2$, and $3$.
As discussed in
Refs.~[\onlinecite{takasu2020,mokhtarijazi2021,carleo2014,cevolani2018}],
the correlation spreading in 2D
can be highly anisotropic in general.
As for the low-energy physics, 
dynamics of the system is dominated by long wavelength excitations,
and the correlation front propagates
according to the Euclidean metrics.
However, when the system is quenched,
various wavelengths are mixed.
In this situation,
short wavelength excitations also participate,
and
the Manhattan metrics are
more relevant to the propagation of the
correlation front.
Reflecting this fact,
the first peaks of the correlation functions   
for $|\bm{r}|=2$ and $|\bm{r}|=\sqrt{2}$,
having the same Manhattan distance,
appear nearly simultaneously
(see Fig.~\ref{fig:extract_v_dd}).

Estimated group and phase velocities for $2\le U/J\le 22$
are summarized in the main text.


\begin{figure}[!]
\centering
\includegraphics[width=.95\columnwidth]{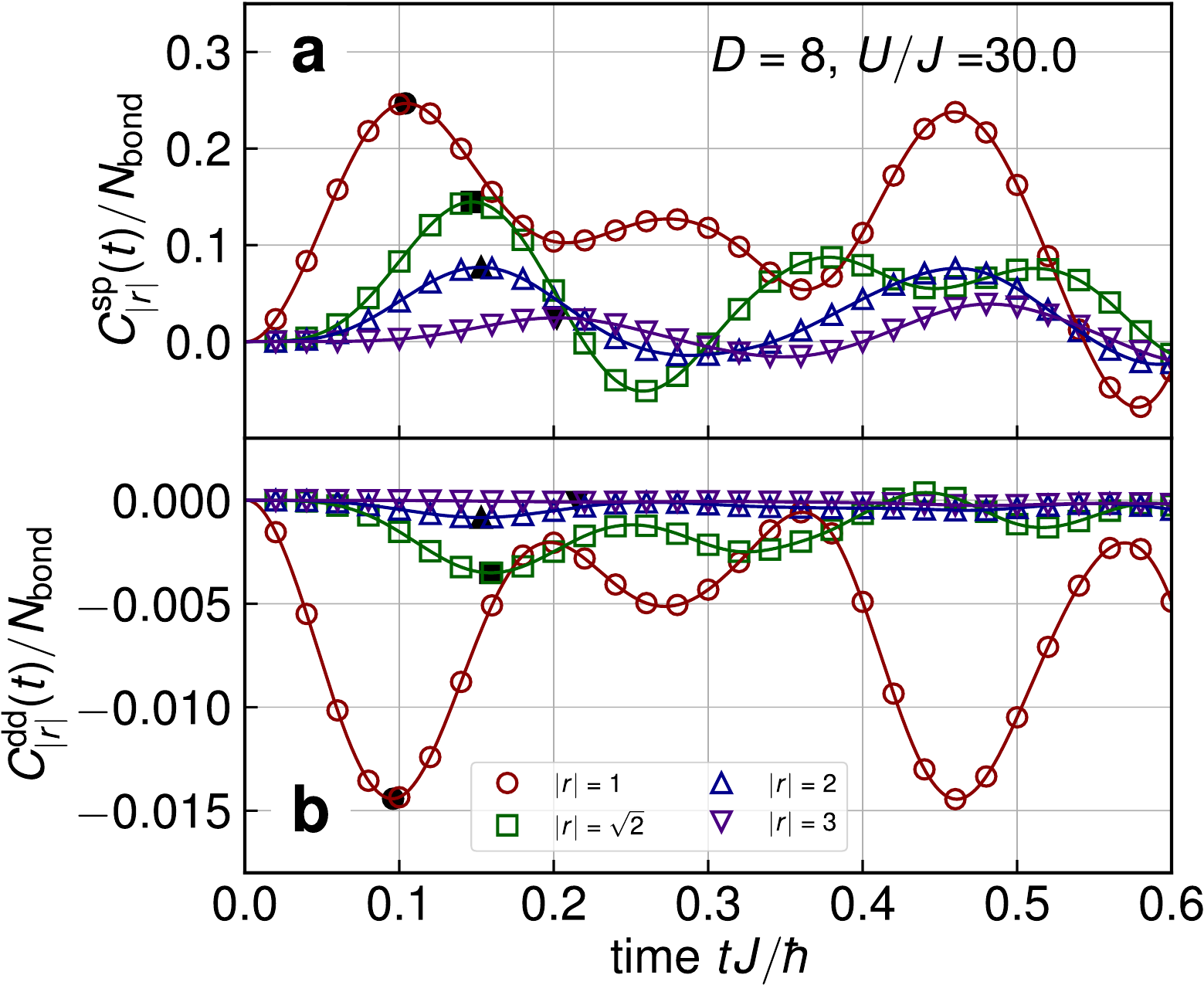}
\caption{(a) Single-particle and (b) density-density
correlation functions per bond at $U/J=30$
for the bond dimension $D=8$.}
\label{fig:correlations_U30}
\end{figure}

\begin{figure}[!]
\centering
\includegraphics[width=.95\columnwidth]{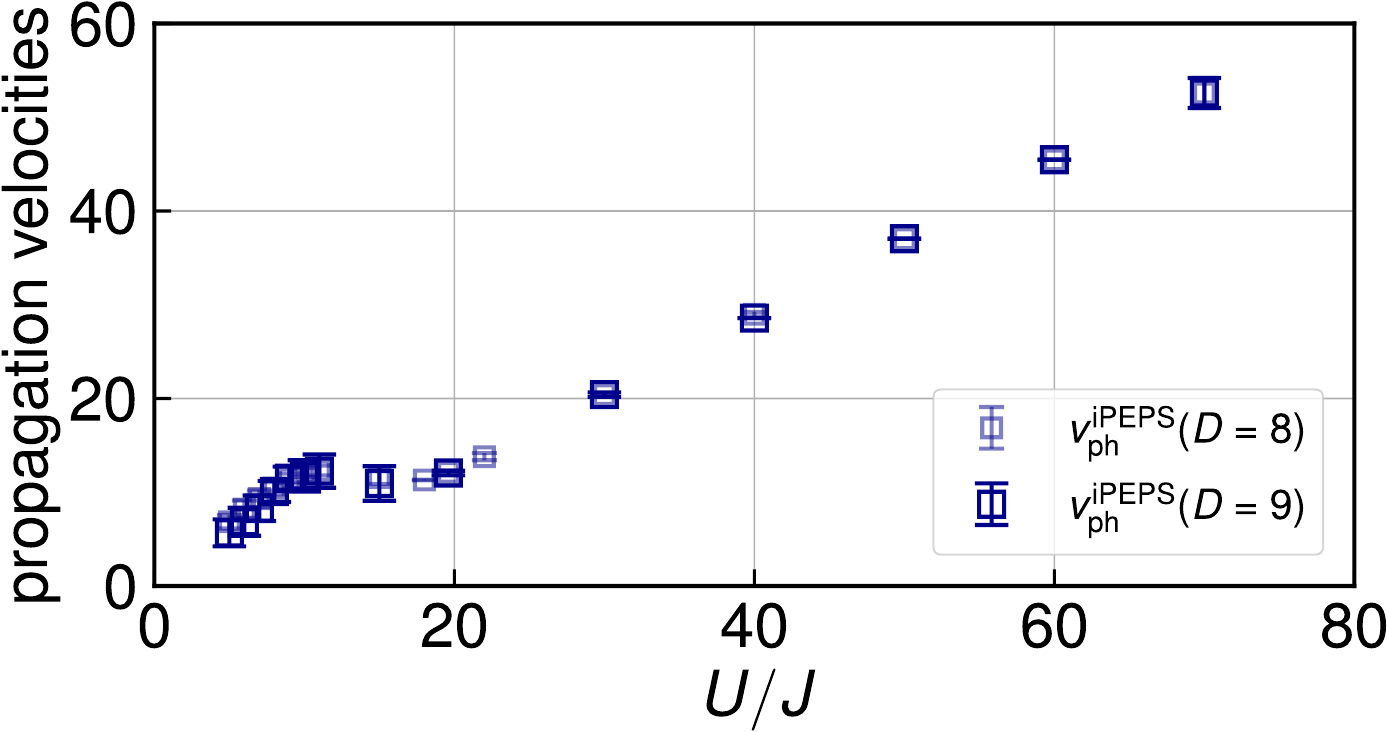}
\caption{Estimated phase velocity from the single-particle correlation
functions for the bond dimensions $D=8$ and $D=9$.
}
\label{fig:vph_large_U}
\end{figure}

\begin{figure}[!]
\centering
\includegraphics[width=.95\columnwidth]{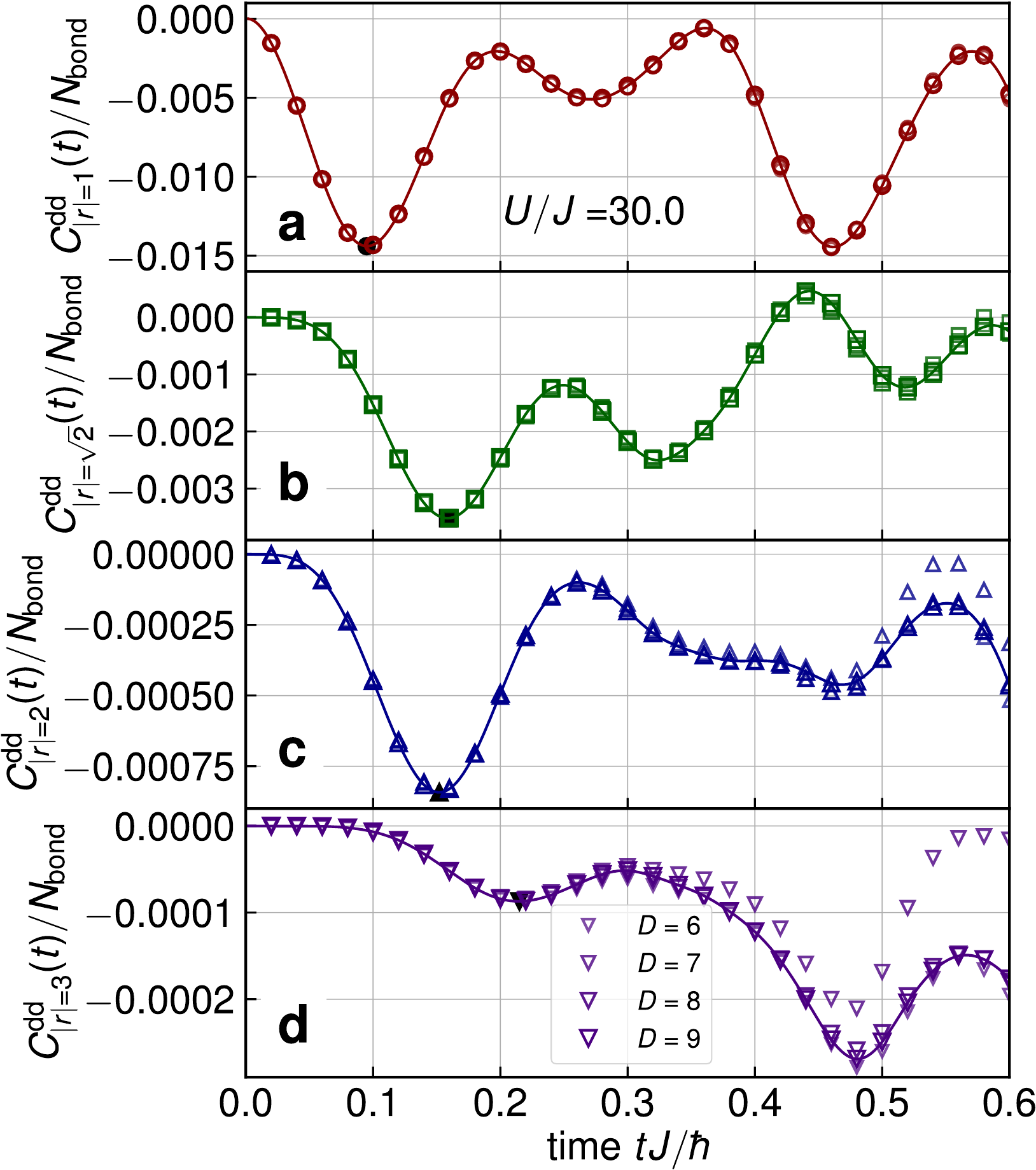}
\caption{Density-density correlation functions per bond at $U/J=30$
that are the same as those in Fig.~\ref{fig:correlations_U30},
but the vertical axes are magnified.}
\label{fig:correlations_U30_magnified}
\end{figure}


\section{Extraction of propagation velocities
for much stronger interaction}

As for much stronger interaction,
we successfully estimate the phase velocity
using the correlation functions obtained by the iPEPS simulations.
On the other hand,
it is much harder to extract the group velocity.
In this section,
we summarize our results for much stronger interaction.

As in the moderate interaction region,
we extract the phase velocity from the single-particle correlation
functions.
They behave similarly
between the moderate and stronger interaction regions
although each first-peak time of the single-particle correlation shifts
to an earlier time as interaction increases
(compare, for example,
Fig.~\ref{fig:correlations_U20}
and
Fig.~\ref{fig:correlations_U30}).
From the first-peak-time dependence of distance,
we extract the phase velocity
as shown in Fig.~\ref{fig:vph_large_U}.
The phase velocity is dominated by the energy scale of interaction
and is nearly a linear function of $U/J$.
The estimated value $v_{\rm ph}\sim 30J/\hbar$ at $U/J\sim 40$
is comparable to the value obtained by the 2PISC method in
Ref.~[\onlinecite{mokhtarijazi2021}].

On the other hand,
within the accessible simulation time,
we are not able to 
extract the group velocity for very large $U/J$.
The group and phase velocities start to deviate 
even when we use the density-density correlation functions
to extract the group velocity.
To accurately estimate the group velocity,
which is slower than the phase velocity,
we have to carefully follow the envelope of the correlation function
for slightly longer time.
To examine how the envelope should look like,
we plot the typical behavior of the density-density correlation
functions for $U/J=30$ in Fig.~\ref{fig:correlations_U30_magnified}.
When we focus on the correlation function at $|\bm{r}|=3$,
the first-peak time is located at $tJ/\hbar\sim 0.2$.
On the other hand,
because this peak $|C^{\rm dd}_{|\bm{r}|=3}(tJ/\hbar\sim 0.2)|$ seems to
be smaller than the second peak at $tJ/\hbar\sim 0.5$,
it is likely that the peak of the true envelope of the correlation
function is at the later time than $tJ/\hbar\sim 0.2$.
However, because the energy is nearly conserved up to a short time
$tJ/\hbar \sim 0.4$,
it is not possible to estimate the peak of the envelope,
which should be determined by connecting at least three peak points.


\begin{figure}[!]
\centering
\includegraphics[width=.95\columnwidth]{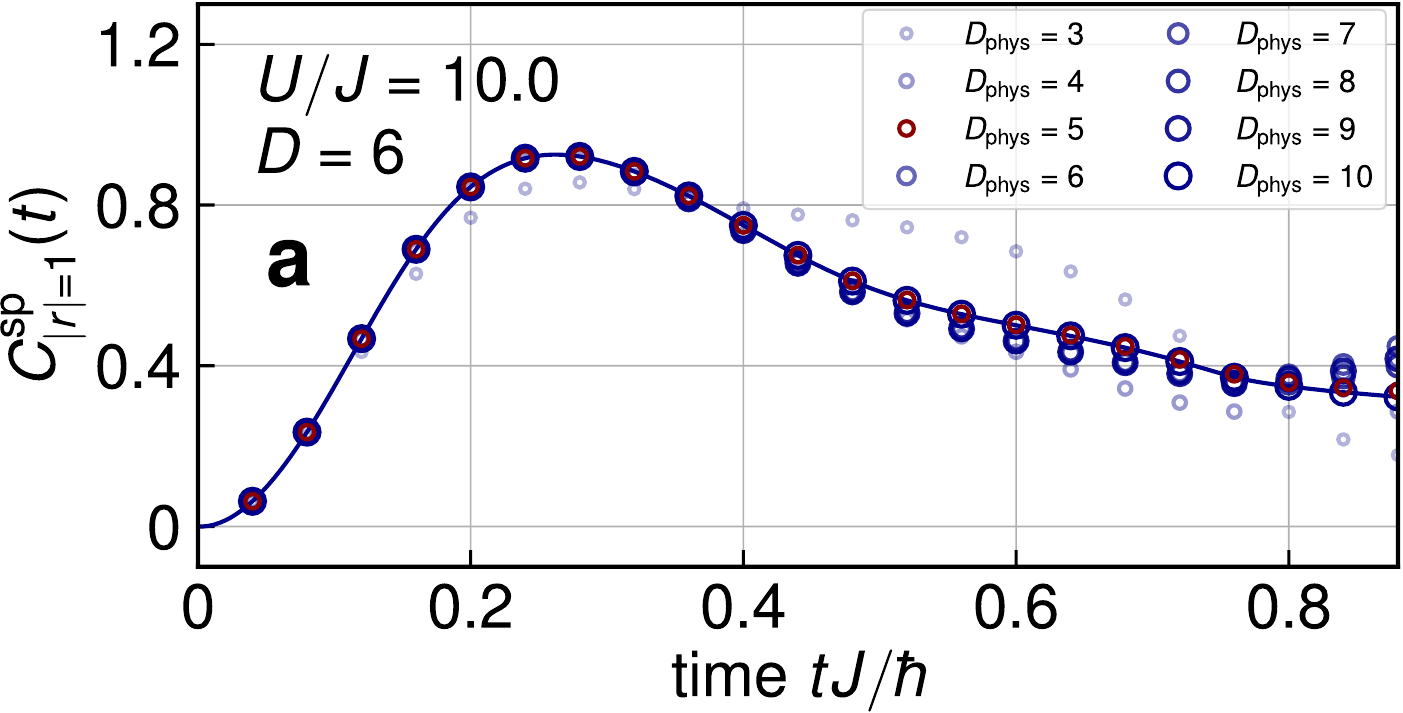}
\\
\includegraphics[width=.95\columnwidth]{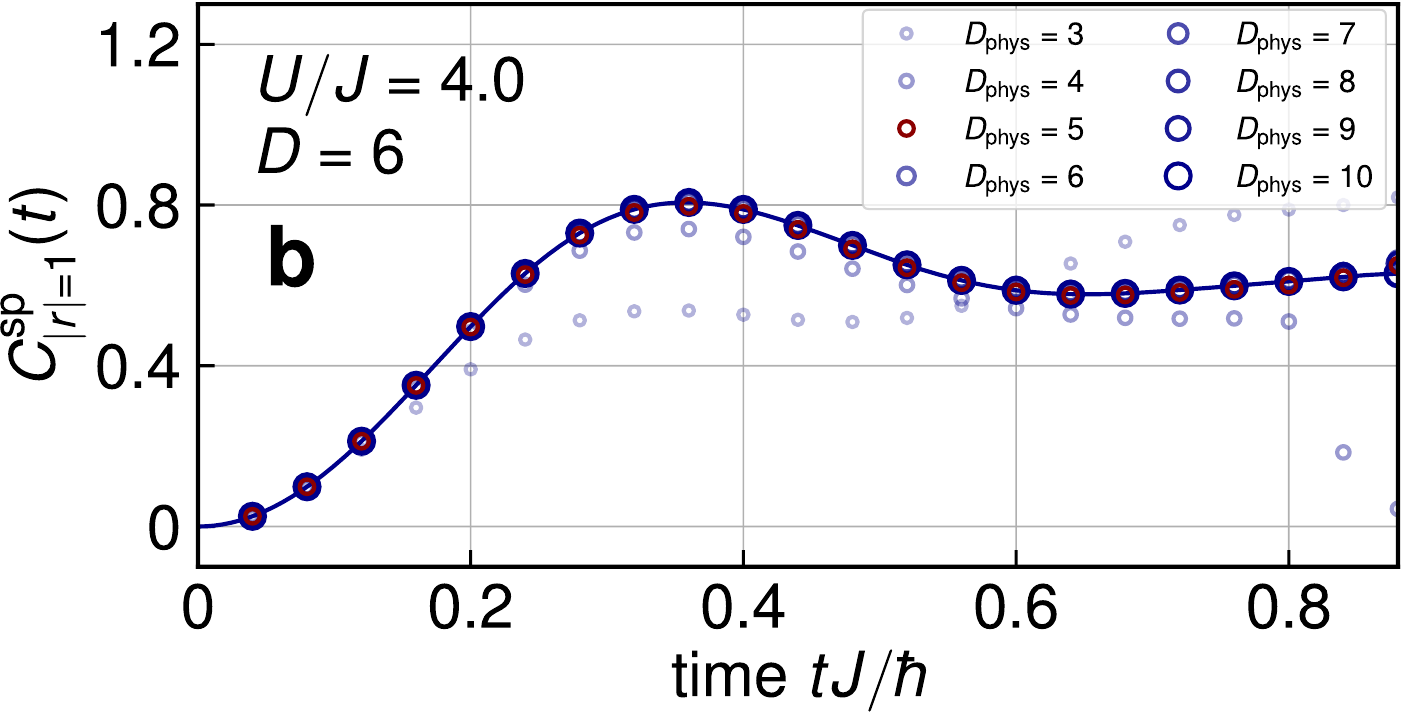}
\\
\includegraphics[width=.95\columnwidth]{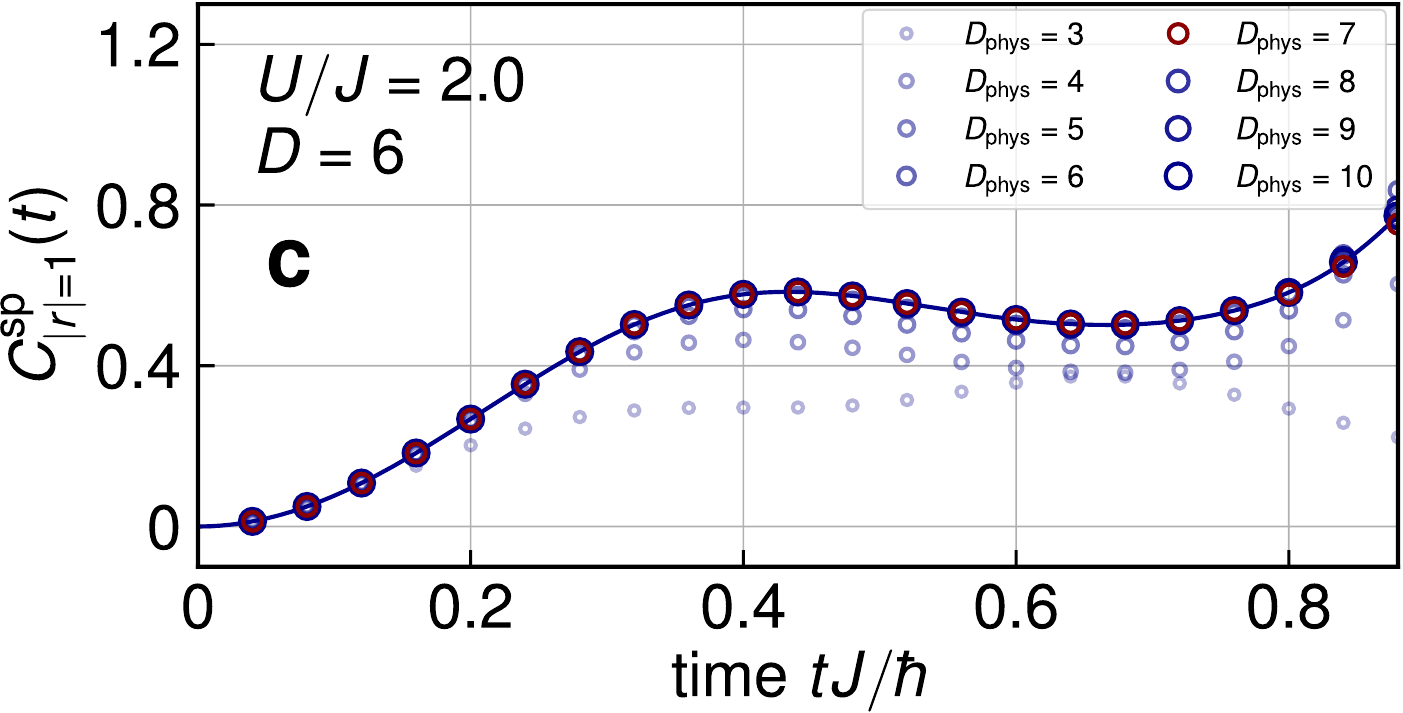}
\caption{Single-particle correlation functions
as functions of the dimension of the local Hilbert space
$D_{\rm phys}$ for (a) $U/J=10$, (b) $U/J=4$, and (c) $U/J=2$.
}
\label{fig:dphys_dep_corr}
\end{figure}


\section{Choice of the dimension of the local Hilbert space}

To determine the appropriate dimension of the local Hilbert space
$D_{\rm phys}$,
we calculate correlation functions up to $D_{\rm phys}=10$
for several interaction parameters.
For a quench to the Mott insulating region,
choosing $D_{\rm phys}=3$ reproduces the experimental
and other numerical results very well.
Here we focus on a quench to the superfluid parameter region.
Because qualitative behavior of correlation functions
is nearly converged for $D\ge 6$,
we examine the $D_{\rm phys}$ dependence at $D=6$.
We show the $D_{\rm phys}$ dependence of
the single-particle correlation functions
for selected interaction parameters
in Fig.~\ref{fig:dphys_dep_corr}.
(The density-density correlation functions
and correlation functions for $|\bm{r}|>1$
also behave in a qualitatively similar manner 
and are not shown here.)

For $U/J=10$,
the first-peak time is nearly identical for $D_{\rm phys}\ge 3$.
The correlation functions nearly converge to the same curve
for $D_{\rm phys}\ge 4$.
Even if we lower the interaction strength to $U/J=4$,
the correlation functions appear to converge very well
for $D_{\rm phys}\ge 5$.
Therefore, we have chosen $D_{\rm phys}=5$
for $U/J\ge 4$.

On the other hand, for $U/J=2$,
the correlation function for $D_{\rm phys}=5$
is slightly smaller than the converged curve.
Although the data for $D_{\rm phys}=5$
and those for $D_{\rm phys}>5$ are not so different,
we have chosen $D_{\rm phys}=7$
particularly for $U/J=2$
to maximize safety.


\def\bibsection{\section*{Supplementary References}}

\input{supp.bbl.back}

\onecolumngrid

%% file: main.bbl.back
%

%% file: main.bbl
\begin{thebibliography}{86}%
\makeatletter
\providecommand \@ifxundefined [1]{%
 \@ifx{#1\undefined}
}%
\providecommand \@ifnum [1]{%
 \ifnum #1\expandafter \@firstoftwo
 \else \expandafter \@secondoftwo
 \fi
}%
\providecommand \@ifx [1]{%
 \ifx #1\expandafter \@firstoftwo
 \else \expandafter \@secondoftwo
 \fi
}%
\providecommand \natexlab [1]{#1}%
\providecommand \enquote  [1]{``#1''}%
\providecommand \bibnamefont  [1]{#1}%
\providecommand \bibfnamefont [1]{#1}%
\providecommand \citenamefont [1]{#1}%
\providecommand \href@noop [0]{\@secondoftwo}%
\providecommand \href [0]{\begingroup \@sanitize@url \@href}%
\providecommand \@href[1]{\@@startlink{#1}\@@href}%
\providecommand \@@href[1]{\endgroup#1\@@endlink}%
\providecommand \@sanitize@url [0]{\catcode `\\12\catcode `\$12\catcode
  `\&12\catcode `\#12\catcode `\^12\catcode `\_12\catcode `\%12\relax}%
\providecommand \@@startlink[1]{}%
\providecommand \@@endlink[0]{}%
\providecommand \url  [0]{\begingroup\@sanitize@url \@url }%
\providecommand \@url [1]{\endgroup\@href {#1}{\urlprefix }}%
\providecommand \urlprefix  [0]{URL }%
\providecommand \Eprint [0]{\href }%
\providecommand \doibase [0]{https://doi.org/}%
\providecommand \selectlanguage [0]{\@gobble}%
\providecommand \bibinfo  [0]{\@secondoftwo}%
\providecommand \bibfield  [0]{\@secondoftwo}%
\providecommand \translation [1]{[#1]}%
\providecommand \BibitemOpen [0]{}%
\providecommand \bibitemStop [0]{}%
\providecommand \bibitemNoStop [0]{.\EOS\space}%
\providecommand \EOS [0]{\spacefactor3000\relax}%
\providecommand \BibitemShut  [1]{\csname bibitem#1\endcsname}%
\let\auto@bib@innerbib\@empty
\bibitem [{\citenamefont {Cheneau}\ \emph {et~al.}(2012)\citenamefont
  {Cheneau}, \citenamefont {Barmettler}, \citenamefont {Poletti}, \citenamefont
  {Endres}, \citenamefont {Schau{\ss}}, \citenamefont {Fukuhara}, \citenamefont
  {Gross}, \citenamefont {Bloch}, \citenamefont {Kollath},\ and\ \citenamefont
  {Kuhr}}]{cheneau2012}%
  \BibitemOpen
  \bibfield  {author} {\bibinfo {author} {\bibfnamefont {M.}~\bibnamefont
  {Cheneau}}, \bibinfo {author} {\bibfnamefont {P.}~\bibnamefont {Barmettler}},
  \bibinfo {author} {\bibfnamefont {D.}~\bibnamefont {Poletti}}, \bibinfo
  {author} {\bibfnamefont {M.}~\bibnamefont {Endres}}, \bibinfo {author}
  {\bibfnamefont {P.}~\bibnamefont {Schau{\ss}}}, \bibinfo {author}
  {\bibfnamefont {T.}~\bibnamefont {Fukuhara}}, \bibinfo {author}
  {\bibfnamefont {C.}~\bibnamefont {Gross}}, \bibinfo {author} {\bibfnamefont
  {I.}~\bibnamefont {Bloch}}, \bibinfo {author} {\bibfnamefont
  {C.}~\bibnamefont {Kollath}},\ and\ \bibinfo {author} {\bibfnamefont
  {S.}~\bibnamefont {Kuhr}},\ }\bibfield  {title} {\emph {\bibinfo {title}
  {{Light-cone-like spreading of correlations in a quantum many-body
  system}}},\ }\href {https://doi.org/10.1038/nature10748} {\bibfield
  {journal} {\bibinfo  {journal} {Nature}\ }\textbf {\bibinfo {volume} {481}},\
  \bibinfo {pages} {484} (\bibinfo {year} {2012})}\BibitemShut {NoStop}%
\bibitem [{\citenamefont {Jurcevic}\ \emph {et~al.}(2014)\citenamefont
  {Jurcevic}, \citenamefont {Lanyon}, \citenamefont {Hauke}, \citenamefont
  {Hempel}, \citenamefont {Zoller}, \citenamefont {Blatt},\ and\ \citenamefont
  {Roos}}]{jurcevic2014}%
  \BibitemOpen
  \bibfield  {author} {\bibinfo {author} {\bibfnamefont {P.}~\bibnamefont
  {Jurcevic}}, \bibinfo {author} {\bibfnamefont {B.~P.}\ \bibnamefont
  {Lanyon}}, \bibinfo {author} {\bibfnamefont {P.}~\bibnamefont {Hauke}},
  \bibinfo {author} {\bibfnamefont {C.}~\bibnamefont {Hempel}}, \bibinfo
  {author} {\bibfnamefont {P.}~\bibnamefont {Zoller}}, \bibinfo {author}
  {\bibfnamefont {R.}~\bibnamefont {Blatt}},\ and\ \bibinfo {author}
  {\bibfnamefont {C.~F.}\ \bibnamefont {Roos}},\ }\bibfield  {title} {\emph
  {\bibinfo {title} {{Quasiparticle engineering and entanglement propagation in
  a quantum many-body system}}},\ }\href {https://doi.org/10.1038/nature13461}
  {\bibfield  {journal} {\bibinfo  {journal} {Nature}\ }\textbf {\bibinfo
  {volume} {511}},\ \bibinfo {pages} {202} (\bibinfo {year}
  {2014})}\BibitemShut {NoStop}%
\bibitem [{\citenamefont {Takasu}\ \emph {et~al.}(2020)\citenamefont {Takasu},
  \citenamefont {Yagami}, \citenamefont {Asaka}, \citenamefont {Fukushima},
  \citenamefont {Nagao}, \citenamefont {Goto}, \citenamefont {Danshita},\ and\
  \citenamefont {Takahashi}}]{takasu2020}%
  \BibitemOpen
  \bibfield  {author} {\bibinfo {author} {\bibfnamefont {Y.}~\bibnamefont
  {Takasu}}, \bibinfo {author} {\bibfnamefont {T.}~\bibnamefont {Yagami}},
  \bibinfo {author} {\bibfnamefont {H.}~\bibnamefont {Asaka}}, \bibinfo
  {author} {\bibfnamefont {Y.}~\bibnamefont {Fukushima}}, \bibinfo {author}
  {\bibfnamefont {K.}~\bibnamefont {Nagao}}, \bibinfo {author} {\bibfnamefont
  {S.}~\bibnamefont {Goto}}, \bibinfo {author} {\bibfnamefont {I.}~\bibnamefont
  {Danshita}},\ and\ \bibinfo {author} {\bibfnamefont {Y.}~\bibnamefont
  {Takahashi}},\ }\bibfield  {title} {\emph {\bibinfo {title} {{Energy
  redistribution and spatio-temporal evolution of correlations after a sudden
  quench of the Bose-Hubbard model}}},\ }\href
  {https://doi.org/10.1126/sciadv.aba9255} {\bibfield  {journal} {\bibinfo
  {journal} {Sci. Adv.}\ }\textbf {\bibinfo {volume} {6}},\ \bibinfo {pages}
  {eaba9255} (\bibinfo {year} {2020})}\BibitemShut {NoStop}%
\bibitem [{\citenamefont {Trotzky}\ \emph {et~al.}(2012)\citenamefont
  {Trotzky}, \citenamefont {Chen}, \citenamefont {Flesch}, \citenamefont
  {McCulloch}, \citenamefont {Schollw{\"o}ck}, \citenamefont {Eisert},\ and\
  \citenamefont {Bloch}}]{trotzky2012}%
  \BibitemOpen
  \bibfield  {author} {\bibinfo {author} {\bibfnamefont {S.}~\bibnamefont
  {Trotzky}}, \bibinfo {author} {\bibfnamefont {Y.-A.}\ \bibnamefont {Chen}},
  \bibinfo {author} {\bibfnamefont {A.}~\bibnamefont {Flesch}}, \bibinfo
  {author} {\bibfnamefont {I.~P.}\ \bibnamefont {McCulloch}}, \bibinfo {author}
  {\bibfnamefont {U.}~\bibnamefont {Schollw{\"o}ck}}, \bibinfo {author}
  {\bibfnamefont {J.}~\bibnamefont {Eisert}},\ and\ \bibinfo {author}
  {\bibfnamefont {I.}~\bibnamefont {Bloch}},\ }\bibfield  {title} {\emph
  {\bibinfo {title} {{Probing the relaxation towards equilibrium in an isolated
  strongly correlated one-dimensional Bose gas}}},\ }\href
  {https://doi.org/10.1038/nphys2232} {\bibfield  {journal} {\bibinfo
  {journal} {Nat. Phys.}\ }\textbf {\bibinfo {volume} {8}},\ \bibinfo {pages}
  {325} (\bibinfo {year} {2012})}\BibitemShut {NoStop}%
\bibitem [{\citenamefont {Langen}\ \emph {et~al.}(2015)\citenamefont {Langen},
  \citenamefont {Erne}, \citenamefont {Geiger}, \citenamefont {Rauer},
  \citenamefont {Schweigler}, \citenamefont {Kuhnert}, \citenamefont
  {Rohringer}, \citenamefont {Mazets}, \citenamefont {Gasenzer},\ and\
  \citenamefont {Schmiedmayer}}]{langen2015}%
  \BibitemOpen
  \bibfield  {author} {\bibinfo {author} {\bibfnamefont {T.}~\bibnamefont
  {Langen}}, \bibinfo {author} {\bibfnamefont {S.}~\bibnamefont {Erne}},
  \bibinfo {author} {\bibfnamefont {R.}~\bibnamefont {Geiger}}, \bibinfo
  {author} {\bibfnamefont {B.}~\bibnamefont {Rauer}}, \bibinfo {author}
  {\bibfnamefont {T.}~\bibnamefont {Schweigler}}, \bibinfo {author}
  {\bibfnamefont {M.}~\bibnamefont {Kuhnert}}, \bibinfo {author} {\bibfnamefont
  {W.}~\bibnamefont {Rohringer}}, \bibinfo {author} {\bibfnamefont {I.~E.}\
  \bibnamefont {Mazets}}, \bibinfo {author} {\bibfnamefont {T.}~\bibnamefont
  {Gasenzer}},\ and\ \bibinfo {author} {\bibfnamefont {J.}~\bibnamefont
  {Schmiedmayer}},\ }\bibfield  {title} {\emph {\bibinfo {title} {{Experimental
  observation of a generalized Gibbs ensemble}}},\ }\href
  {https://doi.org/10.1126/science.1257026} {\bibfield  {journal} {\bibinfo
  {journal} {Science}\ }\textbf {\bibinfo {volume} {348}},\ \bibinfo {pages}
  {207} (\bibinfo {year} {2015})}\BibitemShut {NoStop}%
\bibitem [{\citenamefont {Kaufman}\ \emph {et~al.}(2016)\citenamefont
  {Kaufman}, \citenamefont {Tai}, \citenamefont {Lukin}, \citenamefont
  {Rispoli}, \citenamefont {Schittko}, \citenamefont {Preiss},\ and\
  \citenamefont {Greiner}}]{kaufman2016}%
  \BibitemOpen
  \bibfield  {author} {\bibinfo {author} {\bibfnamefont {A.~M.}\ \bibnamefont
  {Kaufman}}, \bibinfo {author} {\bibfnamefont {M.~E.}\ \bibnamefont {Tai}},
  \bibinfo {author} {\bibfnamefont {A.}~\bibnamefont {Lukin}}, \bibinfo
  {author} {\bibfnamefont {M.}~\bibnamefont {Rispoli}}, \bibinfo {author}
  {\bibfnamefont {R.}~\bibnamefont {Schittko}}, \bibinfo {author}
  {\bibfnamefont {P.~M.}\ \bibnamefont {Preiss}},\ and\ \bibinfo {author}
  {\bibfnamefont {M.}~\bibnamefont {Greiner}},\ }\bibfield  {title} {\emph
  {\bibinfo {title} {{Quantum thermalization through entanglement in an
  isolated many-body system}}},\ }\href
  {https://doi.org/10.1126/science.aaf6725} {\bibfield  {journal} {\bibinfo
  {journal} {Science}\ }\textbf {\bibinfo {volume} {353}},\ \bibinfo {pages}
  {794} (\bibinfo {year} {2016})}\BibitemShut {NoStop}%
\bibitem [{\citenamefont {Schreiber}\ \emph {et~al.}(2015)\citenamefont
  {Schreiber}, \citenamefont {Hodgman}, \citenamefont {Bordia}, \citenamefont
  {L{\"u}schen}, \citenamefont {Fischer}, \citenamefont {Vosk}, \citenamefont
  {Altman}, \citenamefont {Schneider},\ and\ \citenamefont
  {Bloch}}]{schreiber2015}%
  \BibitemOpen
  \bibfield  {author} {\bibinfo {author} {\bibfnamefont {M.}~\bibnamefont
  {Schreiber}}, \bibinfo {author} {\bibfnamefont {S.~S.}\ \bibnamefont
  {Hodgman}}, \bibinfo {author} {\bibfnamefont {P.}~\bibnamefont {Bordia}},
  \bibinfo {author} {\bibfnamefont {H.~P.}\ \bibnamefont {L{\"u}schen}},
  \bibinfo {author} {\bibfnamefont {M.~H.}\ \bibnamefont {Fischer}}, \bibinfo
  {author} {\bibfnamefont {R.}~\bibnamefont {Vosk}}, \bibinfo {author}
  {\bibfnamefont {E.}~\bibnamefont {Altman}}, \bibinfo {author} {\bibfnamefont
  {U.}~\bibnamefont {Schneider}},\ and\ \bibinfo {author} {\bibfnamefont
  {I.}~\bibnamefont {Bloch}},\ }\bibfield  {title} {\emph {\bibinfo {title}
  {{Observation of many-body localization of interacting fermions in a
  quasirandom optical lattice}}},\ }\href
  {https://doi.org/10.1126/science.aaa7432} {\bibfield  {journal} {\bibinfo
  {journal} {Science}\ }\textbf {\bibinfo {volume} {349}},\ \bibinfo {pages}
  {842} (\bibinfo {year} {2015})}\BibitemShut {NoStop}%
\bibitem [{\citenamefont {Choi}\ \emph {et~al.}(2016)\citenamefont {Choi},
  \citenamefont {Hild}, \citenamefont {Zeiher}, \citenamefont {Schau{\ss}},
  \citenamefont {Rubio-Abadal}, \citenamefont {Yefsah}, \citenamefont
  {Khemani}, \citenamefont {Huse}, \citenamefont {Bloch},\ and\ \citenamefont
  {Gross}}]{choi2016}%
  \BibitemOpen
  \bibfield  {author} {\bibinfo {author} {\bibfnamefont {J.-y.}\ \bibnamefont
  {Choi}}, \bibinfo {author} {\bibfnamefont {S.}~\bibnamefont {Hild}}, \bibinfo
  {author} {\bibfnamefont {J.}~\bibnamefont {Zeiher}}, \bibinfo {author}
  {\bibfnamefont {P.}~\bibnamefont {Schau{\ss}}}, \bibinfo {author}
  {\bibfnamefont {A.}~\bibnamefont {Rubio-Abadal}}, \bibinfo {author}
  {\bibfnamefont {T.}~\bibnamefont {Yefsah}}, \bibinfo {author} {\bibfnamefont
  {V.}~\bibnamefont {Khemani}}, \bibinfo {author} {\bibfnamefont {D.~A.}\
  \bibnamefont {Huse}}, \bibinfo {author} {\bibfnamefont {I.}~\bibnamefont
  {Bloch}},\ and\ \bibinfo {author} {\bibfnamefont {C.}~\bibnamefont {Gross}},\
  }\bibfield  {title} {\emph {\bibinfo {title} {{Exploring the many-body
  localization transition in two dimensions}}},\ }\href
  {https://doi.org/10.1126/science.aaf8834} {\bibfield  {journal} {\bibinfo
  {journal} {Science}\ }\textbf {\bibinfo {volume} {352}},\ \bibinfo {pages}
  {1547} (\bibinfo {year} {2016})}\BibitemShut {NoStop}%
\bibitem [{\citenamefont {Smith}\ \emph {et~al.}(2016)\citenamefont {Smith},
  \citenamefont {Lee}, \citenamefont {Richerme}, \citenamefont {Neyenhuis},
  \citenamefont {Hess}, \citenamefont {Hauke}, \citenamefont {Heyl},
  \citenamefont {Huse},\ and\ \citenamefont {Monroe}}]{smith2016}%
  \BibitemOpen
  \bibfield  {author} {\bibinfo {author} {\bibfnamefont {J.}~\bibnamefont
  {Smith}}, \bibinfo {author} {\bibfnamefont {A.}~\bibnamefont {Lee}}, \bibinfo
  {author} {\bibfnamefont {P.}~\bibnamefont {Richerme}}, \bibinfo {author}
  {\bibfnamefont {B.}~\bibnamefont {Neyenhuis}}, \bibinfo {author}
  {\bibfnamefont {P.~W.}\ \bibnamefont {Hess}}, \bibinfo {author}
  {\bibfnamefont {P.}~\bibnamefont {Hauke}}, \bibinfo {author} {\bibfnamefont
  {M.}~\bibnamefont {Heyl}}, \bibinfo {author} {\bibfnamefont {D.~A.}\
  \bibnamefont {Huse}},\ and\ \bibinfo {author} {\bibfnamefont
  {C.}~\bibnamefont {Monroe}},\ }\bibfield  {title} {\emph {\bibinfo {title}
  {{Many-body localization in a quantum simulator with programmable random
  disorder}}},\ }\href {https://doi.org/10.1038/nphys3783} {\bibfield
  {journal} {\bibinfo  {journal} {Nat. Phys.}\ }\textbf {\bibinfo {volume}
  {12}},\ \bibinfo {pages} {907} (\bibinfo {year} {2016})}\BibitemShut
  {NoStop}%
\bibitem [{\citenamefont {Bernien}\ \emph {et~al.}(2017)\citenamefont
  {Bernien}, \citenamefont {Schwartz}, \citenamefont {Keesling}, \citenamefont
  {Levine}, \citenamefont {Omran}, \citenamefont {Pichler}, \citenamefont
  {Choi}, \citenamefont {Zibrov}, \citenamefont {Endres}, \citenamefont
  {Greiner}, \citenamefont {Vuleti\'{c}},\ and\ \citenamefont
  {Lukin}}]{bernien2017}%
  \BibitemOpen
  \bibfield  {author} {\bibinfo {author} {\bibfnamefont {H.}~\bibnamefont
  {Bernien}}, \bibinfo {author} {\bibfnamefont {S.}~\bibnamefont {Schwartz}},
  \bibinfo {author} {\bibfnamefont {A.}~\bibnamefont {Keesling}}, \bibinfo
  {author} {\bibfnamefont {H.}~\bibnamefont {Levine}}, \bibinfo {author}
  {\bibfnamefont {A.}~\bibnamefont {Omran}}, \bibinfo {author} {\bibfnamefont
  {H.}~\bibnamefont {Pichler}}, \bibinfo {author} {\bibfnamefont
  {S.}~\bibnamefont {Choi}}, \bibinfo {author} {\bibfnamefont {A.~S.}\
  \bibnamefont {Zibrov}}, \bibinfo {author} {\bibfnamefont {M.}~\bibnamefont
  {Endres}}, \bibinfo {author} {\bibfnamefont {M.}~\bibnamefont {Greiner}},
  \bibinfo {author} {\bibfnamefont {V.}~\bibnamefont {Vuleti\'{c}}},\ and\
  \bibinfo {author} {\bibfnamefont {M.~D.}\ \bibnamefont {Lukin}},\ }\bibfield
  {title} {\emph {\bibinfo {title} {{Probing many-body dynamics on a 51-atom
  quantum simulator}}},\ }\href {https://doi.org/10.1038/nature24622}
  {\bibfield  {journal} {\bibinfo  {journal} {Nature}\ }\textbf {\bibinfo
  {volume} {551}},\ \bibinfo {pages} {579} (\bibinfo {year}
  {2017})}\BibitemShut {NoStop}%
\bibitem [{\citenamefont {Turner}\ \emph {et~al.}(2018)\citenamefont {Turner},
  \citenamefont {Michailidis}, \citenamefont {Abanin}, \citenamefont {Serbyn},\
  and\ \citenamefont {Papi{\'c}}}]{turner2018}%
  \BibitemOpen
  \bibfield  {author} {\bibinfo {author} {\bibfnamefont {C.~J.}\ \bibnamefont
  {Turner}}, \bibinfo {author} {\bibfnamefont {A.~A.}\ \bibnamefont
  {Michailidis}}, \bibinfo {author} {\bibfnamefont {D.~A.}\ \bibnamefont
  {Abanin}}, \bibinfo {author} {\bibfnamefont {M.}~\bibnamefont {Serbyn}},\
  and\ \bibinfo {author} {\bibfnamefont {Z.}~\bibnamefont {Papi{\'c}}},\
  }\bibfield  {title} {\emph {\bibinfo {title} {{Weak ergodicity breaking from
  quantum many-body scars}}},\ }\href
  {https://doi.org/10.1038/s41567-018-0137-5} {\bibfield  {journal} {\bibinfo
  {journal} {Nat. Phys.}\ }\textbf {\bibinfo {volume} {14}},\ \bibinfo {pages}
  {745} (\bibinfo {year} {2018})}\BibitemShut {NoStop}%
\bibitem [{\citenamefont {Barmettler}\ \emph {et~al.}(2012)\citenamefont
  {Barmettler}, \citenamefont {Poletti}, \citenamefont {Cheneau},\ and\
  \citenamefont {Kollath}}]{barmettler2012}%
  \BibitemOpen
  \bibfield  {author} {\bibinfo {author} {\bibfnamefont {P.}~\bibnamefont
  {Barmettler}}, \bibinfo {author} {\bibfnamefont {D.}~\bibnamefont {Poletti}},
  \bibinfo {author} {\bibfnamefont {M.}~\bibnamefont {Cheneau}},\ and\ \bibinfo
  {author} {\bibfnamefont {C.}~\bibnamefont {Kollath}},\ }\bibfield  {title}
  {\emph {\bibinfo {title} {{Propagation front of correlations in an
  interacting Bose gas}}},\ }\href {https://doi.org/10.1103/PhysRevA.85.053625}
  {\bibfield  {journal} {\bibinfo  {journal} {Phys. Rev. A}\ }\textbf {\bibinfo
  {volume} {85}},\ \bibinfo {pages} {053625} (\bibinfo {year}
  {2012})}\BibitemShut {NoStop}%
\bibitem [{\citenamefont {Nagao}\ \emph {et~al.}(2019)\citenamefont {Nagao},
  \citenamefont {Kunimi}, \citenamefont {Takasu}, \citenamefont {Takahashi},\
  and\ \citenamefont {Danshita}}]{nagao2019}%
  \BibitemOpen
  \bibfield  {author} {\bibinfo {author} {\bibfnamefont {K.}~\bibnamefont
  {Nagao}}, \bibinfo {author} {\bibfnamefont {M.}~\bibnamefont {Kunimi}},
  \bibinfo {author} {\bibfnamefont {Y.}~\bibnamefont {Takasu}}, \bibinfo
  {author} {\bibfnamefont {Y.}~\bibnamefont {Takahashi}},\ and\ \bibinfo
  {author} {\bibfnamefont {I.}~\bibnamefont {Danshita}},\ }\bibfield  {title}
  {\emph {\bibinfo {title} {{Semiclassical quench dynamics of Bose gases in
  optical lattices}}},\ }\href {https://doi.org/10.1103/PhysRevA.99.023622}
  {\bibfield  {journal} {\bibinfo  {journal} {Phys. Rev. A}\ }\textbf {\bibinfo
  {volume} {99}},\ \bibinfo {pages} {023622} (\bibinfo {year}
  {2019})}\BibitemShut {NoStop}%
\bibitem [{\citenamefont {Nagao}\ \emph {et~al.}(2021)\citenamefont {Nagao},
  \citenamefont {Takasu}, \citenamefont {Takahashi},\ and\ \citenamefont
  {Danshita}}]{nagao2021}%
  \BibitemOpen
  \bibfield  {author} {\bibinfo {author} {\bibfnamefont {K.}~\bibnamefont
  {Nagao}}, \bibinfo {author} {\bibfnamefont {Y.}~\bibnamefont {Takasu}},
  \bibinfo {author} {\bibfnamefont {Y.}~\bibnamefont {Takahashi}},\ and\
  \bibinfo {author} {\bibfnamefont {I.}~\bibnamefont {Danshita}},\ }\bibfield
  {title} {\emph {\bibinfo {title} {{SU(3) truncated Wigner approximation for
  strongly interacting Bose gases}}},\ }\href
  {https://doi.org/10.1103/PhysRevResearch.3.043091} {\bibfield  {journal}
  {\bibinfo  {journal} {Phys. Rev. Research}\ }\textbf {\bibinfo {volume}
  {3}},\ \bibinfo {pages} {043091} (\bibinfo {year} {2021})}\BibitemShut
  {NoStop}%
\bibitem [{\citenamefont {Mokhtari-Jazi}\ \emph {et~al.}(2021)\citenamefont
  {Mokhtari-Jazi}, \citenamefont {Fitzpatrick},\ and\ \citenamefont
  {Kennett}}]{mokhtarijazi2021}%
  \BibitemOpen
  \bibfield  {author} {\bibinfo {author} {\bibfnamefont {A.}~\bibnamefont
  {Mokhtari-Jazi}}, \bibinfo {author} {\bibfnamefont {M.~R.~C.}\ \bibnamefont
  {Fitzpatrick}},\ and\ \bibinfo {author} {\bibfnamefont {M.~P.}\ \bibnamefont
  {Kennett}},\ }\bibfield  {title} {\emph {\bibinfo {title} {{Phase and group
  velocities for correlation spreading in the Mott phase of the Bose-Hubbard
  model in dimensions greater than one}}},\ }\href
  {https://doi.org/10.1103/PhysRevA.103.023334} {\bibfield  {journal} {\bibinfo
   {journal} {Phys. Rev. A}\ }\textbf {\bibinfo {volume} {103}},\ \bibinfo
  {pages} {023334} (\bibinfo {year} {2021})}\BibitemShut {NoStop}%
\bibitem [{\citenamefont {Greiner}\ \emph {et~al.}(2002)\citenamefont
  {Greiner}, \citenamefont {Mandel}, \citenamefont {Esslinger}, \citenamefont
  {H{\"a}nsch},\ and\ \citenamefont {Bloch}}]{greiner2002}%
  \BibitemOpen
  \bibfield  {author} {\bibinfo {author} {\bibfnamefont {M.}~\bibnamefont
  {Greiner}}, \bibinfo {author} {\bibfnamefont {O.}~\bibnamefont {Mandel}},
  \bibinfo {author} {\bibfnamefont {T.}~\bibnamefont {Esslinger}}, \bibinfo
  {author} {\bibfnamefont {T.~W.}\ \bibnamefont {H{\"a}nsch}},\ and\ \bibinfo
  {author} {\bibfnamefont {I.}~\bibnamefont {Bloch}},\ }\bibfield  {title}
  {\emph {\bibinfo {title} {{Quantum phase transition from a superfluid to a
  Mott insulator in a gas of ultracold atoms}}},\ }\href
  {https://doi.org/10.1038/415039a} {\bibfield  {journal} {\bibinfo  {journal}
  {Nature}\ }\textbf {\bibinfo {volume} {415}},\ \bibinfo {pages} {39}
  (\bibinfo {year} {2002})}\BibitemShut {NoStop}%
\bibitem [{\citenamefont {L{\"a}uchli}\ and\ \citenamefont
  {Kollath}(2008)}]{lauchli2008}%
  \BibitemOpen
  \bibfield  {author} {\bibinfo {author} {\bibfnamefont {A.~M.}\ \bibnamefont
  {L{\"a}uchli}}\ and\ \bibinfo {author} {\bibfnamefont {C.}~\bibnamefont
  {Kollath}},\ }\bibfield  {title} {\emph {\bibinfo {title} {{Spreading of
  correlations and entanglement after a quench in the one-dimensional
  Bose--Hubbard model}}},\ }\href
  {https://doi.org/10.1088/1742-5468/2008/05/P05018} {\bibfield  {journal}
  {\bibinfo  {journal} {J. Stat. Mech. Theory Exp.}\ }\textbf {\bibinfo
  {volume} {2008}},\ \bibinfo {pages} {P05018} (\bibinfo {year}
  {2008})}\BibitemShut {NoStop}%
\bibitem [{\citenamefont {Schuch}\ \emph {et~al.}(2011)\citenamefont {Schuch},
  \citenamefont {Harrison}, \citenamefont {Osborne},\ and\ \citenamefont
  {Eisert}}]{schuch2011}%
  \BibitemOpen
  \bibfield  {author} {\bibinfo {author} {\bibfnamefont {N.}~\bibnamefont
  {Schuch}}, \bibinfo {author} {\bibfnamefont {S.~K.}\ \bibnamefont
  {Harrison}}, \bibinfo {author} {\bibfnamefont {T.~J.}\ \bibnamefont
  {Osborne}},\ and\ \bibinfo {author} {\bibfnamefont {J.}~\bibnamefont
  {Eisert}},\ }\bibfield  {title} {\emph {\bibinfo {title} {{Information
  propagation for interacting-particle systems}}},\ }\href
  {https://doi.org/10.1103/PhysRevA.84.032309} {\bibfield  {journal} {\bibinfo
  {journal} {Phys. Rev. A}\ }\textbf {\bibinfo {volume} {84}},\ \bibinfo
  {pages} {032309} (\bibinfo {year} {2011})}\BibitemShut {NoStop}%
\bibitem [{\citenamefont {Carleo}\ \emph {et~al.}(2014)\citenamefont {Carleo},
  \citenamefont {Becca}, \citenamefont {Sanchez-Palencia}, \citenamefont
  {Sorella},\ and\ \citenamefont {Fabrizio}}]{carleo2014}%
  \BibitemOpen
  \bibfield  {author} {\bibinfo {author} {\bibfnamefont {G.}~\bibnamefont
  {Carleo}}, \bibinfo {author} {\bibfnamefont {F.}~\bibnamefont {Becca}},
  \bibinfo {author} {\bibfnamefont {L.}~\bibnamefont {Sanchez-Palencia}},
  \bibinfo {author} {\bibfnamefont {S.}~\bibnamefont {Sorella}},\ and\ \bibinfo
  {author} {\bibfnamefont {M.}~\bibnamefont {Fabrizio}},\ }\bibfield  {title}
  {\emph {\bibinfo {title} {{Light-cone effect and supersonic correlations in
  one- and two-dimensional bosonic superfluids}}},\ }\href
  {https://doi.org/10.1103/PhysRevA.89.031602} {\bibfield  {journal} {\bibinfo
  {journal} {Phys. Rev. A}\ }\textbf {\bibinfo {volume} {89}},\ \bibinfo
  {pages} {031602(R)} (\bibinfo {year} {2014})}\BibitemShut {NoStop}%
\bibitem [{\citenamefont {Cevolani}\ \emph {et~al.}(2018)\citenamefont
  {Cevolani}, \citenamefont {Despres}, \citenamefont {Carleo}, \citenamefont
  {Tagliacozzo},\ and\ \citenamefont {Sanchez-Palencia}}]{cevolani2018}%
  \BibitemOpen
  \bibfield  {author} {\bibinfo {author} {\bibfnamefont {L.}~\bibnamefont
  {Cevolani}}, \bibinfo {author} {\bibfnamefont {J.}~\bibnamefont {Despres}},
  \bibinfo {author} {\bibfnamefont {G.}~\bibnamefont {Carleo}}, \bibinfo
  {author} {\bibfnamefont {L.}~\bibnamefont {Tagliacozzo}},\ and\ \bibinfo
  {author} {\bibfnamefont {L.}~\bibnamefont {Sanchez-Palencia}},\ }\bibfield
  {title} {\emph {\bibinfo {title} {{Universal scaling laws for correlation
  spreading in quantum systems with short- and long-range interactions}}},\
  }\href {https://doi.org/10.1103/PhysRevB.98.024302} {\bibfield  {journal}
  {\bibinfo  {journal} {Phys. Rev. B}\ }\textbf {\bibinfo {volume} {98}},\
  \bibinfo {pages} {024302} (\bibinfo {year} {2018})}\BibitemShut {NoStop}%
\bibitem [{\citenamefont {Fitzpatrick}\ and\ \citenamefont
  {Kennett}(2018)}]{fitzpatrick2018}%
  \BibitemOpen
  \bibfield  {author} {\bibinfo {author} {\bibfnamefont {M.~R.~C.}\
  \bibnamefont {Fitzpatrick}}\ and\ \bibinfo {author} {\bibfnamefont {M.~P.}\
  \bibnamefont {Kennett}},\ }\bibfield  {title} {\emph {\bibinfo {title}
  {{Light-cone-like spreading of single-particle correlations in the
  Bose-Hubbard model after a quantum quench in the strong-coupling regime}}},\
  }\href {https://doi.org/10.1103/PhysRevA.98.053618} {\bibfield  {journal}
  {\bibinfo  {journal} {Phys. Rev. A}\ }\textbf {\bibinfo {volume} {98}},\
  \bibinfo {pages} {053618} (\bibinfo {year} {2018})}\BibitemShut {NoStop}%
\bibitem [{\citenamefont {Despres}\ \emph {et~al.}(2019)\citenamefont
  {Despres}, \citenamefont {Villa},\ and\ \citenamefont
  {Sanchez-Palencia}}]{despres2019}%
  \BibitemOpen
  \bibfield  {author} {\bibinfo {author} {\bibfnamefont {J.}~\bibnamefont
  {Despres}}, \bibinfo {author} {\bibfnamefont {L.}~\bibnamefont {Villa}},\
  and\ \bibinfo {author} {\bibfnamefont {L.}~\bibnamefont {Sanchez-Palencia}},\
  }\bibfield  {title} {\emph {\bibinfo {title} {{Twofold correlation spreading
  in a strongly correlated lattice Bose gas}}},\ }\href
  {https://doi.org/10.1038/s41598-019-40679-3} {\bibfield  {journal} {\bibinfo
  {journal} {Sci. Rep.}\ }\textbf {\bibinfo {volume} {9}},\ \bibinfo {pages}
  {4135} (\bibinfo {year} {2019})}\BibitemShut {NoStop}%
\bibitem [{\citenamefont {Wang}\ and\ \citenamefont
  {Hazzard}(2020)}]{wang2020}%
  \BibitemOpen
  \bibfield  {author} {\bibinfo {author} {\bibfnamefont {Z.}~\bibnamefont
  {Wang}}\ and\ \bibinfo {author} {\bibfnamefont {K.~R.~A.}\ \bibnamefont
  {Hazzard}},\ }\bibfield  {title} {\emph {\bibinfo {title} {{Tightening the
  Lieb-Robinson Bound in Locally Interacting Systems}}},\ }\href
  {https://doi.org/10.1103/PRXQuantum.1.010303} {\bibfield  {journal} {\bibinfo
   {journal} {PRX Quantum}\ }\textbf {\bibinfo {volume} {1}},\ \bibinfo {pages}
  {010303} (\bibinfo {year} {2020})}\BibitemShut {NoStop}%
\bibitem [{\citenamefont {Kuwahara}\ and\ \citenamefont
  {Saito}(2021)}]{kuwahara2021}%
  \BibitemOpen
  \bibfield  {author} {\bibinfo {author} {\bibfnamefont {T.}~\bibnamefont
  {Kuwahara}}\ and\ \bibinfo {author} {\bibfnamefont {K.}~\bibnamefont
  {Saito}},\ }\bibfield  {title} {\emph {\bibinfo {title} {{Lieb-Robinson Bound
  and Almost-Linear Light Cone in Interacting Boson Systems}}},\ }\href
  {https://doi.org/10.1103/PhysRevLett.127.070403} {\bibfield  {journal}
  {\bibinfo  {journal} {Phys. Rev. Lett.}\ }\textbf {\bibinfo {volume} {127}},\
  \bibinfo {pages} {070403} (\bibinfo {year} {2021})}\BibitemShut {NoStop}%
\bibitem [{\citenamefont {Yin}\ and\ \citenamefont {Lucas}()}]{yin2021_arxiv}%
  \BibitemOpen
  \bibfield  {author} {\bibinfo {author} {\bibfnamefont {C.}~\bibnamefont
  {Yin}}\ and\ \bibinfo {author} {\bibfnamefont {A.}~\bibnamefont {Lucas}},\
  }\href@noop {} {\bibfield  {title} {\emph {\bibinfo {title} {{Finite speed of
  quantum information in models of interacting bosons at finite density}}}},\
  }\Eprint {https://arxiv.org/abs/2106.09726} {arXiv:2106.09726} \BibitemShut
  {NoStop}%
\bibitem [{\citenamefont {Faupin}\ \emph {et~al.}({\natexlab{a}})\citenamefont
  {Faupin}, \citenamefont {Lemm},\ and\ \citenamefont
  {Sigal}}]{faupin2021a_arxiv}%
  \BibitemOpen
  \bibfield  {author} {\bibinfo {author} {\bibfnamefont {J.}~\bibnamefont
  {Faupin}}, \bibinfo {author} {\bibfnamefont {M.}~\bibnamefont {Lemm}},\ and\
  \bibinfo {author} {\bibfnamefont {I.~M.}\ \bibnamefont {Sigal}},\ }\href@noop
  {} {\bibfield  {title} {\emph {\bibinfo {title} {{On Lieb-Robinson bounds for
  the Bose-Hubbard model}}}},\ }\Eprint {https://arxiv.org/abs/2109.04103}
  {arXiv:2109.04103} \BibitemShut {NoStop}%
\bibitem [{\citenamefont {Faupin}\ \emph {et~al.}({\natexlab{b}})\citenamefont
  {Faupin}, \citenamefont {Lemm},\ and\ \citenamefont
  {Sigal}}]{faupin2021b_arxiv}%
  \BibitemOpen
  \bibfield  {author} {\bibinfo {author} {\bibfnamefont {J.}~\bibnamefont
  {Faupin}}, \bibinfo {author} {\bibfnamefont {M.}~\bibnamefont {Lemm}},\ and\
  \bibinfo {author} {\bibfnamefont {I.~M.}\ \bibnamefont {Sigal}},\ }\href@noop
  {} {\bibfield  {title} {\emph {\bibinfo {title} {{Maximal speed for
  macroscopic particle transport in the Bose-Hubbard model}}}},\ }\Eprint
  {https://arxiv.org/abs/2110.04313} {arXiv:2110.04313} \BibitemShut {NoStop}%
\bibitem [{\citenamefont {Mart\'{\i}n-Delgado}\ \emph
  {et~al.}(2001)\citenamefont {Mart\'{\i}n-Delgado}, \citenamefont
  {Roncaglia},\ and\ \citenamefont {Sierra}}]{martin-delgado2001}%
  \BibitemOpen
  \bibfield  {author} {\bibinfo {author} {\bibfnamefont {M.~A.}\ \bibnamefont
  {Mart\'{\i}n-Delgado}}, \bibinfo {author} {\bibfnamefont {M.}~\bibnamefont
  {Roncaglia}},\ and\ \bibinfo {author} {\bibfnamefont {G.}~\bibnamefont
  {Sierra}},\ }\bibfield  {title} {\emph {\bibinfo {title} {{Stripe ans\"atze
  from exactly solved models}}},\ }\href
  {https://doi.org/10.1103/PhysRevB.64.075117} {\bibfield  {journal} {\bibinfo
  {journal} {Phys. Rev. B}\ }\textbf {\bibinfo {volume} {64}},\ \bibinfo
  {pages} {075117} (\bibinfo {year} {2001})}\BibitemShut {NoStop}%
\bibitem [{\citenamefont {Verstraete}\ and\ \citenamefont
  {Cirac}()}]{verstraete2004a_arxiv}%
  \BibitemOpen
  \bibfield  {author} {\bibinfo {author} {\bibfnamefont {F.}~\bibnamefont
  {Verstraete}}\ and\ \bibinfo {author} {\bibfnamefont {J.~I.}\ \bibnamefont
  {Cirac}},\ }\href@noop {} {\bibfield  {title} {\emph {\bibinfo {title}
  {{Renormalization algorithms for quantum-many body systems in two and higher
  dimensions}}}},\ }\Eprint {https://arxiv.org/abs/cond-mat/0407066}
  {arXiv:cond-mat/0407066} \BibitemShut {NoStop}%
\bibitem [{\citenamefont {Verstraete}\ and\ \citenamefont
  {Cirac}(2004)}]{verstraete2004b}%
  \BibitemOpen
  \bibfield  {author} {\bibinfo {author} {\bibfnamefont {F.}~\bibnamefont
  {Verstraete}}\ and\ \bibinfo {author} {\bibfnamefont {J.~I.}\ \bibnamefont
  {Cirac}},\ }\bibfield  {title} {\emph {\bibinfo {title} {{Valence-bond states
  for quantum computation}}},\ }\href
  {https://doi.org/10.1103/PhysRevA.70.060302} {\bibfield  {journal} {\bibinfo
  {journal} {Phys. Rev. A}\ }\textbf {\bibinfo {volume} {70}},\ \bibinfo
  {pages} {060302(R)} (\bibinfo {year} {2004})}\BibitemShut {NoStop}%
\bibitem [{\citenamefont {Verstraete}\ \emph {et~al.}(2008)\citenamefont
  {Verstraete}, \citenamefont {Murg},\ and\ \citenamefont
  {Cirac}}]{verstraete2008}%
  \BibitemOpen
  \bibfield  {author} {\bibinfo {author} {\bibfnamefont {F.}~\bibnamefont
  {Verstraete}}, \bibinfo {author} {\bibfnamefont {V.}~\bibnamefont {Murg}},\
  and\ \bibinfo {author} {\bibfnamefont {J.~I.}\ \bibnamefont {Cirac}},\
  }\bibfield  {title} {\emph {\bibinfo {title} {{Matrix product states,
  projected entangled pair states, and variational renormalization group
  methods for quantum spin systems}}},\ }\href
  {https://doi.org/10.1080/14789940801912366} {\bibfield  {journal} {\bibinfo
  {journal} {Adv. Phys.}\ }\textbf {\bibinfo {volume} {57}},\ \bibinfo {pages}
  {143} (\bibinfo {year} {2008})}\BibitemShut {NoStop}%
\bibitem [{\citenamefont {Jordan}\ \emph {et~al.}(2008)\citenamefont {Jordan},
  \citenamefont {Or\'us}, \citenamefont {Vidal}, \citenamefont {Verstraete},\
  and\ \citenamefont {Cirac}}]{jordan2008}%
  \BibitemOpen
  \bibfield  {author} {\bibinfo {author} {\bibfnamefont {J.}~\bibnamefont
  {Jordan}}, \bibinfo {author} {\bibfnamefont {R.}~\bibnamefont {Or\'us}},
  \bibinfo {author} {\bibfnamefont {G.}~\bibnamefont {Vidal}}, \bibinfo
  {author} {\bibfnamefont {F.}~\bibnamefont {Verstraete}},\ and\ \bibinfo
  {author} {\bibfnamefont {J.~I.}\ \bibnamefont {Cirac}},\ }\bibfield  {title}
  {\emph {\bibinfo {title} {{Classical Simulation of Infinite-Size Quantum
  Lattice Systems in Two Spatial Dimensions}}},\ }\href
  {https://doi.org/10.1103/PhysRevLett.101.250602} {\bibfield  {journal}
  {\bibinfo  {journal} {Phys. Rev. Lett.}\ }\textbf {\bibinfo {volume} {101}},\
  \bibinfo {pages} {250602} (\bibinfo {year} {2008})}\BibitemShut {NoStop}%
\bibitem [{\citenamefont {Phien}\ \emph {et~al.}(2015)\citenamefont {Phien},
  \citenamefont {Bengua}, \citenamefont {Tuan}, \citenamefont {Corboz},\ and\
  \citenamefont {Or\'us}}]{phien2015}%
  \BibitemOpen
  \bibfield  {author} {\bibinfo {author} {\bibfnamefont {H.~N.}\ \bibnamefont
  {Phien}}, \bibinfo {author} {\bibfnamefont {J.~A.}\ \bibnamefont {Bengua}},
  \bibinfo {author} {\bibfnamefont {H.~D.}\ \bibnamefont {Tuan}}, \bibinfo
  {author} {\bibfnamefont {P.}~\bibnamefont {Corboz}},\ and\ \bibinfo {author}
  {\bibfnamefont {R.}~\bibnamefont {Or\'us}},\ }\bibfield  {title} {\emph
  {\bibinfo {title} {{Infinite projected entangled pair states algorithm
  improved: Fast full update and gauge fixing}}},\ }\href
  {https://doi.org/10.1103/PhysRevB.92.035142} {\bibfield  {journal} {\bibinfo
  {journal} {Phys. Rev. B}\ }\textbf {\bibinfo {volume} {92}},\ \bibinfo
  {pages} {035142} (\bibinfo {year} {2015})}\BibitemShut {NoStop}%
\bibitem [{\citenamefont {Or{\'u}s}(2014)}]{orus2014}%
  \BibitemOpen
  \bibfield  {author} {\bibinfo {author} {\bibfnamefont {R.}~\bibnamefont
  {Or{\'u}s}},\ }\bibfield  {title} {\emph {\bibinfo {title} {{A practical
  introduction to tensor networks: Matrix product states and projected
  entangled pair states}}},\ }\href {https://doi.org/10.1016/j.aop.2014.06.013}
  {\bibfield  {journal} {\bibinfo  {journal} {Ann. Phys. (N.Y.)}\ }\textbf
  {\bibinfo {volume} {349}},\ \bibinfo {pages} {117} (\bibinfo {year}
  {2014})}\BibitemShut {NoStop}%
\bibitem [{\citenamefont {Or{\'u}s}(2019)}]{orus2019}%
  \BibitemOpen
  \bibfield  {author} {\bibinfo {author} {\bibfnamefont {R.}~\bibnamefont
  {Or{\'u}s}},\ }\bibfield  {title} {\emph {\bibinfo {title} {{Tensor networks
  for complex quantum systems}}},\ }\href
  {https://doi.org/10.1038/s42254-019-0086-7} {\bibfield  {journal} {\bibinfo
  {journal} {Nat. Rev. Phys.}\ }\textbf {\bibinfo {volume} {1}},\ \bibinfo
  {pages} {538} (\bibinfo {year} {2019})}\BibitemShut {NoStop}%
\bibitem [{\citenamefont {Hieida}\ \emph {et~al.}(1999)\citenamefont {Hieida},
  \citenamefont {Okunishi},\ and\ \citenamefont {Akutsu}}]{hieida1999}%
  \BibitemOpen
  \bibfield  {author} {\bibinfo {author} {\bibfnamefont {Y.}~\bibnamefont
  {Hieida}}, \bibinfo {author} {\bibfnamefont {K.}~\bibnamefont {Okunishi}},\
  and\ \bibinfo {author} {\bibfnamefont {Y.}~\bibnamefont {Akutsu}},\
  }\bibfield  {title} {\emph {\bibinfo {title} {{Numerical renormalization
  approach to two-dimensional quantum antiferromagnets with valence-bond-solid
  type ground state}}},\ }\href {https://doi.org/10.1088/1367-2630/1/1/007}
  {\bibfield  {journal} {\bibinfo  {journal} {New J. Phys.}\ }\textbf {\bibinfo
  {volume} {1}},\ \bibinfo {pages} {7} (\bibinfo {year} {1999})}\BibitemShut
  {NoStop}%
\bibitem [{\citenamefont {Okunishi}\ and\ \citenamefont
  {Nishino}(2000)}]{okunishi2000}%
  \BibitemOpen
  \bibfield  {author} {\bibinfo {author} {\bibfnamefont {K.}~\bibnamefont
  {Okunishi}}\ and\ \bibinfo {author} {\bibfnamefont {T.}~\bibnamefont
  {Nishino}},\ }\bibfield  {title} {\emph {\bibinfo {title} {{Kramers-Wannier
  approximation for the 3D Ising model}}},\ }\href
  {https://doi.org/10.1143/PTP.103.541} {\bibfield  {journal} {\bibinfo
  {journal} {Prog. Theor. Phys.}\ }\textbf {\bibinfo {volume} {103}},\ \bibinfo
  {pages} {541} (\bibinfo {year} {2000})}\BibitemShut {NoStop}%
\bibitem [{\citenamefont {Nishino}\ \emph {et~al.}(2001)\citenamefont
  {Nishino}, \citenamefont {Hieida}, \citenamefont {Okunishi}, \citenamefont
  {Maeshima}, \citenamefont {Akutsu},\ and\ \citenamefont
  {Gendiar}}]{nishino2001}%
  \BibitemOpen
  \bibfield  {author} {\bibinfo {author} {\bibfnamefont {T.}~\bibnamefont
  {Nishino}}, \bibinfo {author} {\bibfnamefont {Y.}~\bibnamefont {Hieida}},
  \bibinfo {author} {\bibfnamefont {K.}~\bibnamefont {Okunishi}}, \bibinfo
  {author} {\bibfnamefont {N.}~\bibnamefont {Maeshima}}, \bibinfo {author}
  {\bibfnamefont {Y.}~\bibnamefont {Akutsu}},\ and\ \bibinfo {author}
  {\bibfnamefont {A.}~\bibnamefont {Gendiar}},\ }\bibfield  {title} {\emph
  {\bibinfo {title} {{Two-dimensional tensor product variational
  formulation}}},\ }\href {https://doi.org/10.1143/PTP.105.409} {\bibfield
  {journal} {\bibinfo  {journal} {Prog. Theor. Phys.}\ }\textbf {\bibinfo
  {volume} {105}},\ \bibinfo {pages} {409} (\bibinfo {year}
  {2001})}\BibitemShut {NoStop}%
\bibitem [{\citenamefont {Maeshima}\ \emph {et~al.}(2001)\citenamefont
  {Maeshima}, \citenamefont {Hieida}, \citenamefont {Akutsu}, \citenamefont
  {Nishino},\ and\ \citenamefont {Okunishi}}]{maeshima2001}%
  \BibitemOpen
  \bibfield  {author} {\bibinfo {author} {\bibfnamefont {N.}~\bibnamefont
  {Maeshima}}, \bibinfo {author} {\bibfnamefont {Y.}~\bibnamefont {Hieida}},
  \bibinfo {author} {\bibfnamefont {Y.}~\bibnamefont {Akutsu}}, \bibinfo
  {author} {\bibfnamefont {T.}~\bibnamefont {Nishino}},\ and\ \bibinfo {author}
  {\bibfnamefont {K.}~\bibnamefont {Okunishi}},\ }\bibfield  {title} {\emph
  {\bibinfo {title} {{Vertical density matrix algorithm: A higher-dimensional
  numerical renormalization scheme based on the tensor product state
  ansatz}}},\ }\href {https://doi.org/10.1103/PhysRevE.64.016705} {\bibfield
  {journal} {\bibinfo  {journal} {Phys. Rev. E}\ }\textbf {\bibinfo {volume}
  {64}},\ \bibinfo {pages} {016705} (\bibinfo {year} {2001})}\BibitemShut
  {NoStop}%
\bibitem [{\citenamefont {Nishio}\ \emph {et~al.}()\citenamefont {Nishio},
  \citenamefont {Maeshima}, \citenamefont {Gendiar},\ and\ \citenamefont
  {Nishino}}]{nishio2004_arxiv}%
  \BibitemOpen
  \bibfield  {author} {\bibinfo {author} {\bibfnamefont {Y.}~\bibnamefont
  {Nishio}}, \bibinfo {author} {\bibfnamefont {N.}~\bibnamefont {Maeshima}},
  \bibinfo {author} {\bibfnamefont {A.}~\bibnamefont {Gendiar}},\ and\ \bibinfo
  {author} {\bibfnamefont {T.}~\bibnamefont {Nishino}},\ }\href@noop {}
  {\bibfield  {title} {\emph {\bibinfo {title} {{Tensor product variational
  formulation for quantum systems}}}},\ }\Eprint
  {https://arxiv.org/abs/cond-mat/0401115} {arXiv:cond-mat/0401115}
  \BibitemShut {NoStop}%
\bibitem [{\citenamefont {Kshetrimayum}\ \emph {et~al.}(2017)\citenamefont
  {Kshetrimayum}, \citenamefont {Weimer},\ and\ \citenamefont
  {Or{\'u}s}}]{kshetrimayum2017}%
  \BibitemOpen
  \bibfield  {author} {\bibinfo {author} {\bibfnamefont {A.}~\bibnamefont
  {Kshetrimayum}}, \bibinfo {author} {\bibfnamefont {H.}~\bibnamefont
  {Weimer}},\ and\ \bibinfo {author} {\bibfnamefont {R.}~\bibnamefont
  {Or{\'u}s}},\ }\bibfield  {title} {\emph {\bibinfo {title} {{A simple tensor
  network algorithm for two-dimensional steady states}}},\ }\href
  {https://doi.org/10.1038/s41467-017-01511-6} {\bibfield  {journal} {\bibinfo
  {journal} {Nat. Commun.}\ }\textbf {\bibinfo {volume} {8}},\ \bibinfo {pages}
  {1} (\bibinfo {year} {2017})}\BibitemShut {NoStop}%
\bibitem [{\citenamefont {Czarnik}\ \emph {et~al.}(2019)\citenamefont
  {Czarnik}, \citenamefont {Dziarmaga},\ and\ \citenamefont
  {Corboz}}]{czarnik2019}%
  \BibitemOpen
  \bibfield  {author} {\bibinfo {author} {\bibfnamefont {P.}~\bibnamefont
  {Czarnik}}, \bibinfo {author} {\bibfnamefont {J.}~\bibnamefont {Dziarmaga}},\
  and\ \bibinfo {author} {\bibfnamefont {P.}~\bibnamefont {Corboz}},\
  }\bibfield  {title} {\emph {\bibinfo {title} {{Time evolution of an infinite
  projected entangled pair state: An efficient algorithm}}},\ }\href
  {https://doi.org/10.1103/PhysRevB.99.035115} {\bibfield  {journal} {\bibinfo
  {journal} {Phys. Rev. B}\ }\textbf {\bibinfo {volume} {99}},\ \bibinfo
  {pages} {035115} (\bibinfo {year} {2019})}\BibitemShut {NoStop}%
\bibitem [{\citenamefont {Hubig}\ and\ \citenamefont
  {Cirac}(2019)}]{hubig2019}%
  \BibitemOpen
  \bibfield  {author} {\bibinfo {author} {\bibfnamefont {C.}~\bibnamefont
  {Hubig}}\ and\ \bibinfo {author} {\bibfnamefont {J.~I.}\ \bibnamefont
  {Cirac}},\ }\bibfield  {title} {\emph {\bibinfo {title} {{Time-dependent
  study of disordered models with infinite projected entangled pair states}}},\
  }\href {https://doi.org/10.21468/SciPostPhys.6.3.031} {\bibfield  {journal}
  {\bibinfo  {journal} {SciPost Phys.}\ }\textbf {\bibinfo {volume} {6}},\
  \bibinfo {pages} {31} (\bibinfo {year} {2019})}\BibitemShut {NoStop}%
\bibitem [{\citenamefont {Kshetrimayum}\ \emph {et~al.}(2020)\citenamefont
  {Kshetrimayum}, \citenamefont {Goihl},\ and\ \citenamefont
  {Eisert}}]{kshetrimayum2020}%
  \BibitemOpen
  \bibfield  {author} {\bibinfo {author} {\bibfnamefont {A.}~\bibnamefont
  {Kshetrimayum}}, \bibinfo {author} {\bibfnamefont {M.}~\bibnamefont
  {Goihl}},\ and\ \bibinfo {author} {\bibfnamefont {J.}~\bibnamefont
  {Eisert}},\ }\bibfield  {title} {\emph {\bibinfo {title} {{Time evolution of
  many-body localized systems in two spatial dimensions}}},\ }\href
  {https://doi.org/10.1103/PhysRevB.102.235132} {\bibfield  {journal} {\bibinfo
   {journal} {Phys. Rev. B}\ }\textbf {\bibinfo {volume} {102}},\ \bibinfo
  {pages} {235132} (\bibinfo {year} {2020})}\BibitemShut {NoStop}%
\bibitem [{\citenamefont {Kshetrimayum}\ \emph {et~al.}(2021)\citenamefont
  {Kshetrimayum}, \citenamefont {Goihl}, \citenamefont {Kennes},\ and\
  \citenamefont {Eisert}}]{kshetrimayum2021}%
  \BibitemOpen
  \bibfield  {author} {\bibinfo {author} {\bibfnamefont {A.}~\bibnamefont
  {Kshetrimayum}}, \bibinfo {author} {\bibfnamefont {M.}~\bibnamefont {Goihl}},
  \bibinfo {author} {\bibfnamefont {D.~M.}\ \bibnamefont {Kennes}},\ and\
  \bibinfo {author} {\bibfnamefont {J.}~\bibnamefont {Eisert}},\ }\bibfield
  {title} {\emph {\bibinfo {title} {{Quantum time crystals with programmable
  disorder in higher dimensions}}},\ }\href
  {https://doi.org/10.1103/PhysRevB.103.224205} {\bibfield  {journal} {\bibinfo
   {journal} {Phys. Rev. B}\ }\textbf {\bibinfo {volume} {103}},\ \bibinfo
  {pages} {224205} (\bibinfo {year} {2021})}\BibitemShut {NoStop}%
\bibitem [{\citenamefont {Hubig}\ \emph {et~al.}(2020)\citenamefont {Hubig},
  \citenamefont {Bohrdt}, \citenamefont {Knap}, \citenamefont {Grusdt},\ and\
  \citenamefont {Cirac}}]{hubig2020}%
  \BibitemOpen
  \bibfield  {author} {\bibinfo {author} {\bibfnamefont {C.}~\bibnamefont
  {Hubig}}, \bibinfo {author} {\bibfnamefont {A.}~\bibnamefont {Bohrdt}},
  \bibinfo {author} {\bibfnamefont {M.}~\bibnamefont {Knap}}, \bibinfo {author}
  {\bibfnamefont {F.}~\bibnamefont {Grusdt}},\ and\ \bibinfo {author}
  {\bibfnamefont {J.~I.}\ \bibnamefont {Cirac}},\ }\bibfield  {title} {\emph
  {\bibinfo {title} {{Evaluation of time-dependent correlators after a local
  quench in iPEPS: hole motion in the t-J model}}},\ }\href
  {https://doi.org/10.21468/SciPostPhys.8.2.021} {\bibfield  {journal}
  {\bibinfo  {journal} {SciPost Phys.}\ }\textbf {\bibinfo {volume} {8}},\
  \bibinfo {pages} {21} (\bibinfo {year} {2020})}\BibitemShut {NoStop}%
\bibitem [{\citenamefont {Alhambra}\ and\ \citenamefont
  {Cirac}(2021)}]{alhambra2021}%
  \BibitemOpen
  \bibfield  {author} {\bibinfo {author} {\bibfnamefont {A.~M.}\ \bibnamefont
  {Alhambra}}\ and\ \bibinfo {author} {\bibfnamefont {J.~I.}\ \bibnamefont
  {Cirac}},\ }\bibfield  {title} {\emph {\bibinfo {title} {{Locally Accurate
  Tensor Networks for Thermal States and Time Evolution}}},\ }\href
  {https://doi.org/10.1103/PRXQuantum.2.040331} {\bibfield  {journal} {\bibinfo
   {journal} {PRX Quantum}\ }\textbf {\bibinfo {volume} {2}},\ \bibinfo {pages}
  {040331} (\bibinfo {year} {2021})}\BibitemShut {NoStop}%
\bibitem [{\citenamefont {Schmitt}\ \emph {et~al.}()\citenamefont {Schmitt},
  \citenamefont {Rams}, \citenamefont {Dziarmaga}, \citenamefont {Heyl},\ and\
  \citenamefont {Zurek}}]{schmitt2021_arxiv}%
  \BibitemOpen
  \bibfield  {author} {\bibinfo {author} {\bibfnamefont {M.}~\bibnamefont
  {Schmitt}}, \bibinfo {author} {\bibfnamefont {M.~M.}\ \bibnamefont {Rams}},
  \bibinfo {author} {\bibfnamefont {J.}~\bibnamefont {Dziarmaga}}, \bibinfo
  {author} {\bibfnamefont {M.}~\bibnamefont {Heyl}},\ and\ \bibinfo {author}
  {\bibfnamefont {W.~H.}\ \bibnamefont {Zurek}},\ }\href@noop {} {\bibfield
  {title} {\emph {\bibinfo {title} {{Quantum phase transition dynamics in the
  two-dimensional transverse-field Ising model}}}},\ }\Eprint
  {https://arxiv.org/abs/2106.09046} {arXiv:2106.09046} \BibitemShut {NoStop}%
\bibitem [{\citenamefont {Dziarmaga}(2021)}]{dziarmaga2021}%
  \BibitemOpen
  \bibfield  {author} {\bibinfo {author} {\bibfnamefont {J.}~\bibnamefont
  {Dziarmaga}},\ }\bibfield  {title} {\emph {\bibinfo {title} {{Time evolution
  of an infinite projected entangled pair state: Neighborhood tensor
  update}}},\ }\href {https://doi.org/10.1103/PhysRevB.104.094411} {\bibfield
  {journal} {\bibinfo  {journal} {Phys. Rev. B}\ }\textbf {\bibinfo {volume}
  {104}},\ \bibinfo {pages} {094411} (\bibinfo {year} {2021})}\BibitemShut
  {NoStop}%
\bibitem [{\citenamefont {Weimer}\ \emph {et~al.}(2021)\citenamefont {Weimer},
  \citenamefont {Kshetrimayum},\ and\ \citenamefont {Or\'us}}]{weimer2021}%
  \BibitemOpen
  \bibfield  {author} {\bibinfo {author} {\bibfnamefont {H.}~\bibnamefont
  {Weimer}}, \bibinfo {author} {\bibfnamefont {A.}~\bibnamefont
  {Kshetrimayum}},\ and\ \bibinfo {author} {\bibfnamefont {R.}~\bibnamefont
  {Or\'us}},\ }\bibfield  {title} {\emph {\bibinfo {title} {{Simulation methods
  for open quantum many-body systems}}},\ }\href
  {https://doi.org/10.1103/RevModPhys.93.015008} {\bibfield  {journal}
  {\bibinfo  {journal} {Rev. Mod. Phys.}\ }\textbf {\bibinfo {volume} {93}},\
  \bibinfo {pages} {015008} (\bibinfo {year} {2021})}\BibitemShut {NoStop}%
\bibitem [{\citenamefont {Mc~Keever}\ and\ \citenamefont
  {Szyma\ifmmode~\acute{n}\else \'{n}\fi{}ska}(2021)}]{keever2021}%
  \BibitemOpen
  \bibfield  {author} {\bibinfo {author} {\bibfnamefont {C.}~\bibnamefont
  {Mc~Keever}}\ and\ \bibinfo {author} {\bibfnamefont {M.~H.}\ \bibnamefont
  {Szyma\ifmmode~\acute{n}\else \'{n}\fi{}ska}},\ }\bibfield  {title} {\emph
  {\bibinfo {title} {{Stable iPEPO Tensor-Network Algorithm for Dynamics of
  Two-Dimensional Open Quantum Lattice Models}}},\ }\href
  {https://doi.org/10.1103/PhysRevX.11.021035} {\bibfield  {journal} {\bibinfo
  {journal} {Phys. Rev. X}\ }\textbf {\bibinfo {volume} {11}},\ \bibinfo
  {pages} {021035} (\bibinfo {year} {2021})}\BibitemShut {NoStop}%
\bibitem [{\citenamefont {Kilda}\ \emph {et~al.}(2021)\citenamefont {Kilda},
  \citenamefont {Biella}, \citenamefont {Schiro}, \citenamefont {Fazio},\ and\
  \citenamefont {Keeling}}]{kilda2021}%
  \BibitemOpen
  \bibfield  {author} {\bibinfo {author} {\bibfnamefont {D.}~\bibnamefont
  {Kilda}}, \bibinfo {author} {\bibfnamefont {A.}~\bibnamefont {Biella}},
  \bibinfo {author} {\bibfnamefont {M.}~\bibnamefont {Schiro}}, \bibinfo
  {author} {\bibfnamefont {R.}~\bibnamefont {Fazio}},\ and\ \bibinfo {author}
  {\bibfnamefont {J.}~\bibnamefont {Keeling}},\ }\bibfield  {title} {\emph
  {\bibinfo {title} {{On the stability of the infinite Projected Entangled Pair
  Operator ansatz for driven-dissipative 2D lattices}}},\ }\href
  {https://doi.org/10.21468/SciPostPhysCore.4.1.005} {\bibfield  {journal}
  {\bibinfo  {journal} {SciPost Phys. Core}\ }\textbf {\bibinfo {volume} {4}},\
  \bibinfo {pages} {5} (\bibinfo {year} {2021})}\BibitemShut {NoStop}%
\bibitem [{\citenamefont {Fisher}\ \emph {et~al.}(1989)\citenamefont {Fisher},
  \citenamefont {Weichman}, \citenamefont {Grinstein},\ and\ \citenamefont
  {Fisher}}]{fisher1989}%
  \BibitemOpen
  \bibfield  {author} {\bibinfo {author} {\bibfnamefont {M.~P.~A.}\
  \bibnamefont {Fisher}}, \bibinfo {author} {\bibfnamefont {P.~B.}\
  \bibnamefont {Weichman}}, \bibinfo {author} {\bibfnamefont {G.}~\bibnamefont
  {Grinstein}},\ and\ \bibinfo {author} {\bibfnamefont {D.~S.}\ \bibnamefont
  {Fisher}},\ }\bibfield  {title} {\emph {\bibinfo {title} {{Boson localization
  and the superfluid-insulator transition}}},\ }\href
  {https://doi.org/10.1103/PhysRevB.40.546} {\bibfield  {journal} {\bibinfo
  {journal} {Phys. Rev. B}\ }\textbf {\bibinfo {volume} {40}},\ \bibinfo
  {pages} {546} (\bibinfo {year} {1989})}\BibitemShut {NoStop}%
\bibitem [{\citenamefont {Jaksch}\ \emph {et~al.}(1998)\citenamefont {Jaksch},
  \citenamefont {Bruder}, \citenamefont {Cirac}, \citenamefont {Gardiner},\
  and\ \citenamefont {Zoller}}]{Jaksch1998}%
  \BibitemOpen
  \bibfield  {author} {\bibinfo {author} {\bibfnamefont {D.}~\bibnamefont
  {Jaksch}}, \bibinfo {author} {\bibfnamefont {C.}~\bibnamefont {Bruder}},
  \bibinfo {author} {\bibfnamefont {J.~I.}\ \bibnamefont {Cirac}}, \bibinfo
  {author} {\bibfnamefont {C.~W.}\ \bibnamefont {Gardiner}},\ and\ \bibinfo
  {author} {\bibfnamefont {P.}~\bibnamefont {Zoller}},\ }\bibfield  {title}
  {\emph {\bibinfo {title} {{Cold Bosonic Atoms in Optical Lattices}}},\ }\href
  {https://doi.org/10.1103/PhysRevLett.81.3108} {\bibfield  {journal} {\bibinfo
   {journal} {Phys. Rev. Lett.}\ }\textbf {\bibinfo {volume} {81}},\ \bibinfo
  {pages} {3108} (\bibinfo {year} {1998})}\BibitemShut {NoStop}%
\bibitem [{\citenamefont {Weinberg}\ and\ \citenamefont
  {Bukov}(2017)}]{weinberg2017}%
  \BibitemOpen
  \bibfield  {author} {\bibinfo {author} {\bibfnamefont {P.}~\bibnamefont
  {Weinberg}}\ and\ \bibinfo {author} {\bibfnamefont {M.}~\bibnamefont
  {Bukov}},\ }\bibfield  {title} {\emph {\bibinfo {title} {{QuSpin: a Python
  Package for Dynamics and Exact Diagonalisation of Quantum Many Body Systems
  part I: spin chains}}},\ }\href
  {https://doi.org/10.21468/SciPostPhys.2.1.003} {\bibfield  {journal}
  {\bibinfo  {journal} {SciPost Phys.}\ }\textbf {\bibinfo {volume} {2}},\
  \bibinfo {pages} {003} (\bibinfo {year} {2017})}\BibitemShut {NoStop}%
\bibitem [{\citenamefont {Weinberg}\ and\ \citenamefont
  {Bukov}(2019)}]{weinberg2019}%
  \BibitemOpen
  \bibfield  {author} {\bibinfo {author} {\bibfnamefont {P.}~\bibnamefont
  {Weinberg}}\ and\ \bibinfo {author} {\bibfnamefont {M.}~\bibnamefont
  {Bukov}},\ }\bibfield  {title} {\emph {\bibinfo {title} {{QuSpin: a Python
  Package for Dynamics and Exact Diagonalisation of Quantum Many Body Systems.
  Part II: bosons, fermions and higher spins}}},\ }\href
  {https://doi.org/10.21468/SciPostPhys.7.2.020} {\bibfield  {journal}
  {\bibinfo  {journal} {SciPost Phys.}\ }\textbf {\bibinfo {volume} {7}},\
  \bibinfo {pages} {20} (\bibinfo {year} {2019})}\BibitemShut {NoStop}%
\bibitem [{\citenamefont {Lieb}\ and\ \citenamefont
  {Robinson}(1972)}]{lieb1972}%
  \BibitemOpen
  \bibfield  {author} {\bibinfo {author} {\bibfnamefont {E.~H.}\ \bibnamefont
  {Lieb}}\ and\ \bibinfo {author} {\bibfnamefont {D.~W.}\ \bibnamefont
  {Robinson}},\ }\bibfield  {title} {\emph {\bibinfo {title} {{The finite group
  velocity of quantum spin systems}}},\ }\href
  {https://doi.org/10.1007/BF01645779} {\bibfield  {journal} {\bibinfo
  {journal} {Commun. Math. Phys.}\ }\textbf {\bibinfo {volume} {28}},\ \bibinfo
  {pages} {251} (\bibinfo {year} {1972})}\BibitemShut {NoStop}%
\bibitem [{\citenamefont {Hastings}()}]{hastings2010_arxiv}%
  \BibitemOpen
  \bibfield  {author} {\bibinfo {author} {\bibfnamefont {M.~B.}\ \bibnamefont
  {Hastings}},\ }\href@noop {} {\bibfield  {title} {\emph {\bibinfo {title}
  {{Locality in Quantum Systems}}}},\ }\Eprint
  {https://arxiv.org/abs/1008.5137} {arXiv:1008.5137} \BibitemShut {NoStop}%
\bibitem [{\citenamefont {Elstner}\ and\ \citenamefont
  {Monien}(1999)}]{elstner1999}%
  \BibitemOpen
  \bibfield  {author} {\bibinfo {author} {\bibfnamefont {N.}~\bibnamefont
  {Elstner}}\ and\ \bibinfo {author} {\bibfnamefont {H.}~\bibnamefont
  {Monien}},\ }\bibfield  {title} {\emph {\bibinfo {title} {{Dynamics and
  thermodynamics of the Bose-Hubbard model}}},\ }\href
  {https://doi.org/10.1103/PhysRevB.59.12184} {\bibfield  {journal} {\bibinfo
  {journal} {Phys. Rev. B}\ }\textbf {\bibinfo {volume} {59}},\ \bibinfo
  {pages} {12184} (\bibinfo {year} {1999})}\BibitemShut {NoStop}%
\bibitem [{\citenamefont {Capogrosso-Sansone}\ \emph
  {et~al.}(2008)\citenamefont {Capogrosso-Sansone}, \citenamefont {S\"oyler},
  \citenamefont {Prokof'ev},\ and\ \citenamefont
  {Svistunov}}]{capogrosso-sansone2008}%
  \BibitemOpen
  \bibfield  {author} {\bibinfo {author} {\bibfnamefont {B.}~\bibnamefont
  {Capogrosso-Sansone}}, \bibinfo {author} {\bibfnamefont {{\c{S}}.~G.}\
  \bibnamefont {S\"oyler}}, \bibinfo {author} {\bibfnamefont {N.}~\bibnamefont
  {Prokof'ev}},\ and\ \bibinfo {author} {\bibfnamefont {B.}~\bibnamefont
  {Svistunov}},\ }\bibfield  {title} {\emph {\bibinfo {title} {{Monte Carlo
  study of the two-dimensional Bose-Hubbard model}}},\ }\href
  {https://doi.org/10.1103/PhysRevA.77.015602} {\bibfield  {journal} {\bibinfo
  {journal} {Phys. Rev. A}\ }\textbf {\bibinfo {volume} {77}},\ \bibinfo
  {pages} {015602} (\bibinfo {year} {2008})}\BibitemShut {NoStop}%
\bibitem [{\citenamefont {Krutitsky}(2016)}]{krutitsky2016}%
  \BibitemOpen
  \bibfield  {author} {\bibinfo {author} {\bibfnamefont {K.~V.}\ \bibnamefont
  {Krutitsky}},\ }\bibfield  {title} {\emph {\bibinfo {title} {{Ultracold
  bosons with short-range interaction in regular optical lattices}}},\ }\href
  {https://doi.org/10.1016/j.physrep.2015.10.004} {\bibfield  {journal}
  {\bibinfo  {journal} {Phys. Rep.}\ }\textbf {\bibinfo {volume} {607}},\
  \bibinfo {pages} {1} (\bibinfo {year} {2016})}\BibitemShut {NoStop}%
\bibitem [{\citenamefont {Huber}\ \emph {et~al.}(2007)\citenamefont {Huber},
  \citenamefont {Altman}, \citenamefont {B\"uchler},\ and\ \citenamefont
  {Blatter}}]{huber2007}%
  \BibitemOpen
  \bibfield  {author} {\bibinfo {author} {\bibfnamefont {S.~D.}\ \bibnamefont
  {Huber}}, \bibinfo {author} {\bibfnamefont {E.}~\bibnamefont {Altman}},
  \bibinfo {author} {\bibfnamefont {H.~P.}\ \bibnamefont {B\"uchler}},\ and\
  \bibinfo {author} {\bibfnamefont {G.}~\bibnamefont {Blatter}},\ }\bibfield
  {title} {\emph {\bibinfo {title} {{Dynamical properties of ultracold bosons
  in an optical lattice}}},\ }\href
  {https://doi.org/10.1103/PhysRevB.75.085106} {\bibfield  {journal} {\bibinfo
  {journal} {Phys. Rev. B}\ }\textbf {\bibinfo {volume} {75}},\ \bibinfo
  {pages} {085106} (\bibinfo {year} {2007})}\BibitemShut {NoStop}%
\bibitem [{\citenamefont {Davidson}\ and\ \citenamefont
  {Polkovnikov}(2015)}]{davidson2015}%
  \BibitemOpen
  \bibfield  {author} {\bibinfo {author} {\bibfnamefont {S.~M.}\ \bibnamefont
  {Davidson}}\ and\ \bibinfo {author} {\bibfnamefont {A.}~\bibnamefont
  {Polkovnikov}},\ }\bibfield  {title} {\emph {\bibinfo {title} {{SU(3)
  Semiclassical Representation of Quantum Dynamics of Interacting Spins}}},\
  }\href {https://doi.org/10.1103/PhysRevLett.114.045701} {\bibfield  {journal}
  {\bibinfo  {journal} {Phys. Rev. Lett.}\ }\textbf {\bibinfo {volume} {114}},\
  \bibinfo {pages} {045701} (\bibinfo {year} {2015})}\BibitemShut {NoStop}%
\bibitem [{\citenamefont {Murg}\ \emph {et~al.}(2007)\citenamefont {Murg},
  \citenamefont {Verstraete},\ and\ \citenamefont {Cirac}}]{murg2007}%
  \BibitemOpen
  \bibfield  {author} {\bibinfo {author} {\bibfnamefont {V.}~\bibnamefont
  {Murg}}, \bibinfo {author} {\bibfnamefont {F.}~\bibnamefont {Verstraete}},\
  and\ \bibinfo {author} {\bibfnamefont {J.~I.}\ \bibnamefont {Cirac}},\
  }\bibfield  {title} {\emph {\bibinfo {title} {{Variational study of hard-core
  bosons in a two-dimensional optical lattice using projected entangled pair
  states}}},\ }\href {https://doi.org/10.1103/PhysRevA.75.033605} {\bibfield
  {journal} {\bibinfo  {journal} {Phys. Rev. A}\ }\textbf {\bibinfo {volume}
  {75}},\ \bibinfo {pages} {033605} (\bibinfo {year} {2007})}\BibitemShut
  {NoStop}%
\bibitem [{\citenamefont {Jordan}\ \emph {et~al.}(2009)\citenamefont {Jordan},
  \citenamefont {Or\'us},\ and\ \citenamefont {Vidal}}]{jordan2009}%
  \BibitemOpen
  \bibfield  {author} {\bibinfo {author} {\bibfnamefont {J.}~\bibnamefont
  {Jordan}}, \bibinfo {author} {\bibfnamefont {R.}~\bibnamefont {Or\'us}},\
  and\ \bibinfo {author} {\bibfnamefont {G.}~\bibnamefont {Vidal}},\ }\bibfield
   {title} {\emph {\bibinfo {title} {{Numerical study of the hard-core
  Bose-Hubbard model on an infinite square lattice}}},\ }\href
  {https://doi.org/10.1103/PhysRevB.79.174515} {\bibfield  {journal} {\bibinfo
  {journal} {Phys. Rev. B}\ }\textbf {\bibinfo {volume} {79}},\ \bibinfo
  {pages} {174515} (\bibinfo {year} {2009})}\BibitemShut {NoStop}%
\bibitem [{\citenamefont {Kshetrimayum}\ \emph {et~al.}(2019)\citenamefont
  {Kshetrimayum}, \citenamefont {Rizzi}, \citenamefont {Eisert},\ and\
  \citenamefont {Or\'us}}]{kshetrimayum2019}%
  \BibitemOpen
  \bibfield  {author} {\bibinfo {author} {\bibfnamefont {A.}~\bibnamefont
  {Kshetrimayum}}, \bibinfo {author} {\bibfnamefont {M.}~\bibnamefont {Rizzi}},
  \bibinfo {author} {\bibfnamefont {J.}~\bibnamefont {Eisert}},\ and\ \bibinfo
  {author} {\bibfnamefont {R.}~\bibnamefont {Or\'us}},\ }\bibfield  {title}
  {\emph {\bibinfo {title} {{Tensor Network Annealing Algorithm for
  Two-Dimensional Thermal States}}},\ }\href
  {https://doi.org/10.1103/PhysRevLett.122.070502} {\bibfield  {journal}
  {\bibinfo  {journal} {Phys. Rev. Lett.}\ }\textbf {\bibinfo {volume} {122}},\
  \bibinfo {pages} {070502} (\bibinfo {year} {2019})}\BibitemShut {NoStop}%
\bibitem [{\citenamefont {Jahromi}\ and\ \citenamefont
  {Or\'us}(2019)}]{jahromi2019}%
  \BibitemOpen
  \bibfield  {author} {\bibinfo {author} {\bibfnamefont {S.~S.}\ \bibnamefont
  {Jahromi}}\ and\ \bibinfo {author} {\bibfnamefont {R.}~\bibnamefont
  {Or\'us}},\ }\bibfield  {title} {\emph {\bibinfo {title} {{Universal
  tensor-network algorithm for any infinite lattice}}},\ }\href
  {https://doi.org/10.1103/PhysRevB.99.195105} {\bibfield  {journal} {\bibinfo
  {journal} {Phys. Rev. B}\ }\textbf {\bibinfo {volume} {99}},\ \bibinfo
  {pages} {195105} (\bibinfo {year} {2019})}\BibitemShut {NoStop}%
\bibitem [{\citenamefont {Jahromi}\ and\ \citenamefont
  {Or{\'u}s}(2020)}]{jahromi2020}%
  \BibitemOpen
  \bibfield  {author} {\bibinfo {author} {\bibfnamefont {S.~S.}\ \bibnamefont
  {Jahromi}}\ and\ \bibinfo {author} {\bibfnamefont {R.}~\bibnamefont
  {Or{\'u}s}},\ }\bibfield  {title} {\emph {\bibinfo {title} {{Thermal bosons
  in 3d optical lattices via tensor networks}}},\ }\href
  {https://doi.org/10.1038/s41598-020-75548-x} {\bibfield  {journal} {\bibinfo
  {journal} {Sci. Rep.}\ }\textbf {\bibinfo {volume} {10}},\ \bibinfo {pages}
  {19051} (\bibinfo {year} {2020})}\BibitemShut {NoStop}%
\bibitem [{\citenamefont {Schmoll}\ \emph {et~al.}(2020)\citenamefont
  {Schmoll}, \citenamefont {Jahromi}, \citenamefont {H\"ormann}, \citenamefont
  {M\"uhlhauser}, \citenamefont {Schmidt},\ and\ \citenamefont
  {Or\'us}}]{schmoll2020}%
  \BibitemOpen
  \bibfield  {author} {\bibinfo {author} {\bibfnamefont {P.}~\bibnamefont
  {Schmoll}}, \bibinfo {author} {\bibfnamefont {S.~S.}\ \bibnamefont
  {Jahromi}}, \bibinfo {author} {\bibfnamefont {M.}~\bibnamefont {H\"ormann}},
  \bibinfo {author} {\bibfnamefont {M.}~\bibnamefont {M\"uhlhauser}}, \bibinfo
  {author} {\bibfnamefont {K.~P.}\ \bibnamefont {Schmidt}},\ and\ \bibinfo
  {author} {\bibfnamefont {R.}~\bibnamefont {Or\'us}},\ }\bibfield  {title}
  {\emph {\bibinfo {title} {{Fine Grained Tensor Network Methods}}},\ }\href
  {https://doi.org/10.1103/PhysRevLett.124.200603} {\bibfield  {journal}
  {\bibinfo  {journal} {Phys. Rev. Lett.}\ }\textbf {\bibinfo {volume} {124}},\
  \bibinfo {pages} {200603} (\bibinfo {year} {2020})}\BibitemShut {NoStop}%
\bibitem [{\citenamefont {Tu}\ \emph {et~al.}(2020)\citenamefont {Tu},
  \citenamefont {Wu},\ and\ \citenamefont {Suzuki}}]{tu2020}%
  \BibitemOpen
  \bibfield  {author} {\bibinfo {author} {\bibfnamefont {W.-L.}\ \bibnamefont
  {Tu}}, \bibinfo {author} {\bibfnamefont {H.-K.}\ \bibnamefont {Wu}},\ and\
  \bibinfo {author} {\bibfnamefont {T.}~\bibnamefont {Suzuki}},\ }\bibfield
  {title} {\emph {\bibinfo {title} {{Frustration-induced supersolid phases of
  extended Bose-Hubbard model in the hard-core limit}}},\ }\href
  {https://doi.org/10.1088/1361-648x/aba383} {\bibfield  {journal} {\bibinfo
  {journal} {J. Phys.: Cond. Mat.}\ }\textbf {\bibinfo {volume} {32}},\
  \bibinfo {pages} {455401} (\bibinfo {year} {2020})}\BibitemShut {NoStop}%
\bibitem [{\citenamefont {Wu}\ and\ \citenamefont {Tu}(2020)}]{wu2020}%
  \BibitemOpen
  \bibfield  {author} {\bibinfo {author} {\bibfnamefont {H.-K.}\ \bibnamefont
  {Wu}}\ and\ \bibinfo {author} {\bibfnamefont {W.-L.}\ \bibnamefont {Tu}},\
  }\bibfield  {title} {\emph {\bibinfo {title} {{Competing quantum phases of
  hard-core bosons with tilted dipole-dipole interaction}}},\ }\href
  {https://doi.org/10.1103/PhysRevA.102.053306} {\bibfield  {journal} {\bibinfo
   {journal} {Phys. Rev. A}\ }\textbf {\bibinfo {volume} {102}},\ \bibinfo
  {pages} {053306} (\bibinfo {year} {2020})}\BibitemShut {NoStop}%
\bibitem [{\citenamefont {Vlaar}\ and\ \citenamefont
  {Corboz}(2021)}]{vlaar2021}%
  \BibitemOpen
  \bibfield  {author} {\bibinfo {author} {\bibfnamefont {P.~C.~G.}\
  \bibnamefont {Vlaar}}\ and\ \bibinfo {author} {\bibfnamefont
  {P.}~\bibnamefont {Corboz}},\ }\bibfield  {title} {\emph {\bibinfo {title}
  {{Simulation of three-dimensional quantum systems with projected
  entangled-pair states}}},\ }\href
  {https://doi.org/10.1103/PhysRevB.103.205137} {\bibfield  {journal} {\bibinfo
   {journal} {Phys. Rev. B}\ }\textbf {\bibinfo {volume} {103}},\ \bibinfo
  {pages} {205137} (\bibinfo {year} {2021})}\BibitemShut {NoStop}%
\bibitem [{\citenamefont {Trotter}(1959)}]{trotter1959}%
  \BibitemOpen
  \bibfield  {author} {\bibinfo {author} {\bibfnamefont {H.~F.}\ \bibnamefont
  {Trotter}},\ }\bibfield  {title} {\emph {\bibinfo {title} {{On the product of
  semi-groups of operators}}},\ }\href
  {https://doi.org/10.1090/S0002-9939-1959-0108732-6} {\bibfield  {journal}
  {\bibinfo  {journal} {Proc. Amer. Math. Soc.}\ }\textbf {\bibinfo {volume}
  {10}},\ \bibinfo {pages} {545} (\bibinfo {year} {1959})}\BibitemShut
  {NoStop}%
\bibitem [{\citenamefont {Suzuki}(1966)}]{suzuki1966}%
  \BibitemOpen
  \bibfield  {author} {\bibinfo {author} {\bibfnamefont {M.}~\bibnamefont
  {Suzuki}},\ }\bibfield  {title} {\emph {\bibinfo {title} {{Pair-product model
  of Heisenberg ferromagnets}}},\ }\href {https://doi.org/10.1143/JPSJ.21.2274}
  {\bibfield  {journal} {\bibinfo  {journal} {J. Phys. Soc. Jpn.}\ }\textbf
  {\bibinfo {volume} {21}},\ \bibinfo {pages} {2274} (\bibinfo {year}
  {1966})}\BibitemShut {NoStop}%
\bibitem [{\citenamefont {Suzuki}(1976)}]{suzuki1976}%
  \BibitemOpen
  \bibfield  {author} {\bibinfo {author} {\bibfnamefont {M.}~\bibnamefont
  {Suzuki}},\ }\bibfield  {title} {\emph {\bibinfo {title} {{Relationship
  between d-dimensional quantal spin systems and (d+1)-dimensional ising
  systems: Equivalence, critical exponents and systematic approximants of the
  partition function and spin correlations}}},\ }\href
  {https://doi.org/10.1143/PTP.56.1454} {\bibfield  {journal} {\bibinfo
  {journal} {Prog. Theor. Phys.}\ }\textbf {\bibinfo {volume} {56}},\ \bibinfo
  {pages} {1454} (\bibinfo {year} {1976})}\BibitemShut {NoStop}%
\bibitem [{\citenamefont {Jiang}\ \emph {et~al.}(2008)\citenamefont {Jiang},
  \citenamefont {Weng},\ and\ \citenamefont {Xiang}}]{jiang2008}%
  \BibitemOpen
  \bibfield  {author} {\bibinfo {author} {\bibfnamefont {H.~C.}\ \bibnamefont
  {Jiang}}, \bibinfo {author} {\bibfnamefont {Z.~Y.}\ \bibnamefont {Weng}},\
  and\ \bibinfo {author} {\bibfnamefont {T.}~\bibnamefont {Xiang}},\ }\bibfield
   {title} {\emph {\bibinfo {title} {{Accurate Determination of Tensor Network
  State of Quantum Lattice Models in Two Dimensions}}},\ }\href
  {https://doi.org/10.1103/PhysRevLett.101.090603} {\bibfield  {journal}
  {\bibinfo  {journal} {Phys. Rev. Lett.}\ }\textbf {\bibinfo {volume} {101}},\
  \bibinfo {pages} {090603} (\bibinfo {year} {2008})}\BibitemShut {NoStop}%
\bibitem [{\citenamefont {Motoyama}\ \emph {et~al.}()\citenamefont {Motoyama},
  \citenamefont {Okubo}, \citenamefont {Yoshimi}, \citenamefont {Morita},
  \citenamefont {Kato},\ and\ \citenamefont {Kawashima}}]{motoyama2021_arxiv}%
  \BibitemOpen
  \bibfield  {author} {\bibinfo {author} {\bibfnamefont {Y.}~\bibnamefont
  {Motoyama}}, \bibinfo {author} {\bibfnamefont {T.}~\bibnamefont {Okubo}},
  \bibinfo {author} {\bibfnamefont {K.}~\bibnamefont {Yoshimi}}, \bibinfo
  {author} {\bibfnamefont {S.}~\bibnamefont {Morita}}, \bibinfo {author}
  {\bibfnamefont {T.}~\bibnamefont {Kato}},\ and\ \bibinfo {author}
  {\bibfnamefont {N.}~\bibnamefont {Kawashima}},\ }\href@noop {} {\bibfield
  {title} {\emph {\bibinfo {title} {{TeNeS: Tensor Network Solver for Quantum
  Lattice Systems}}}},\ }\Eprint {https://arxiv.org/abs/2112.13184}
  {arXiv:2112.13184} \BibitemShut {NoStop}%
\bibitem [{ten()}]{tenes}%
  \BibitemOpen
  \href {https://github.com/issp-center-dev/TeNeS} {\bibinfo {title} {{TeNeS:
  https://github.com/issp-center-dev/TeNeS}}}\BibitemShut {NoStop}%
\bibitem [{ptn()}]{ptns}%
  \BibitemOpen
  \href {https://github.com/TsuyoshiOkubo/pTNS} {\bibinfo {title} {{pTNS:
  https://github.com/TsuyoshiOkubo/pTNS}}}\BibitemShut {NoStop}%
\bibitem [{\citenamefont {Nishino}\ and\ \citenamefont
  {Okunishi}(1996)}]{nishino1996}%
  \BibitemOpen
  \bibfield  {author} {\bibinfo {author} {\bibfnamefont {T.}~\bibnamefont
  {Nishino}}\ and\ \bibinfo {author} {\bibfnamefont {K.}~\bibnamefont
  {Okunishi}},\ }\bibfield  {title} {\emph {\bibinfo {title} {{Corner Transfer
  Matrix Renormalization Group Method}}},\ }\href
  {https://doi.org/10.1143/JPSJ.65.891} {\bibfield  {journal} {\bibinfo
  {journal} {J. Phys. Soc. Jpn.}\ }\textbf {\bibinfo {volume} {65}},\ \bibinfo
  {pages} {891} (\bibinfo {year} {1996})}\BibitemShut {NoStop}%
\bibitem [{\citenamefont {Nishino}\ and\ \citenamefont
  {Okunishi}(1997)}]{nishino1997}%
  \BibitemOpen
  \bibfield  {author} {\bibinfo {author} {\bibfnamefont {T.}~\bibnamefont
  {Nishino}}\ and\ \bibinfo {author} {\bibfnamefont {K.}~\bibnamefont
  {Okunishi}},\ }\bibfield  {title} {\emph {\bibinfo {title} {{Corner transfer
  matrix algorithm for classical renormalization group}}},\ }\href
  {https://doi.org/10.1143/JPSJ.66.3040} {\bibfield  {journal} {\bibinfo
  {journal} {J. Phys. Soc. Jpn.}\ }\textbf {\bibinfo {volume} {66}},\ \bibinfo
  {pages} {3040} (\bibinfo {year} {1997})}\BibitemShut {NoStop}%
\bibitem [{\citenamefont {Nishino}\ \emph {et~al.}(1999)\citenamefont
  {Nishino}, \citenamefont {Hikihara}, \citenamefont {Okunishi},\ and\
  \citenamefont {Hieida}}]{nishino1999}%
  \BibitemOpen
  \bibfield  {author} {\bibinfo {author} {\bibfnamefont {T.}~\bibnamefont
  {Nishino}}, \bibinfo {author} {\bibfnamefont {T.}~\bibnamefont {Hikihara}},
  \bibinfo {author} {\bibfnamefont {K.}~\bibnamefont {Okunishi}},\ and\
  \bibinfo {author} {\bibfnamefont {Y.}~\bibnamefont {Hieida}},\ }\bibfield
  {title} {\emph {\bibinfo {title} {{Density matrix renormalization group:
  Introduction from a variational point of view}}},\ }\href
  {https://doi.org/10.1142/S0217979299000023} {\bibfield  {journal} {\bibinfo
  {journal} {Int. J. Mod. Phys. B}\ }\textbf {\bibinfo {volume} {13}},\
  \bibinfo {pages} {1} (\bibinfo {year} {1999})}\BibitemShut {NoStop}%
\bibitem [{\citenamefont {Or\'us}\ and\ \citenamefont
  {Vidal}(2009)}]{orus2009}%
  \BibitemOpen
  \bibfield  {author} {\bibinfo {author} {\bibfnamefont {R.}~\bibnamefont
  {Or\'us}}\ and\ \bibinfo {author} {\bibfnamefont {G.}~\bibnamefont {Vidal}},\
  }\bibfield  {title} {\emph {\bibinfo {title} {{Simulation of two-dimensional
  quantum systems on an infinite lattice revisited: Corner transfer matrix for
  tensor contraction}}},\ }\href {https://doi.org/10.1103/PhysRevB.80.094403}
  {\bibfield  {journal} {\bibinfo  {journal} {Phys. Rev. B}\ }\textbf {\bibinfo
  {volume} {80}},\ \bibinfo {pages} {094403} (\bibinfo {year}
  {2009})}\BibitemShut {NoStop}%
\bibitem [{\citenamefont {Corboz}\ \emph {et~al.}(2010)\citenamefont {Corboz},
  \citenamefont {Jordan},\ and\ \citenamefont {Vidal}}]{corboz2010}%
  \BibitemOpen
  \bibfield  {author} {\bibinfo {author} {\bibfnamefont {P.}~\bibnamefont
  {Corboz}}, \bibinfo {author} {\bibfnamefont {J.}~\bibnamefont {Jordan}},\
  and\ \bibinfo {author} {\bibfnamefont {G.}~\bibnamefont {Vidal}},\ }\bibfield
   {title} {\emph {\bibinfo {title} {{Simulation of fermionic lattice models in
  two dimensions with projected entangled-pair states: Next-nearest neighbor
  Hamiltonians}}},\ }\href {https://doi.org/10.1103/PhysRevB.82.245119}
  {\bibfield  {journal} {\bibinfo  {journal} {Phys. Rev. B}\ }\textbf {\bibinfo
  {volume} {82}},\ \bibinfo {pages} {245119} (\bibinfo {year}
  {2010})}\BibitemShut {NoStop}%
\bibitem [{\citenamefont {Corboz}\ \emph {et~al.}(2011)\citenamefont {Corboz},
  \citenamefont {White}, \citenamefont {Vidal},\ and\ \citenamefont
  {Troyer}}]{corboz2011}%
  \BibitemOpen
  \bibfield  {author} {\bibinfo {author} {\bibfnamefont {P.}~\bibnamefont
  {Corboz}}, \bibinfo {author} {\bibfnamefont {S.~R.}\ \bibnamefont {White}},
  \bibinfo {author} {\bibfnamefont {G.}~\bibnamefont {Vidal}},\ and\ \bibinfo
  {author} {\bibfnamefont {M.}~\bibnamefont {Troyer}},\ }\bibfield  {title}
  {\emph {\bibinfo {title} {{Stripes in the two-dimensional $t$-$J$ model with
  infinite projected entangled-pair states}}},\ }\href
  {https://doi.org/10.1103/PhysRevB.84.041108} {\bibfield  {journal} {\bibinfo
  {journal} {Phys. Rev. B}\ }\textbf {\bibinfo {volume} {84}},\ \bibinfo
  {pages} {041108(R)} (\bibinfo {year} {2011})}\BibitemShut {NoStop}%
\bibitem [{\citenamefont {Corboz}\ \emph {et~al.}(2014)\citenamefont {Corboz},
  \citenamefont {Rice},\ and\ \citenamefont {Troyer}}]{corboz2014}%
  \BibitemOpen
  \bibfield  {author} {\bibinfo {author} {\bibfnamefont {P.}~\bibnamefont
  {Corboz}}, \bibinfo {author} {\bibfnamefont {T.~M.}\ \bibnamefont {Rice}},\
  and\ \bibinfo {author} {\bibfnamefont {M.}~\bibnamefont {Troyer}},\
  }\bibfield  {title} {\emph {\bibinfo {title} {{Competing States in the
  $t$-$J$ Model: Uniform $d$-Wave State versus Stripe State}}},\ }\href
  {https://doi.org/10.1103/PhysRevLett.113.046402} {\bibfield  {journal}
  {\bibinfo  {journal} {Phys. Rev. Lett.}\ }\textbf {\bibinfo {volume} {113}},\
  \bibinfo {pages} {046402} (\bibinfo {year} {2014})}\BibitemShut {NoStop}%
\end{thebibliography}
